\documentclass[12pt]{article}
\usepackage{indentfirst}
\usepackage{yfonts}
\usepackage{color}
\usepackage{mhchem}
\usepackage{xcolor}
\usepackage{harpoon}
\usepackage{cite}
\usepackage{hyperref}
\usepackage{cancel}
\hypersetup{colorlinks=true,linkcolor=red,anchorcolor=black,citecolor=green}
\usepackage[toc,page]{appendix}
\usepackage{amsfonts}
\usepackage{bbold}
\usepackage{textcomp}
\usepackage[DIV13]{typearea}
\usepackage{amsmath, amsthm, amssymb, mathtools,empheq,latexsym,dsfont}
\usepackage{bbm}
\usepackage{slashed, simplewick}
\usepackage[utf8]{inputenc}
\usepackage{graphicx,placeins}
\usepackage[export]{adjustbox} 
\usepackage{makeidx}
\usepackage[font=normalsize,labelfont=bf]{caption}
\usepackage{nicefrac}
\usepackage{subfigure}
\usepackage{array, bigdelim,multirow,multicol}
\usepackage[integrals]{wasysym}
\usepackage[normalem]{ulem}
\usepackage{fancybox}
\usepackage{bm}
\usepackage{float}
\usepackage{rotating}
\usepackage{colortbl}
\usepackage{booktabs}
\usepackage[top=2cm,textwidth=16.6cm,textheight=22.75cm]{geometry}
\usepackage{doi}
\usepackage{extarrows}
\usepackage{pifont}
\usepackage{mathrsfs}
\usepackage{lmodern}

\usepackage{braket}

\usepackage{tikz}

\usetikzlibrary{quotes,angles}
\graphicspath{{immagini/}}
\usetikzlibrary{arrows.meta}
\usetikzlibrary{decorations}
\usepackage{tikz-3dplot}
\usepackage{tikz-feynman}
\usepackage{mathrsfs}
\usepackage{enumerate}
\usetikzlibrary {arrows.meta}

\usepackage{framed}
\usepackage{longtable}
\usepackage{makecell}
\usepackage{arydshln}
\usepackage{tabularx}
\usepackage{makecell}
\usepackage{diagbox}
\allowdisplaybreaks  
\graphicspath{{0nbb/}}

\definecolor{Gray}{gray}{0.92}

\newcommand{\be}{\begin{equation}}
        \newcommand{\ee}{\end{equation}}
\newcommand{\bea}{\begin{eqnarray}}
        \newcommand{\eea}{\end{eqnarray}}

\def\lsi{\raise0.3ex\hbox{$<$\kern-0.75em\raise-1.1ex\hbox{$\sim$}}}
\def\gsi{\raise0.3ex\hbox{$>$\kern-0.75em\raise-1.1ex\hbox{$\sim$}}}
\newcommand{\lsim}{\mathop{\lsi}}

\newcommand{\ignore}[1]{}

\makeatletter
\renewcommand*{\@fnsymbol}[1]{\ensuremath{\ifcase#1\or *\or  \mathsection\or \ddagger\or
                \dagger\or \mathparagraph\or \|\or **\or \dagger\dagger
                \or \ddagger\ddagger \else\@ctrerr\fi}}
\makeatother

\makeatletter
\@addtoreset{equation}{section}
\makeatother

\graphicspath{{STm/}}
\date{}
\begin{document}
\title{\begin{center}
{\Large\bf  Probing the singlet--triplet scotogenic model with lepton flavor violation, gravitational waves and dark matter}
\end{center}}
\date{}
\author{Shi-Yu Li$^{a}$\footnote{E-mail: {\tt lishiyu@mail.ustc.edu.cn}}\,, \ Jun-Nan Lu$^{a}$\footnote{E-mail: {\tt
junnanlu@ustc.edu.cn}}\,,  \
Yang Zhang$^{b}$ \footnote{E-mail: {\tt
zhangyang2025@htu.edu.cn}}\,,\
Gui-Jun Ding$^{a,c}$\footnote{E-mail: {\tt
dinggj@ustc.edu.cn}}\ \\*[20pt]
\centerline{\begin{minipage}{\linewidth}
\begin{center}
$^a${\it \small Department of Modern Physics, and Anhui Center for fundamental sciences in theoretical physics,\\
University of Science and Technology of China, Hefei, Anhui 230026, China}\\[2mm]
$^b${\it \small  School of Physics, Henan Normal University, Xinxiang 453007, China}\\[2mm]
$^c${\it \small  College of Physics, Guizhou University, Guiyang 550025, China}\\[2mm]
\end{center}
\end{minipage}} \\[10mm]}
\maketitle
\begin{abstract}

We perform a comprehensive phenomenological analysis of neutrino masses, lepton-flavor violation, gravitational-wave signature induced by the first-order electroweak phase transition, and scalar dark matter within the framework of the singlet--triplet scotogenic model. We derive general analytical expressions for the one-loop Wilson coefficients of the penguin and box contributions to lepton-flavor-violating processes, considering generic scalar and fermion mediators. These general results are then applied to the singlet--triplet scotogenic model to calculate the branching ratios of $\mu\to e\gamma$ and $\mu\to e\overline{e}e$, as well as the coherent $\mu$--$e$ conversion rate. We systematically investigate the lepton-flavor-violation phenomenology and the constraints from current and future experiments. It is found that the coupling $\lambda_5$ plays a key role in connecting neutrino masses, lepton flavor violation, and dark matter direct detection. We also study the finite-temperature effective potential and the gravitational-wave signatures of the resulting first-order phase transition. By analyzing the thermal history of the model, we identify the viable parameter space allowing for a first-order electroweak phase transition in the early Universe and find that the portal parameters $\lambda_1^\Omega$ and $\mu_1$ play a crucial role in determining the transition strength. The corresponding stochastic gravitational-wave spectra can be probed by future gravitational-wave observatories. As a representative example, we identify a benchmark point that simultaneously satisfies the current constraints from lepton-flavor-violation and dark-matter experiments and generates a stochastic gravitational-wave spectrum that may be probed by the Laser Interferometer Space Antenna, the Big Bang Observer, and the Ultimate Deci-hertz Interferometer Gravitational-wave Observatory.  
\end{abstract}



\clearpage

\section{Introduction}

The Standard Model (SM) of particle physics has been remarkably successful in describing a wide range of experimental observations. But it does not provide a complete description of nature. In particular, numerous neutrino oscillation experiments have established nonzero neutrino masses and lepton flavor mixing, with the measured mass-squared splittings requiring at least two massive light neutrinos~\cite{McDonald:2016ixn,Kajita:2016cak}.  Cosmological observations also provide strong evidence for a nonbaryonic dark matter (DM) component~\cite{Planck:2018vyg,Bertone:2010zza,Cirelli:2024ssz}. However, the minimal renormalizable formulation of the SM offers no mechanism for generating the observed neutrino masses and no viable particle candidate for the DM of the Universe. 

To address these shortcomings of the SM, the scotogenic mechanism was proposed to link neutrino mass generation to DM~\cite{Tao:1996vb,Ma:2006km}. In this framework, neutrino masses arise radiatively through interactions with a dark sector, while an exact $\mathbb{Z}_2$ symmetry forbids their generation at tree level and ensures DM stability, see Refs.~\cite{Cai:2017jrq,Avila:2025qsc} for reviews.
However, in the simplest realization, loop effects may induce spontaneous breaking of the $\mathbb{Z}_2$ symmetry, compromising both DM stability and the radiative origin of neutrino masses~\cite{Bouchand:2012dx,Merle:2015ica,Merle:2015gea}. This potential problem motivates extending the particle content, as in the singlet--triplet scotogenic model~\cite{Hirsch:2013ola}, which introduces a Majorana fermion triplet $\Sigma$ and a real scalar triplet $\Omega$ in addition to the Majorana fermion singlet $N$ and the inert scalar doublet $\eta$. In this extension, the $\mathbb{Z}_2$ symmetry can remain unbroken up to high energies under the renormalization group evolution of the scalar-sector couplings~\cite{Merle:2016scw}. 
In addition to providing a framework for neutrino masses and DM, the enlarged particle content leads to a rich phenomenology, including lepton-flavor violation (LFV) and distinctive collider signatures and so on~\cite{Hirsch:2013ola,Merle:2016scw,Rocha-Moran:2016enp,Diaz:2016udz,Choubey:2017yyn,Restrepo:2019ilz,Avila:2019hhv,Batra:2022org,Karan:2023adm,Singh:2023eye,Lozano:2025tst}.

The DM phenomenology of the singlet--triplet scotogenic model has been extensively studied, with scalar and fermionic DM realizations explored in Refs.~\cite{Diaz:2016udz,Avila:2019hhv,Batra:2022org} and Refs.~\cite{Hirsch:2013ola,Restrepo:2019ilz,Karan:2023adm}, respectively. The Yukawa interactions responsible for radiative neutrino masses also induce charged LFV. The corresponding phenomenology was investigated in Ref.~\cite{Rocha-Moran:2016enp}, which considered $\mu\to e\gamma$, $\mu\to e\overline{e}e$, and coherent $\mu$--$e$ conversion and examined their dependence on neutrino mixing parameters and new-particle masses. The explicit analytical expressions for the Wilson coefficients arising from the photon-penguin diagrams were provided in~\cite{Rocha-Moran:2016enp}, while those for the $Z$-penguin and box coefficients were not presented. Such expressions clarify the mass and coupling dependence of LFV amplitudes, and facilitate independent verification and systematic analyses beyond specific numerical benchmarks.  The relic abundance, direct-detection prospects, and collider signatures of this model have been investigated in conjunction with neutrino-mass and LFV constraints~\cite{Diaz:2016udz,Choubey:2017yyn,Restrepo:2019ilz,Avila:2019hhv,Karan:2023adm,Lozano:2025tst}. Moreover, the extended scalar sector modifies the  finite-temperature effective potential of the SM and may allow a first-order phase transition (FOPT) at the electroweak scale, in contrast to the smooth crossover in the SM~\cite{DOnofrio:2015gop,DOnofrio:2014rug}. Such a phase transition could generate a stochastic gravitational-wave (GW) background through bubble nucleation and the subsequent dynamics of the thermal plasma~\cite{Hindmarsh:2013xza,Hindmarsh:2015qta,Hindmarsh:2017gnf,Caprini:2019egz,Hindmarsh:2020hop}, potentially yielding observable signals. This motivates our investigation of the prospects for detecting the resulting GWs.

In this work, we present a combined phenomenological study of LFV, phase-transition gravitational waves (PTGWs), and scalar DM in the singlet--triplet scotogenic model. We first derive complete analytical expressions for the one-loop Wilson coefficients of photon-penguin, $Z$-penguin, and box contributions mediated by generic scalar and fermion fields, and then apply them to the model. Our analytical results extend those of Ref.~\cite{Rocha-Moran:2016enp} by also providing explicit expressions for the $Z$-penguin and box coefficients. In the numerical analysis, we systematically study LFV for both neutrino mass orderings, imposing the relevant theoretical constraints and varying the neutrino oscillation parameters within their respective $3\sigma$ ranges. Specifically, we examine the predictions for $\mu\to e\gamma$, $\mu\to e\overline e e$, and $\mu^-+\mathrm{Ti}\to e^-+\mathrm{Ti}$, focusing on their dependence on the new-particle masses up to $100~\mathrm{TeV}$, the Casas-Ibarra parametrization angle $\varrho$, and the Majorana phase. The predicted rates are assessed against the current and future sensitivities from MEG~II~\cite{MEGII:2025gzr}, SINDRUM~\cite{SINDRUM:1987nra}, Mu3e~\cite{Mu3e:2020gyw,MartinPerez:2023vfv}, SINDRUM~II~\cite{SINDRUMII:1993gxf,SINDRUMII:2006dvw} and PRIME~\cite{prime}. 
A central objective of this work is to assess whether the extended scalar sector can realize an electroweak FOPT via bubble nucleation and generate a PTGW signal accessible to future space-based interferometers, including LISA~\cite{Sathyaprakash:2009xs}, evolved eLISA~\cite{Amaro-Seoane:2012aqc}, ALIA~\cite{Berry:2012im}, BBO and DECIGO~\cite{Yagi:2011wg}, TianQin~\cite{Hu:2024vvu}, Taiji~\cite{Luo:2021qji}, and Ultimate DECIGO~\cite{Takahashi:2004yr}. To address this question, we derive the zero- and finite-temperature effective potentials,  determine the nucleation temperature and the parameters characterizing the transition, and calculate the resulting stochastic GW spectra. Furthermore we identify the dominant scalar interactions that control the transition dynamics and the characteristic amplitudes and frequencies of the PTGW spectra.  To test the compatibility of the LFV and PTGW results with DM phenomenology, we present a representative benchmark that can accommodate the experimental data of neutrino oscillation and LFV observations  and produces a potentially observable PTGW signal. At the same parameter point, we evaluate the thermal relic abundance and spin-independent (SI) DM--nucleon scattering cross section of the scalar DM candidate and compare them with cosmological and direct-detection data. This benchmark illustrates that neutrino oscillation data, LFV constraints, potentially observable PTGWs, and viable scalar DM can be accommodated simultaneously in the singlet--triplet scotogenic model.

The paper is organized as follows.
Section~\ref{sec:model} introduces the field content, interactions, and particle mass spectra of the singlet--triplet scotogenic model, together with the one-loop neutrino mass matrix and its parametrization in terms of neutrino oscillation observables.
We give the analytical expressions of the Wilson coefficients and LFV observables relevant for the $e_j\to e_i\gamma$, $e_j\to e_i\overline{e}_ie_i$ and $\mu-e$ conversion in section~\ref{sec:LFV}.  The effective potential of this model and the dynamics of the phase-transition are discussed in section~\ref{sec:effV_PTGW}, and the approximate formulas for the energy density spectrum of the generated GW are presented. We summarize the theoretical and experimental constraints on the model parameters in section~\ref{sec:constraint}.  The comprehensive numerical analysis is performed for the phenomenology of LFV, PTGWs, and DM in section~\ref{sec:ph}. Finally we draw the conclusion in section~\ref{sec:con}. Appendix~\ref{app:lfv-observables} collects the effective amplitudes and generic LFV formulae. Appendices~\ref{app:general-penguins} and~\ref{app:general-box} derive the analytical formulae for the one-loop Wilson coefficients of the penguin and box contributions to LFV processes, respectively, with scalar and fermion mediators. The loop functions used in this article are listed in Appendix~\ref{ap:loop-function}.

\section{The singlet--triplet scotogenic model\label{sec:model}}
\begin{table}[htb!]
\centering
\begin{tabular}{| c | c | c | c | c | c | c | c | }
\hline\hline
& \multicolumn{3}{ c |  }{\quad SM \quad} &  \multicolumn{2}{ c | }{\quad new fermions \quad}  & \multicolumn{2}{ c |}{\quad new scalars \quad}  \\
\hline
    &   $L$   & $e$   & $\phi$   & $\Sigma$ &    $N$    & $\eta$ & $\Omega$ \\
\hline
Generations & \quad  3 & \quad  3 & \quad  1 & \quad  1 & \quad  1 & \quad  1 & \quad 1 \\
\hline
\hline
$\rm SU(3)_C$  & \quad   1    & \quad  1    & \quad   1     & \quad    1     &   \quad  1    & \quad    1    &   \quad  1     \\
$\rm SU(2)_L$  & \quad   2    & \quad  1    &  \quad   2     & \quad    3     &   \quad  1    & \quad    2    &   \quad  3     \\
$\rm U(1)_Y$        & \quad  $-1/2$  & \quad $-1$   & \quad    $1/2$   & \quad    0     &   \quad  0    & \quad    $1/2$  &  \quad   0     \\
$\mathbb{Z} _{2}$      & \quad  $+$   & \quad $+$   & \quad   $+$    & \quad   $-$    & \quad    $-$   & \quad   $-$   &  \quad  $+$    \\
\hline
$\rm L$        & \quad   1    & \quad  1    &   \quad  0     & \quad    0     & \quad    0    & \quad    $-1$    &  \quad   0     \\
\hline\hline
\end{tabular}
\caption{Particle content and quantum numbers of the singlet--triplet scotogenic model. The charge assignments of the fields under the global lepton number symmetry ($\rm L$) are also shown. }
\label{tab:fields_charges_1}
\end{table}

The singlet--triplet scotogenic model~\cite{Hirsch:2013ola} extends the SM by introducing two new fermionic fields, $\Sigma$ and $N$, together with two additional scalar multiplets, $\eta$ and $\Omega$, as summarized in table~\ref{tab:fields_charges_1}. Notice that $\phi$ is the SM Higgs doublet. The $SU(2)_L$ triplet $\Sigma$ and the singlet $N$ are both Majorana fermions with vanishing hypercharge ($Y=0$). The scalar doublet $\eta$ carries hypercharge $+1/2$, while the real scalar triplet $\Omega$ is neutral under $U(1)_Y$. Besides the gauge symmetry $SU(3)_{\rm c}\otimes SU(2)_L\otimes U(1)_Y$, the model implements a discrete $\mathbb{Z}_2$ symmetry that guarantees the radiative origin of neutrino masses and stabilizes the DM candidate. Under this symmetry, the fields $\Sigma$, $N$, and $\eta$ are assigned odd parity ($\mathbb{Z}_2=-1$), whereas $\Omega$ and all SM fields transform with even parity ($\mathbb{Z}_2=+1$). The $SU(2)_L$ triplet fields $\Omega$, $\Sigma$, and its 
charge-conjugated counterpart $\Sigma^c$ can be written in matrix 
form as
\begin{equation}
\Omega = \left( \begin{array}{cc}
\frac{\widetilde{\Omega}^0}{\sqrt{2}} ~&~ \Omega^+ \\
\Omega^- ~&~ -\frac{\widetilde{\Omega}^0}{\sqrt{2}}
\end{array} \right)\,,\qquad 
\Sigma =
\begin{pmatrix}
\frac{\Sigma^0}{\sqrt{2}} ~&~ \Sigma^+ \\
\Sigma^- ~&~ -\frac{\Sigma^0}{\sqrt{2}}
\end{pmatrix}\,,\qquad
\Sigma^c =
\begin{pmatrix}
\frac{\Sigma^{0c}}{\sqrt{2}} ~&~ \Sigma^{+c}\\
\Sigma^{-c} ~&~ -\frac{\Sigma^{0c}}{\sqrt{2}}
\end{pmatrix}\,,
\label{eq:triplets}
\end{equation}
where $\widetilde{\Omega}^{0}$ is assumed to be a real field and $\Omega^{+}=(\Omega^{-})^{*}$. With the field content and symmetry assignments specified above, the fermionic sector of the Lagrangian, including the Hermitian conjugate (h.c.), reads
\begin{eqnarray}
\label{eq:laginteraction}
\mathcal{L} &\supset& 
-y_{ij}\,\overline{L_i}\, e_j\, \phi
- Y_{N,i}\, (\overline{L_i}\,\tilde{\eta})\,N
- Y_{\Sigma,i}\, \overline{L_i^c}\, i\sigma_{2}\Sigma\, \eta
- Y_{\Omega}\, {\rm Tr}\!\left[\overline{\Sigma}\,\Omega \right] N \nonumber\\
&-& \frac{1}{2} M_{\Sigma}\, {\rm Tr}\!\left(i\sigma_{2}\overline{\Sigma^{c}}\, i\sigma_{2}\Sigma \right)
- \frac{1}{2}M_N\, \overline{N^{c}} N 
+ {\rm h.c.}\,.
\end{eqnarray}
Here $\tilde{\eta}\equiv i\sigma_2\eta^\ast$, and $i,j=1,2,3$ label the lepton flavor indices. The Yukawa couplings $Y_{N,i}$, $Y_{\Sigma,i}$ and $Y_\Omega$ parametrize the interactions between the new fermionic states and the scalar sector, while the last two terms encode the Majorana mass contributions of the fermionic triplet $\Sigma$ and the singlet $N$, respectively. 

\subsection{The scalar potentials}
The tree-level scalar potential $\mathcal{V}$, which is invariant under the $SU(2)_L \otimes U(1)_Y \otimes \mathbb{Z}_2$ symmetry and includes mass terms for the scalar fields $\phi$, $\eta$, and $\Omega$, suggests non-zero vacuum expectation values (VEVs) for $\phi$ and $\Omega$. It is given by:
\begin{eqnarray}
 V &=&-m_{\phi}^2\phi^{\dag}\phi+m_{\eta}^{2}\eta^{\dag}\eta -\frac{m_{\Omega}^{2}}{2}\rm Tr \left(\Omega^{\dag}\Omega\right) + \frac{\lambda_1}{2}\left(\phi^{\dag}\phi\right)^{2}\nonumber\\
&+&\frac{\lambda_2}{2}\left(\eta^{\dag}\eta\right)^{2}+\lambda_{3}\left(\phi^{\dag}\phi\right)\left(\eta^{\dag}\eta\right) + \lambda_{4}\left(\phi^{\dag}\eta\right)\left(\eta^{\dag}\phi\right)+\left[ \frac{\lambda_5}{2}\left(\phi^{\dag}\eta\right)^2 + \text{h.c} \right] \nonumber \\
&+& \mu_{1}\phi^{\dag}\Omega\phi + \mu_{2}\eta^{\dag}\Omega\eta +\frac{\lambda_{1}^{\Omega}}{2}\left(\phi^{\dag}\phi \right) \rm Tr\left(\Omega^{\dag}\Omega\right) \nonumber\\
&+&\frac{\lambda_{2}^{\Omega}}{4}\left[ \rm Tr\left(\Omega^{\dag}\Omega\right) \right]^2
+\frac{\lambda_{3}^{\Omega}}{2}\left(\eta^{\dag}\eta \right) \rm Tr\left(\Omega^{\dag}\Omega\right)\,.
\label{eq:scalarpot}
\end{eqnarray}
Here the $\lambda$'s denote dimensionless quartic couplings and the $\mu$'s are cubic parameters with mass dimension. Although the scalar $\eta$ carries the same gauge quantum numbers as the Higgs doublet $\phi$, its odd parity under the discrete symmetry $\mathbb{Z}_2$ requires that an even number of $\eta$ fields appear in any gauge-invariant interaction. It is worth noting that the $\lambda_5$ term provides the only source of lepton-number violation in this model. The VEVs of the neutral scalar fields are assumed to be  
\begin{equation}
\langle \phi^0 \rangle = \frac{v_\phi}{\sqrt{2}} \,, \qquad 
\langle \widetilde{\Omega}^0 \rangle = v_\Omega\, ,
\label{eq:vevs}
\end{equation}
where $\phi^{0}$ is the neutral component of the Higgs doublet $\phi$. Here 
$v_\phi=246~\text{GeV},\, v_\Omega \neq 0$ and the restriction of $\mathbb{Z}_2$ symmetry implies $\braket{\eta^0}=0$. The scalar multiplets can be expanded around their VEVs as  
\begin{eqnarray}
\phi &=&\begin{pmatrix}
    \phi^{+} \\ \phi^{0}
\end{pmatrix}=
\begin{pmatrix}
\phi^{+} \\
\left( v_{\phi} + h_{0} + i\,\phi_{I} \right)/\sqrt{2}
\end{pmatrix}, \\
\Omega &=&
\begin{pmatrix}
\left( v_{\Omega} + \Omega_{0} \right)/\sqrt{2} & \Omega^{+} \\
\Omega^{-} & -\left( v_{\Omega} + \Omega_{0} \right)/\sqrt{2}
\end{pmatrix}\,,\\
\eta &=& \begin{pmatrix}
    \eta^{+} \\ \eta^{0}
\end{pmatrix}=
\begin{pmatrix}
\eta^{+} \\
\left( \eta_{R} + i\,\eta_{I} \right)/\sqrt{2}
\end{pmatrix}\,.
\end{eqnarray}
Here $\phi^\pm$, $\Omega^\pm$, and $\eta^\pm$ denote the charged partners of the neutral fields $h_0$, $\Omega_0$, and $\eta_R$, which are even under combined charge conjugation and parity (CP), as well as of the neutral CP-odd fields $\phi_I$ and $\eta_I$. For the real triplet field, the reality condition implies $\Omega^{\pm *} = \Omega^{\mp}$. 
These VEVs are restricted by the tadpole equations
\begin{eqnarray}
\frac{\partial V}{\partial \phi^{0*}}\Big|_{\text{vac}} &=& -m_\phi^2 \, v_\phi + \frac{1}{2} \lambda_1 v_\phi^3 + \frac{1}{2} \lambda_1^\Omega v_\phi v_\Omega^2 - \frac{1}{\sqrt{2}} \, v_\phi v_\Omega \, \mu_1 = 0 \, , \label{eq:tad1} \\
\frac{\partial V}{\partial \Omega^{0}}\Big|_{\text{vac}}&=& -m_\Omega^2 \, v_\Omega + \lambda_2^\Omega v_\Omega^3 + \frac{1}{2} \lambda_1^\Omega v_\phi^2 v_\Omega - \frac{1}{2 \sqrt{2}} v_\phi^2 \, \mu_1 = 0 \, , \label{eq:tad2}
\end{eqnarray}
obtained from the scalar potential in Eq.~\eqref{eq:scalarpot}. Then we can obtain
\begin{eqnarray}\label{eq:mphisq}
\nonumber m_{\phi}^{2}&=&\frac{1}{2} \lambda_1 v_\phi^2 + \frac{1}{2} \lambda_1^\Omega  v_\Omega^2 - \frac{1}{\sqrt{2}} \,  v_\Omega \, \mu_1\,,\\
m_\Omega^2&=&\lambda_2^\Omega v_\Omega^2 + \frac{1}{2} \lambda_1^\Omega v_\phi^2 - \frac{1}{2 \sqrt{2}} \frac{v_\phi^2}{v_{\Omega}} \, \mu_1\,.\label{eq:mphimOmg}
\end{eqnarray}

Both $\phi$ and $\Omega$ transform non-trivially under $SU(2)_L$, and their VEVs contribute to the masses of the electroweak gauge bosons, i.e.

\begin{eqnarray}
m_W^2 &=& \frac{1}{4} g^2 \left( v_\phi^2 + 4\, v_\Omega^2 \right), \qquad 
m_Z^2 = \frac{1}{4} (g^2 + g'^2)\, v_\phi^2 \,,\label{eq:mass-ZW}
\end{eqnarray}
where $g$ and $g'$ are the gauge couplings of $SU(2)_L$ and $U(1)_Y$ respectively. Thus, the triplet VEV $v_\Omega$ induces an additional contribution to the $W$-boson mass while leaving the $Z$-boson mass unchanged with respect to the SM prediction. Consequently, the electroweak $\rho$ parameter receives a deviation from unity, which is strongly constrained by precision measurements. The current electroweak fit requires $v_\Omega \lesssim 4~\text{GeV}$ at the $3\sigma$ level.

\subsection{The mass spectra of scalars}
The mass matrix for the neutral CP-even sector, $\left( h_0\, ,\, \Omega_0 \right)$, reads
\begin{eqnarray}
M_S^2 &=& 
\begin{pmatrix}
-m_\phi^2 + \frac{3}{2}\lambda_1 v_\phi^2 + \frac{1}{2}\lambda_1^\Omega v_\Omega^2 
     - \frac{1}{\sqrt{2}} v_\Omega \mu_1 &
\lambda^\Omega_1 v_\phi v_\Omega - \frac{1}{\sqrt{2}} v_\phi \mu_1\\[2mm]
\lambda^\Omega_1 v_\phi v_\Omega - \frac{1}{\sqrt{2}} v_\phi \mu_1 &
-m_\Omega^2 + \frac{1}{2}\lambda^\Omega_1 v_\phi^2 + 3\lambda^\Omega_2 v_\Omega^2
\end{pmatrix}\,.
\end{eqnarray}
The mass term is written as
\begin{equation}
\mathcal{L} \supset \frac{1}{2}(h_0,\Omega_0)\, M_S^2 \,(h_0,\Omega_0)^T\,.
\end{equation}
The diagonalization of matrix $M_S^2$ leads to the mass eigenstates $S_{1,2}$ where the lighter eigenstate $S_2$ corresponds to the observed $125\,\mathrm{GeV}$ Higgs boson, while the heavier eigenstate $S_1$ is a new scalar beyond the Standard Model (BSM). 
The mass eigenstates $S_{1,2}$ are linear combinations of $h_0$ and $\Omega_0$,
\begin{eqnarray}
    \begin{pmatrix}
        h_0\\\Omega_0
    \end{pmatrix}&=\begin{pmatrix}
        \cos\theta_S~&~-\sin\theta_S\\
        \sin\theta_S~&~\cos\theta_S
    \end{pmatrix}\begin{pmatrix}
        S_1\\S_2
    \end{pmatrix}
\end{eqnarray}
where the rotation angle $\theta_S$ is
\begin{eqnarray}
  \tan2\theta_S&=&\frac{2\lambda_1^\Omega v_\phi v_\Omega-\sqrt{2}\mu_1v_\phi}{\hat{m}_\phi^2-\hat{m}_\Omega^2}\,,\nonumber\\
   \sin2\theta_S&=&\frac{2\lambda_1^\Omega v_\phi v_\Omega-\sqrt{2}\mu_1v_\phi}{\sqrt{\left(\hat{m}_\phi^2-\hat{m}_\Omega^2\right)^2+4\left(\lambda_1^\Omega v_\phi v_\Omega-\frac{1}{\sqrt{2}}\mu_1v_\phi\right)^2}}\,,\nonumber\\
    \cos2\theta_S&=&\frac{\hat{m}_\phi^2-\hat{m}_\Omega^2}{\sqrt{\left(\hat{m}_\phi^2-\hat{m}_\Omega^2\right)^2+4\left(\lambda_1^\Omega v_\phi v_\Omega-\frac{1}{\sqrt{2}}\mu_1v_\phi\right)^2}}\,. 
\end{eqnarray}
The mass eigenvalues are given by
\begin{eqnarray}
m_{S_1}^2&=&\frac{1}{2}\Big[\hat{m}_\phi^2+\hat{m}_\Omega^2+\sqrt{\left(\hat{m}_\phi^2-\hat{m}_\Omega^2\right)^2+4\left(\lambda_1^\Omega v_\phi v_\Omega-\frac{1}{\sqrt{2}}\mu_1v_\phi\right)^2}\Big]\,,\nonumber\\
m_{S_2}^2&=&\frac{1}{2}\Big[\hat{m}_\phi^2+\hat{m}_\Omega^2-\sqrt{\left(\hat{m}_\phi^2-\hat{m}_\Omega^2\right)^2+4\left(\lambda_1^\Omega v_\phi v_\Omega-\frac{1}{\sqrt{2}}\mu_1v_\phi\right)^2}\Big]\,,\label{eq:mS}
\end{eqnarray}
with
\begin{equation}
\hat{m}_\phi^2\equiv-m_\phi^2+\frac{3\lambda_1v_\phi^2}{2}+\frac{\lambda_1^\Omega v_\Omega^2}{2}-\frac{\mu_1v_\Omega}{\sqrt{2}}\,,\qquad
 \hat{m}_\Omega^2\equiv-m_\Omega^2+\frac{\lambda_1^\Omega v_\phi^2}{2}+3\lambda_2^\Omega v_\Omega^2\,.
\end{equation}
The charged scalar mass matrix in the basis $(\phi^\pm,\, \Omega^\pm)$ takes the form
\begin{eqnarray}
M_{H^\pm}^2 &=&
\begin{pmatrix}
-m_\phi^2
+ \frac12 \lambda_1 v_\phi^2
+ \frac12 \lambda^\Omega_1 v_\Omega^2
+ \frac{1}{\sqrt{2}} v_\Omega \mu_1
&
\frac{1}{\sqrt{2}}\, v_\phi \mu_1 \\[2mm]
\frac{1}{\sqrt{2}}\, v_\phi \mu_1
&
-m_\Omega^2
+ \frac12 \lambda^\Omega_1 v_\phi^2
+ \lambda^\Omega_2 v_\Omega^2
\end{pmatrix}\,.
\end{eqnarray}
The corresponding mass term is
\begin{equation}
\mathcal{L} \supset (\phi^{-},\Omega^{-})\, M_{H^\pm}^2\, (\phi^{+},\Omega^{+})^{T}\,.
\end{equation}
The mass eigenstates $\kappa_{1,2}^+$ are linear combinations of $\phi^+$ and $\Omega^+$,
\begin{eqnarray}
    \begin{pmatrix}
 \phi^+\\\Omega^+
\end{pmatrix}&=\begin{pmatrix}
\cos\theta_\kappa~&~-\sin\theta_\kappa\\
\sin\theta_\kappa~&~\cos\theta_\kappa
\end{pmatrix}\begin{pmatrix}
\kappa_1^+\\\kappa_2^+
\end{pmatrix}
\end{eqnarray}
where the rotation angle $\theta_\kappa$ is
\begin{eqnarray}
  \tan2\theta_\kappa=\frac{\sqrt{2}\mu_1v_\phi}{\widetilde{m}_\phi^2-\widetilde{m}_\Omega^2}\,,
   \sin2\theta_\kappa=\frac{\sqrt{2}\mu_1v_\phi}{\sqrt{\left(\widetilde{m}_\phi^2-\widetilde{m}_\Omega^2\right)^2+2\left(\mu_1v_\phi\right)^2}}\,,
    \cos2\theta_\kappa=\frac{\widetilde{m}_\phi^2-\widetilde{m}_\Omega^2}{\sqrt{\left(\widetilde{m}_\phi^2-\widetilde{m}_\Omega^2\right)^2+ 2\mu^2_1v^2_\phi}}\,. \nonumber\\
\end{eqnarray}
The mass eigenvalues are determined to be
\begin{eqnarray}
    m_{\kappa_1^+}^2&=&\frac{1}{2}\Big[\widetilde{m}_\phi^2+\widetilde{m}_\Omega^2+\sqrt{\left(\widetilde{m}_\phi^2-\widetilde{m}_\Omega^2\right)^2+2\mu_1^2v_\phi^2}\Big]\,,\nonumber\\
     m_{\kappa_2^+}^2&=&\frac{1}{2}\Big[\widetilde{m}_\phi^2+\widetilde{m}_\Omega^2-\sqrt{\left(\widetilde{m}_\phi^2-\widetilde{m}_\Omega^2\right)^2+2\mu_1^2v_\phi^2}\Big]\,,\label{eq:mkappa}
\end{eqnarray}
with
\begin{equation}
\widetilde{m}_\phi^2\equiv-m_\phi^2+\frac{\lambda_1^\Omega v_\Omega^2}{2}+\frac{\lambda_1v_\phi^2}{2}+\frac{\mu_1v_\Omega}{\sqrt{2}}\,,\qquad
        \widetilde{m}_\Omega^2\equiv-m_\Omega^2+\frac{\lambda_1^\Omega v_\phi^2}{2}+\lambda_2^\Omega v_\Omega^2\,.
\end{equation}
Notice that the field $\kappa^{\pm}_2$ is the Goldstone mode eaten by the $W^\pm$ boson. Moreover, the CP-odd neutral scalar $\phi_I$ has a mass
\begin{equation}
m_{\phi_I}^{2}
=
-m_\phi^2 
+ \frac12 \lambda_1 v_\phi^2
+ \frac12 \lambda_1^\Omega v_\Omega^2
- \frac{1}{\sqrt{2}}\, v_\Omega \mu_1 \, .\label{eq:mphiI}
\end{equation}
This pseudoscalar corresponds to the Goldstone boson eaten by the $Z$ boson. The scalar $\eta$ is odd under $\mathbb{Z}_2$, whereas $\phi$ and $\Omega$ are even. The scalar field $\eta$ doesn't develop VEV and the $\mathbb{Z}_2$ symmetry is still preserved after electroweak symmetry breaking (EWSB). As a consequence, there is no mixture between $\eta$ and $\phi$, $\Omega$.
The masses of the components of the inert doublet $\eta$ are
\begin{eqnarray}
 m_{\eta_R}^2 &=& m_{\eta}^2 
+ \frac12(\lambda_3 + \lambda_4 + \lambda_5)\, v_\phi^2
+ \frac12 \lambda_{3}^{\Omega} v_\Omega^2
- \frac{1}{\sqrt{2}}\, v_\Omega \mu_2\,, 
\label{eq:metaR}
\\[2mm]
m_{\eta_I}^2 &=& m_{\eta}^2 
+ \frac12(\lambda_3 + \lambda_4 - \lambda_5)\, v_\phi^2
+ \frac12 \lambda_{3}^{\Omega} v_\Omega^2
- \frac{1}{\sqrt{2}}\, v_\Omega \mu_2\,, \label{eq:metaI}\\[2mm]
m_{\eta^\pm}^2 &=& m_{\eta}^2 
+ \frac12 \lambda_3 v_\phi^2
+ \frac12 \lambda^\Omega_{3} v_\Omega^2
+ \frac{1}{\sqrt{2}}\, v_\Omega \mu_2\,.\label{eq:metapm}
\end{eqnarray}
The masses of $\eta_R$ and $\eta_I$ are split by the dimensionless coupling $\lambda_5$, which uniquely controls the mass difference between the CP-even and CP-odd neutral components of $\eta$. As a result, the two neutral components of the inert doublet are no longer degenerate with each other after EWSB, allowing the physical CP-even and CP-odd scalars to be clearly distinguished.

\subsection{The mass spectrum of fermions}

In this model, two Majorana fermions, $N$ and $\Sigma$, are introduced.  
The $Z_{2}$-odd neutral fields $\Sigma^0$ and $N$ mix through the Yukawa coupling $Y_\Omega$ and the triplet VEV $v_\Omega$.  
In the basis $\left(\Sigma_0^{c},\, N\right)$, the Majorana mass term can be extracted from Eq.~\eqref{eq:laginteraction} as
\begin{equation}
-\frac{1}{2}\,(\overline{\Sigma_{0}},\,\overline{N^{c}})\,
  M_{\chi}\,
  \begin{pmatrix} \Sigma_{0}^{c} \\ N \end{pmatrix}\,,
\end{equation}
where the symmetric Majorana mass matrix is 
\begin{equation}
M_\chi =
\begin{pmatrix}
M_\Sigma ~&~ Y_\Omega v_\Omega \\
Y_\Omega v_\Omega ~&~ M_N
\end{pmatrix}\,.\label{eq:Mchi}
\end{equation}
For real $Y_{\Omega}$, the matrix $M_\chi$ is diagonalized by an orthogonal transformation $V(\theta)$,
\begin{equation}
  V(\theta)^{T} M_\chi V(\theta)
  = \text{diag}\!\left(m_{\chi_1},\,m_{\chi_2}\right)\,,
\end{equation}
with
\begin{equation}
V(\theta)=\begin{pmatrix}
\cos\theta ~&~ \sin\theta \\
-\sin\theta ~&~ \cos\theta
\end{pmatrix}\,.
\end{equation}
The mixing angle $\theta$ and the eigenvalues $m_{\chi_{1,2}}$ are given by
\begin{eqnarray}
\tan 2\theta
=\frac{2 Y_{\Omega} v_{\Omega}}{M_{N} - M_{\Sigma}}\,, 
\quad\sin 2\theta
=\frac{2 Y_{\Omega} v_{\Omega}}{\sqrt{\left(M_\Sigma - M_N\right)^2 + 4 Y_\Omega^2 v_\Omega^2}}\,,\quad\cos 2\theta
=\frac{M_{N} - M_{\Sigma}}{\sqrt{\left(M_\Sigma - M_N\right)^2 + 4 Y_\Omega^2 v_\Omega^2}}\,,\nonumber\\\label{eq:2th}
\end{eqnarray}
and
\begin{eqnarray}
m_{\chi_{1,2}}
= \frac{1}{2}\left[M_{\Sigma}+M_{N}
\mp \sqrt{\left(M_\Sigma - M_N\right)^2 + 4 Y_\Omega^2 v_\Omega^2}\,\right]\,.
\label{eq:mchi2}
\end{eqnarray}
Accordingly, the Majorana mass eigenstates $\chi_{1,2}$ are related to $\Sigma_0^{c}$ and $N$ through $V(\theta)$:
\begin{equation}
  \label{eq:relation_eigenstates}
  \begin{pmatrix} \chi_1 \\ \chi_2 \end{pmatrix}
  = V(\theta)^{T}
  \begin{pmatrix} \Sigma_{0}^{c} \\ N \end{pmatrix}
  =
  \begin{pmatrix}
  \cos\theta ~&~ -\sin\theta \\
  \sin\theta ~&~ \cos\theta
  \end{pmatrix}
  \begin{pmatrix} \Sigma_{0}^{c} \\ N \end{pmatrix}\,.
\end{equation}
From Eq.~\eqref{eq:laginteraction}, we know that the charged fermion $\Sigma^{\pm}$ has no mixture with other fermions and its mass is equal to $M_{\Sigma}$.

\subsection{Light neutrino masses\label{sec:numass}}
\begin{figure}[htbp]
\centering
\includegraphics[width=1\linewidth]{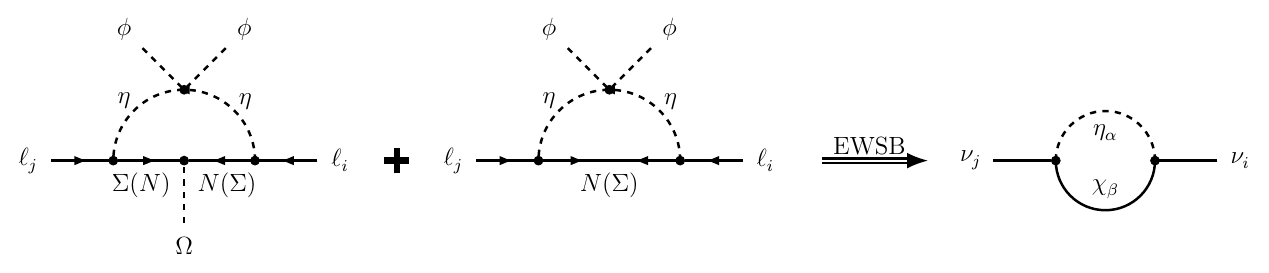}
\caption{Neutrino mass generation mechanism in the singlet--triplet scotogenic model. The left panel displays the one-loop Feynman diagram at the SM scale, while the right panel shows the corresponding one-loop diagrams after EWSB. Here $\eta_\alpha$ represents either $\eta_R$ or $\eta_I$, as their mass splitting arises only after EWSB. The fermionic state $\chi_\beta$ denotes the EWSB-induced mixed eigenstate, corresponding to either $\chi_1$ or $\chi_2$. }
    \label{fig:numass}
\end{figure}
The singlet--triplet scotogenic model generates neutrino masses at one loop, as shown in figure~\ref{fig:numass}. At low energies, the neutrino mass matrix can be expressed as
\begin{eqnarray}
(\mathcal{M}_{\nu})_{ij}&=&\left(h^{*}\Lambda h^{\dagger} \right)_{ij}\,, \label{eq:mnu}
\end{eqnarray}
where $h$ is a $3\times2$ matrix and its elements are given by
\begin{eqnarray}
(h)_{i1}=-\frac{Y_{\Sigma,i}^*}{\sqrt{2}}\cos\theta-Y_{N,i}\sin\theta \,,~~~ (h)_{i2}=-\frac{Y_{\Sigma,i}^*}{\sqrt{2}}\sin\theta+Y_{N,i}\cos\theta\,,\label{eq:h-YV}
\end{eqnarray}
with $i=1, 2, 3$, 
and $\Lambda$ in Eq.~\eqref{eq:mnu} is a diagonal two-dimensional matrix given by
\begin{equation}
\Lambda = \left( \begin{array}{ccc} \Lambda_{1} & 0  \\ 0 & \Lambda_{2} \end{array} \right)\,,\quad \Lambda_\alpha\equiv \frac{m_{\chi_\alpha}}{32\pi^{2}}\left( -\frac{m^{2}_{\eta_{R}}\text{ln}(\frac{m^{2}_{\eta_{R}}}{m^{2}_{\chi_\alpha}})}{m^{2}_{\eta_{R}}-m^{2}_{\chi_\alpha}} + \frac{m^{2}_{\eta_{I}}\text{ln}(\frac{m^{2}_{\eta_{I}}}{m^{2}_{\chi_\alpha}})}{m^{2}_{\eta_{I}}-m^{2}_{\chi_\alpha}}\right)\,.\label{eq:Lambda12}
\end{equation}
We note that in the limit $\lambda_5 \to 0$, the scalar masses 
$m_{\eta_R}$ and $m_{\eta_I}$, given in 
Eqs.~\eqref{eq:metaR} and \eqref{eq:metaI}, become 
degenerate. As a result, the neutrino masses vanish due to an exact 
cancellation between the $\eta_R$ and $\eta_I$ loop contributions. 
This behavior is expected, since the condition 
$m_{\eta_R}^2 = m_{\eta_I}^2$ corresponds to $\lambda_5 = 0$, under 
which a conserved lepton number can be defined. Consequently, a 
small value $\lambda_5 \ll 1$ is technically natural in the sense 
of 't~Hooft~\cite{tHooft:1979rat}, as the limit $\lambda_5 \to 0$ 
enhances the symmetry of the theory. In the limit $\lambda_{5}\ll 1$, we have
\begin{equation}
\Lambda_{\alpha} \approx   \frac{\lambda_{5}v^{2}_{\phi}m_{\chi_\alpha}}{32\pi^{2}}  
\frac{m^{2}_{\chi_\alpha}-m_{\eta_{0}}^{2}+m^{2}_{\chi_\alpha}\ln(\frac{m_{\eta_{0}}^{2}}{m^{2}_{\chi_\alpha}})}{\left(m_{\eta_{0}}^{2}-m^{2}_{\chi_\alpha}\right)^{2}}\,,\label{eq:Lam-lam5}
\end{equation}
with
\begin{eqnarray}
m_{\eta_0}^2 &=& m_{\eta}^2 
+ \frac12(\lambda_3 + \lambda_4 )\, v_\phi^2
+ \frac12 \lambda_{3}^{\Omega} v_\Omega^2
- \frac{1}{\sqrt{2}}\, v_\Omega \mu_2\,.
\end{eqnarray}
If $m_{\chi_{\alpha}}^{2}\gg m^{2}_{\eta_{0}}$, then
\begin{equation}
 \Lambda_{\alpha} \simeq \frac{\lambda_{5}v_{\phi}^{2}}{32\pi^{2}m_{\chi_\alpha}} \left( 
 1+\ln \frac{m_{\eta_{0}}^{2}}{m_{\chi_{\alpha}}^{2}}\right)\,. \label{eq:Lam-chi-gg-eta}  
\end{equation}
If $m_{\chi_{\alpha}}^{2}\ll m^{2}_{\eta_{0}}$, then
\begin{equation}
 \Lambda_{\alpha} \simeq -\frac{\lambda_{5}v_{\phi}^{2}}{32\pi^{2}} \frac{m_{\chi_{\alpha}}}{m_{\eta_{0}}^{2}}\,.  \label{eq:Lam-chi-ll-eta} 
\end{equation}
If $m_{\chi_{\alpha}}^{2}\simeq m^{2}_{\eta_{0}}$, then
\begin{equation}
 \Lambda_{\alpha} \simeq -\frac{\lambda_{5}v_{\phi}^{2}}{64\pi^{2}} \frac{1}{m_{\chi_{\alpha}}}\,.  \label{eq:Lam-chi-eq-eta} 
\end{equation}
Since there are only two fermion fields $N$ and $\Sigma$ mediating the neutrino mass, the lightest neutrino is massless. The neutrino mass matrix $\mathcal{M}_{\nu}$ is diagonalized as
\begin{equation}
  U^{T}\mathcal{M}_{\nu} U= \widehat{\mathcal{M}_{\nu}}=\left( \begin{array}{ccc} m_{1} & 0 & 0 \\ 0 & m_{2} & 0 \\ 0 & 0 & m_{3} \end{array} \right)\,,
\end{equation}
where $m_1=0$ for normal ordering (NO) and $m_3=0$ for inverted ordering (IO), and $U$ is the Pontecorvo--Maki--Nakagawa--Sakata (PMNS) matrix. We adopt the standard parametrization of the lepton mixing matrix~\cite{ParticleDataGroup:2024cfk},
\begin{equation}\label{eq:PMNS}
U=\left(\begin{array}{ccc}
c_{12}c_{13}  &   s_{12}c_{13}   &   s_{13}e^{-i\delta_{CP}}  \\
-s_{12}c_{23}-c_{12}s_{13}s_{23}e^{i\delta_{CP}}   &  c_{12}c_{23}-s_{12}s_{13}s_{23}e^{i\delta_{CP}}  &  c_{13}s_{23}  \\
s_{12}s_{23}-c_{12}s_{13}c_{23}e^{i\delta_{CP}}   & -c_{12}s_{23}-s_{12}s_{13}c_{23}e^{i\delta_{CP}}  &  c_{13}c_{23}
\end{array}\right)\text{diag}(1,e^{i\alpha_{21}},e^{i\alpha_{31}})\,,
\end{equation}
where $c_{ij}\equiv \cos\theta_{ij}$, $s_{ij}\equiv \sin\theta_{ij}$, $\delta_{CP}$ is the Dirac CP violation phase, and $\alpha_{21,31}$ are Majorana CP phases which appear only if neutrinos are Majorana particles. If the lightest neutrino is massless, there is a single Majorana phase $\alpha_{21}$ and the other phase, $\alpha_{31}$, is unphysical. Thus, the diagonal phase matrix in the above equation can be replaced by $\text{diag}(1,e^{i\alpha_{21}},1)$.

A neutrino mass matrix as the one in Eq.~\eqref{eq:mnu} formally resembles that obtained in the standard type-I seesaw with two generations of right-handed neutrinos. In this case we can make use of an adapted Casas-Ibarra parameterization~\cite{Casas:2001sr,Ibarra:2003up} to obtain an expression for the
Yukawa matrix $h$,
\begin{eqnarray}
h=U\sqrt{\widehat{\mathcal{M}}_\nu}R\sqrt{\Lambda}^{-1}\,, \label{eq:CI}
\end{eqnarray}
where $R$ is a $3 \times 2$ complex matrix such that $R R^T =\mathbb{1}$, where $\mathbb{1}=\text{diag}(0,1,1)$ for NO and $\mathbb{1}=\text{diag}(1,1,0)$ for IO.

It has, however, enough freedom to accommodate both neutrino spectra, NO and IO, and the form of the complex $R$ matrix introduced in Eq.~\eqref{eq:CI} depends on this
choice~\cite{Ibarra:2003up},
\begin{eqnarray}
R &=& \left( \begin{array}{cc} 0 &~ 0 \\ \cos \varrho &~ \sin \varrho \\ -\sin \varrho &~ \cos \varrho
\end{array} \right) \quad \text{\bf for NO} \quad (m_1 = 0) \, ,\label{eq:R-NO} \\
R &=& \left( \begin{array}{cc}
\cos \varrho &~ \sin \varrho \\
-\sin \varrho &~ \cos \varrho \\
0 &~ 0
\end{array} \right) \quad \text{\bf for IO} \quad (m_3 = 0) \, .\label{eq:R-IO}
\end{eqnarray}
The two non-vanishing rows of $R$ define a $2\times2$ complex orthogonal block. This block belongs to one of the two disconnected branches of $O(2,\mathbb{C})$, characterized by determinant $\pm1$. In the present work we adopt the representative forms in Eqs.~\eqref{eq:R-NO} and~\eqref{eq:R-IO}, for which the corresponding $2\times2$ block has determinant $+1$. The alternative branch, with determinant $-1$, can be accounted for by shifting the relevant Majorana phase by $\pi$. Therefore, following~\cite{Granelli:2022eru}, we extend the range of the Majorana phase $\alpha_{21}$ from $[0,\pi]$ to $[0,2\pi]$, which effectively covers both branches.

\section{Lepton flavor violation\label{sec:LFV}}

LFV processes, such as $\mu \to e\gamma$, can generically be induced by the new degree of freedom responsible for massive neutrinos~\cite{Cheng:1980tp}.  Since LFV is strictly forbidden in the SM with massless neutrinos, any observation would constitute a clear signal of BSM physics, potentially revealing rich new dynamics in the lepton sector. To date, no definitive evidence of LFV has been observed. For the radiative decay, the MEG~II analysis of the 2021--2022 data yielded the branching ratio bound $\mathrm{BR}(\mu^+\to e^+\gamma)<1.5\times10^{-13}$ at $90\%$ confidence level (CL)~\cite{MEGII:2025gzr}, while the experiment's design sensitivity is $6\times10^{-14}$~\cite{Venturini:2024keu}.  The SINDRUM collaboration reported  $\mathrm{BR}(\mu^+\to e^+e^-e^+)<1.0\times10^{-12}$ at $90\%$ CL~\cite{SINDRUM:1987nra}, and Mu3e aims to reach a sensitivity of order $10^{-16}$ in its second phase~\cite{MartinPerez:2023vfv}. Regarding coherent $\mu$--$e$ conversion in nuclei, SINDRUM~II obtained the conversion rate bound $\mathrm{CR}(\mu^-+\mathrm{Au}\to e^-+\mathrm{Au})<7\times10^{-13}$~\cite{SINDRUMII:2006dvw}. COMET and Mu2e plan to use aluminum targets to probe conversion rates in the $10^{-15}$--$10^{-17}$ range~\cite{Lee:2022moh}. For coherent $\mu$--$e$ conversion in titanium, we adopt the SINDRUM~II bound $\mathrm{CR}(\mu^-+\mathrm{Ti}\to e^-+\mathrm{Ti})<4.3\times10^{-12}$ at $90\%$ CL~\cite{SINDRUMII:1993gxf} and the future sensitivity of order $10^{-18}$ envisaged for PRIME~\cite{prime}.

In light of  both current and previous experimental results, we shall investigate the  charged LFV processes $e_j \to e_i \gamma$, $e_j \to e_i \overline{e}_i e_i$, and $\mu$--$e$ conversion, in order to confront the theoretical predictions of the singlet--triplet scotogenic model with existing experimental bounds. Here $e_i$ and $e_j$ denote the charged leptons of the $i$-th and $j$-th generations, respectively with $i,j=1, 2, 3$ and $i<j$. Our analysis focuses on the leading one-loop contributions and aims to determine the corresponding amplitudes. In the mass eigenstate basis, two distinct sets of fields contribute to these processes: the first set comprises $\chi_i$ and $\eta^\pm$, while the second set includes $\Sigma^\pm$ together with either $\eta_R$ or $\eta_I$. Each diagram involves exclusively one of these two field combinations, contributing through either penguin or box topologies. To obtain analytic expressions, we compute the relevant Feynman diagrams at leading order in the small mass ratio $m/M$, where $m$ denotes the charged-lepton mass scale and $M$ characterizes the heavy BSM sector. The complete analytical results, together with their phenomenological implications, are presented in the subsequent sections, while several useful expressions are collected in appendix~\ref{app:general-penguins} and appendix~\ref{app:general-box}.

\subsection{$e_j \to e_i \gamma$\label{subsec:lfv-radiative}}
\begin{figure}[htbp]
        \centering
        \includegraphics[scale=1]{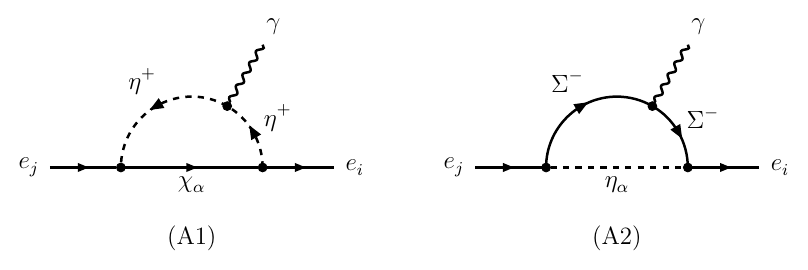}
        \caption{Photon penguin diagrams leading to the dominant Wilson coefficients $K_2^R$\label{fi:mueA}}
\end{figure}
The radiative decay $e_j \to e_i \gamma$ constitutes one of the most sensitive probes of charged LFV. Within the singlet--triplet scotogenic model, this process arises at the one-loop level as depicted in figure~\ref{fi:mueA}. The corresponding transition amplitude can be parameterized in terms of two dipole operators associated with left- and right-handed chiralities. At leading order in the mass ratio $m/M$, the effective amplitude with non-vanishing Wilson coefficients can be written as
\begin{eqnarray}
    \mathcal{M}_{e_j \to e_i \gamma} 
    &=& e\,\overline{u}_i(p_i) 
    \big[
        \left( \gamma^\mu q^2 - q^\mu \slashed{q} \right)
        \left( K_1^L P_L\right) + i m_{e_j}\, \sigma^{\mu\nu} q_\nu
        \left( K_2^L P_L + K_2^R P_R \right)
    \big]u_j(p_j) 
    \epsilon_\mu^* \,.\label{eq:M-megApen}
\end{eqnarray}
Here $e$ denotes the electromagnetic coupling, $P_{L,R}=\frac{1}{2}\left(1\mp\gamma_5\right)$ are the chiral projection operators, $\epsilon_{\mu}$ refers to the photon polarization vector.  The factor $K_1^L$ denotes the Wilson coefficient of the monopole operator. Although this term does not contribute to the radiative decay $e_j \to e_i \gamma$ where the photon is on shell,  it can contribute to the process $e_j \to e_i \overline{e}_i e_i$ via an off-shell photon. The analytic expression for $K_1^L$ will be given later in the discussion of $e_j \to e_i \overline{e}_i e_i$. The factors $K_2^{L,R}$ encode the loop-induced dipole form factors, which depend on the Yukawa couplings and the heavy-particle mass parameters. In the singlet--triplet scotogenic model,  only left-handed leptons are involved in the Yukawa interactions proportional to $Y_{N, \alpha}$ and $Y_{\Sigma, \alpha}$, as shown
in the Lagrangian of Eq.~\eqref{eq:laginteraction}.  From Eqs.~(\ref{eq:ykw-def}, \ref{eq:coeff}, \ref{eq:K2R-res}), we see that the Wilson coefficient $K_2^R$ is proportional to $1$, while $K_2^L$ is proportional to $m_{e_i}/m_{e_j}$. Therefore, in our calculation we retain only the right-handed term $K_2^R$ and neglect $K_2^L$. Since $m_{e_j} \gg m_{e_i}$ for the processes of interest, the contribution from $K_2^L$ is strongly suppressed and can be safely ignored. 

In the present model, both the $\chi_i$--$\eta^\pm$ and $\Sigma^\pm$--$\eta_{R,I}$ loop contributions enter $K_2^R$, albeit with distinct Yukawa structures and mass dependences. The explicit analytical expression for $K_2^R$, incorporating all relevant one-loop contributions, is given by

\begin{eqnarray}
 K_2^R &=& \frac{1}{16\pi^2}\left(D^0 + D^-\right)\,,\label{eq:K2R-exp}
\end{eqnarray}
where $D^0$ and $D^-$ are given by 
\begin{eqnarray}
D^0 &=& \frac{1}{2m_{\eta^+}^2}\sum_{\alpha=1}^{2}Y_{\chi,i\alpha}Y_{\chi,j\alpha}^*F_2(\zeta_\alpha),\\[1ex]
D^- &=& -\frac{Y_{\Sigma,j} Y_{\Sigma,i}^{*}}{4}
\left[\frac{G_2(\rho_R)}{m_{\eta_R}^2}+\frac{G_2(\rho_I)}{m_{\eta_I}^2}\right]\,,\label{eq:D0Dm}
\end{eqnarray}
where the couplings $Y_{\chi, i\alpha}$ with $i=1, 2, 3$ and $\alpha=1, 2$ are given by
\begin{equation}
Y_{\chi,i1}= \frac{Y_{\Sigma,i}^*}{\sqrt{2}}\cos\theta-Y_{N,i}\sin\theta\,,\quad
Y_{\chi,i2}=\frac{Y_{\Sigma,i}^*}{\sqrt{2}}\sin\theta+Y_{N,i}\cos\theta\,,
\end{equation}
and
\begin{eqnarray}
&&\zeta_\alpha = \frac{m_{\chi_\alpha}^2}{m_{\eta^+}^2}\,, \qquad
\rho_R = \frac{M_\Sigma^2}{m_{\eta_R}^2}\,, \qquad
\rho_I = \frac{M_\Sigma^2}{m_{\eta_I}^2}\,.\label{eq:zeta-rho-def}
\end{eqnarray}
From Eqs.~\eqref{eq:metaR} and \eqref{eq:metaI}, we see $m^2_{\eta_R}-m^2_{\eta_I}=\lambda_5 v^2_{\phi}$, the tiny neutrino mass requires very small $\lambda_5$ so that $\eta_R$ and $\eta_I$ are approximately degenerate with $m^2_{\eta_R}\approx m^2_{\eta_I}$. In this limit, our result for $K_2^R$ coincides with that in Ref.~\cite{Rocha-Moran:2016enp}.

\subsection{$e_j \to e_i \bar{e}_i e_i$\label{subsec:lfv-threebody}}

The total amplitude receives contributions from the $\gamma$-penguin, $Z$-penguin, Higgs-penguin, and box diagrams. The Higgs-penguin contributions are neglected because they are suppressed by the small Yukawa couplings of the final-state charged
leptons, which belong to the first two generations. Accordingly, the decay amplitude can be written as
\begin{eqnarray}
    \mathcal{M}(e_j^- \to e_i^- e_i^- e_i^+) = 
    \mathcal{M}_{\gamma\text{-penguin}} + 
    \mathcal{M}_{Z\text{-penguin}} + 
    \mathcal{M}_{\text{Boxes}} \,.
\end{eqnarray}
\begin{figure}[htbp]
\centering
\includegraphics[scale=0.7]{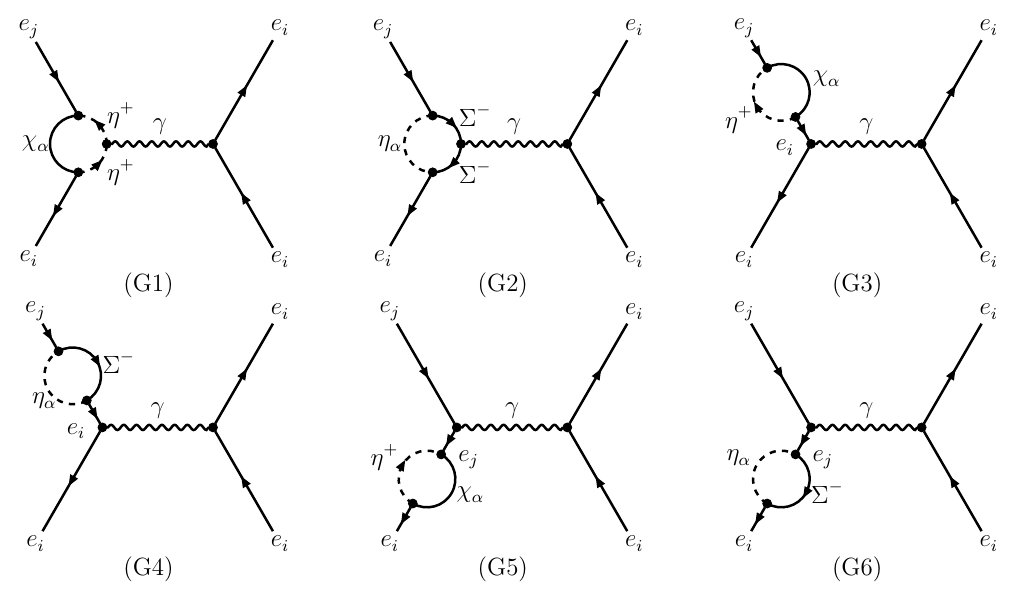}
\caption{$\gamma$-penguin diagrams contributing to $e_j \to e_i \bar{e}_i e_i$, where $\chi_\alpha = \chi_1, \chi_2$ and $\eta_\alpha = \eta_R, \eta_I$. \label{fi:Apenguin}}
\end{figure}
The contributions from photon-mediated diagrams, referred to as $\gamma$-penguin diagrams and displayed in figure~\ref{fi:Apenguin}, can be encoded in the effective amplitude $\mathcal{M}_{\gamma\text{-penguin}}$, which is given by~\cite{Ilakovac:1994kj}:
\begin{eqnarray}
\mathcal{M}_{\gamma\text{-penguin}}&=&\frac{e^2}{q^2}\overline{u}_i(p_1)\left[\left( \gamma^\mu q^2 - q^\mu \slashed{q} \right)
\left(K_1^L P_L\right)+ i \, \sigma^{\mu\nu} q_\nu\left(K_2^L P_L+K_2^R P_R \right)\right]u_j(p)\nonumber\\
&&\times\overline{u}_i(p_2)\gamma^\mu v_i(p_3)-\left(p_1\leftrightarrow p_2\right)\,,\label{eq:STM-Apen}
\end{eqnarray}
where $K_1^L$ and $K_2^{L,R}$ coincide with the coefficients appearing in the effective amplitude for $e_j \to e_i \gamma$ given in Eq.~\eqref{eq:M-megApen}. The analytic expression for the monopole Wilson coefficient $K_1^L$ is determined to be as follows,
\begin{eqnarray}
K_1^L&=&\frac{1}{16\pi^2}\left(M^0+M^-\right)\,,\label{eq:K1L-exp}
\end{eqnarray}
with $M^0$ and $M^-$ given by
\begin{eqnarray}
M^0 &=& \frac{1}{6m_{\eta^+}^2}\sum_{\alpha=1}^{2}Y_{\chi,i\alpha}Y_{\chi,j\alpha}^*F_1(\zeta_\alpha)\,,~~~   M^- = \frac{Y_{\Sigma,j} Y_{\Sigma,i}^{*}}{12}\left[\frac{G_1(\rho_R)}{m_{\eta_R}^2}+\frac{G_1(\rho_I)}{m_{\eta_I}^2}\right]\,.\label{eq:M0Mm}
\end{eqnarray}
In the limit $m_{\eta_R}^2 \simeq m_{\eta_I}^2 \equiv m_{\eta^0}^2$, our expression for $K_1^L$ does not coincide with the corresponding result presented in Ref.~\cite{Rocha-Moran:2016enp}. Specifically, we observe a sign difference in the contribution associated with $M^0$.

\begin{figure}[htbp]
\centering
\includegraphics[scale=0.7]{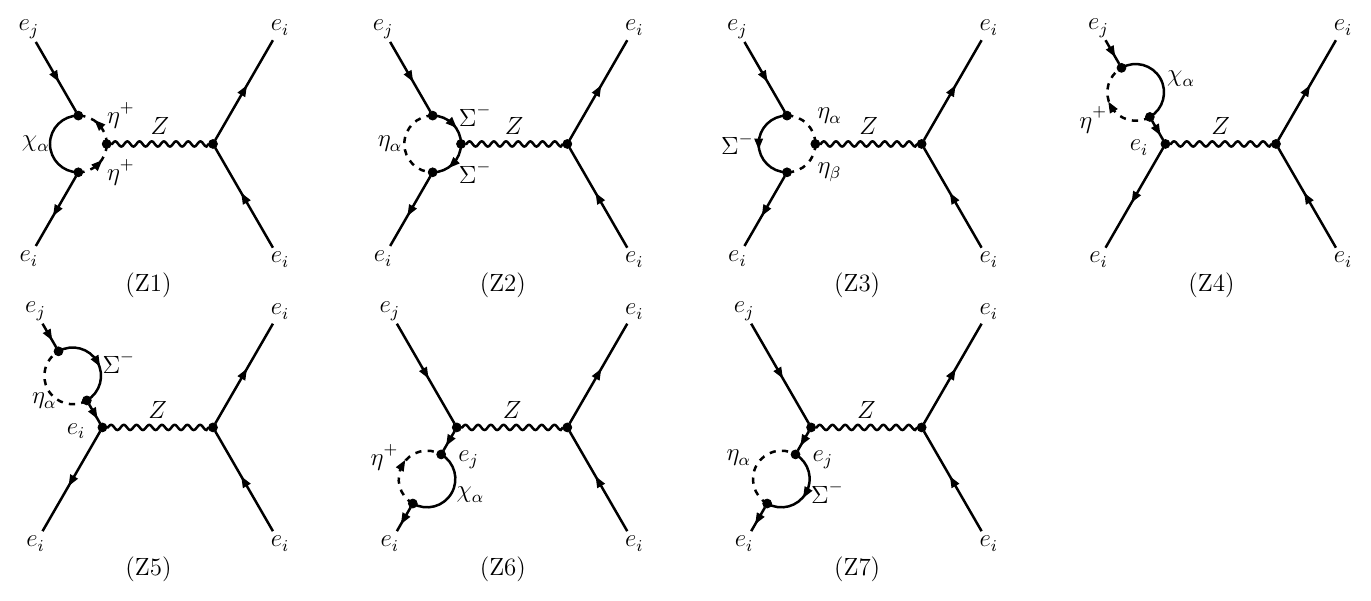}
\caption{$Z$-penguin diagrams leading to the $e_j\to e_i\bar{e}_ie_i$, where $\chi_\alpha = \chi_1, \chi_2$ and $\eta_\alpha = \eta_R, \eta_I$.\label{fi:Zpenguin}}
\end{figure}
The Feynman diagrams mediated by the $Z$ boson are referred to as $Z$-penguin diagrams and are displayed in figure~\ref{fi:Zpenguin}. It is noteworthy that there exists an extra one-loop contribution denoted as Z3 in figure~\ref{fi:Zpenguin}, in comparison with the $\gamma$-penguin diagrams. This contribution arises from the Feynman diagrams with $\eta_R$, $\eta_I$, and $\Sigma^-$ running in the loop, and it has no counterpart in the $\gamma$-penguin diagrams shown in figure~\ref{fi:Apenguin}. The origin of this additional contribution is the non-vanishing coupling between the electrically neutral scalar and the $Z$ gauge boson. By neglecting the momentum dependence of the $Z$ boson propagator, the leading-order effective amplitude of the $Z$-penguin contribution can be expressed as
\begin{eqnarray}
\mathcal{M}_{Z\text{-penguin}}=\frac{1}{m_Z^2}\left[\overline{u}_i(p_1)\gamma_\mu \left(F_LP_L+F_RP_R\right)u_j(p)\right]\left[\overline{u}_i(p_2)\gamma^\mu\left(Z_L^{(e)}P_L+Z_R^{(e)}P_R\right)v_i(p_3)\right]-(p_1\leftrightarrow p_2)\,,\nonumber\\\label{eq:STM-Zpen}
\end{eqnarray}
where
\begin{eqnarray}
Z_L^{(e)}=\frac{g}{\cos\theta_W}\left(\frac{1}{2}-\sin^2\theta_W\right),~~~Z_R^{(e)}=-\frac{g}{\cos\theta_W}\sin^2\theta_W
\end{eqnarray}
with $\theta_W$ being the Weinberg angle. After performing the loop calculation, the Wilson coefficients $F_L$ and $F_R$ take the form
\begin{eqnarray}
F_L&=&-\frac{gY_{\Sigma,j}Y_{\Sigma,i}^{*}}{128\pi^2\cos\theta_W}\left[F_4(\rho_R)+F_4(\rho_I)-F_{13}(\rho_R,\rho_I)\right],\label{eq:FL}\\
F_R&=&-\frac{gm_{e_i}m_{e_j}}{32\pi^2\cos\theta_W}\Bigg\{\frac{1}{m_{\eta^+}^2}\sum_{\alpha=1}^{2}Y_{\chi,i\alpha}Y_{\chi,j\alpha}^*F_2(\zeta_\alpha)\nonumber\\
&&\hspace{1.0cm}+Y_{\Sigma,j}Y_{\Sigma,i}^{*}\Bigg[\frac{F_{14}\left(\rho_R,\rho_I\right)}{2M_\Sigma^2}+\frac{F_2(\rho_R)}{m_{\eta_R}^2}+\frac{F_2(\rho_I)}{m_{\eta_I}^2}\Bigg]\Bigg\}\,.\label{eq:FR}
\end{eqnarray}
In general, $F_R$ scales as $\mathcal{O}(m^2/M^2)$, while $F_L$ is of order $\mathcal{O}(1)$, suggesting that $F_L$ should dominate the $Z$-penguin contribution. However, in the limit $m_{\eta_R} \simeq m_{\eta_I}$, corresponding to $\lambda_5 \simeq 0$, as required for neutrino-mass generation, $F_L$ in Eq.~\eqref{eq:FL} can be approximated as
\begin{eqnarray}
F_L=-\frac{gY_{\Sigma,j}Y_{\Sigma,i}^{*}v_\phi^4\lambda_5^2}{128m_{\eta_0}^4\pi^2\cos\theta_W}F_2(\rho_\eta)\,,\label{eq:FL-0}
\end{eqnarray}
where $\rho_\eta\equiv M_\Sigma^2/m_{\eta_0}^2$. Thus, $F_L$ tends to vanish in $m_{\eta_R} \simeq m_{\eta_I}$ limit. Consequently, the contribution of $F_R$ may dominate over that of $F_L$ when $m_{\eta_R} \simeq m_{\eta_I}$, which is an interesting feature of the model.

\begin{figure}[t!]
\centering
\includegraphics[scale=1]{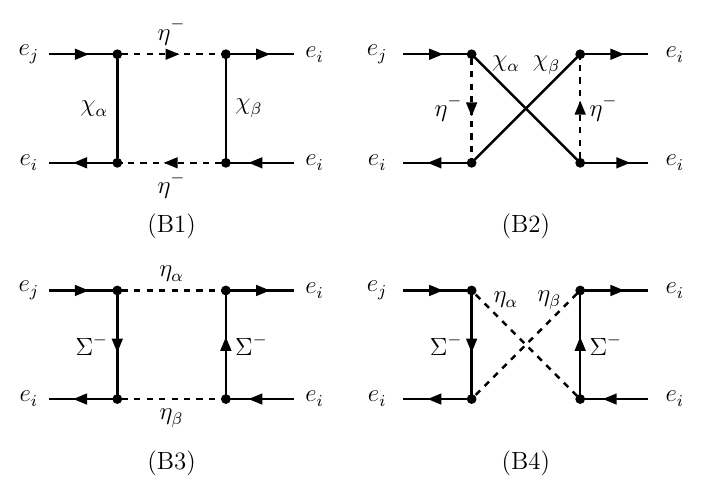}
\caption{Box diagrams contributing to the $e_j \to e_i \bar{e}_i e_i$ process, where $\chi_\alpha = \chi_1, \chi_2$ and $\eta_\alpha = \eta_R, \eta_I$. \label{fi:box}  }
\end{figure}

The box-type Feynman diagrams specific to this model are illustrated in figure~\ref{fi:box}. These diagrams induce a non-zero $B_1^L$ contribution in the complete four-fermion effective amplitude:
\begin{eqnarray}
\mathcal{M}_{\text{boxes}} &=& e^2B_1^L\left[\overline{u}_{e_i}(k_1)\gamma_\mu P_Lu_{e_j}(p)\right]\left[\overline{u}_{e_i}(k_3)\gamma^\mu P_Lv_{e_i}(k_2)\right]\,.\label{eq:STM-box}
\end{eqnarray}
The corresponding leading-order expression for $B_1^L$ is given by
\begin{eqnarray}
e^2 B_1^L &=& 
-\sum_{\alpha=1}^{2}\sum_{\beta=1}^{2}\frac{Y_{\chi,i\beta} Y_{\chi,i\beta} Y_{\chi,i\alpha}^* Y_{\chi,j\alpha}^*}{8\pi^2 m_{\eta^+}^2} 
G_{12}(\zeta_\alpha, \zeta_\beta)
-\sum_{\alpha=1}^{2}\sum_{\beta=1}^{2}\frac{Y_{\chi,i\alpha} Y_{\chi,i\beta} Y_{\chi,i\beta}^* Y_{\chi,j\alpha}^*}{8\pi^2 m_{\eta^+}^2} 
G_{13}(\zeta_\alpha, \zeta_\beta) \nonumber\\
&&-\frac{Y_{\Sigma,i}Y_{\Sigma,j}Y_{\Sigma,i}^*Y_{\Sigma,i}^*}{8\pi^2M_\Sigma^2}G_{14}(\rho_R,\rho_I)\,.\label{eq:B1L-exp}
\end{eqnarray}

\subsection{$\mu$--$e$ conversion \label{subsec:lfv-conversion}}

\begin{figure}[htbp]
\centering
\includegraphics[scale=0.7]{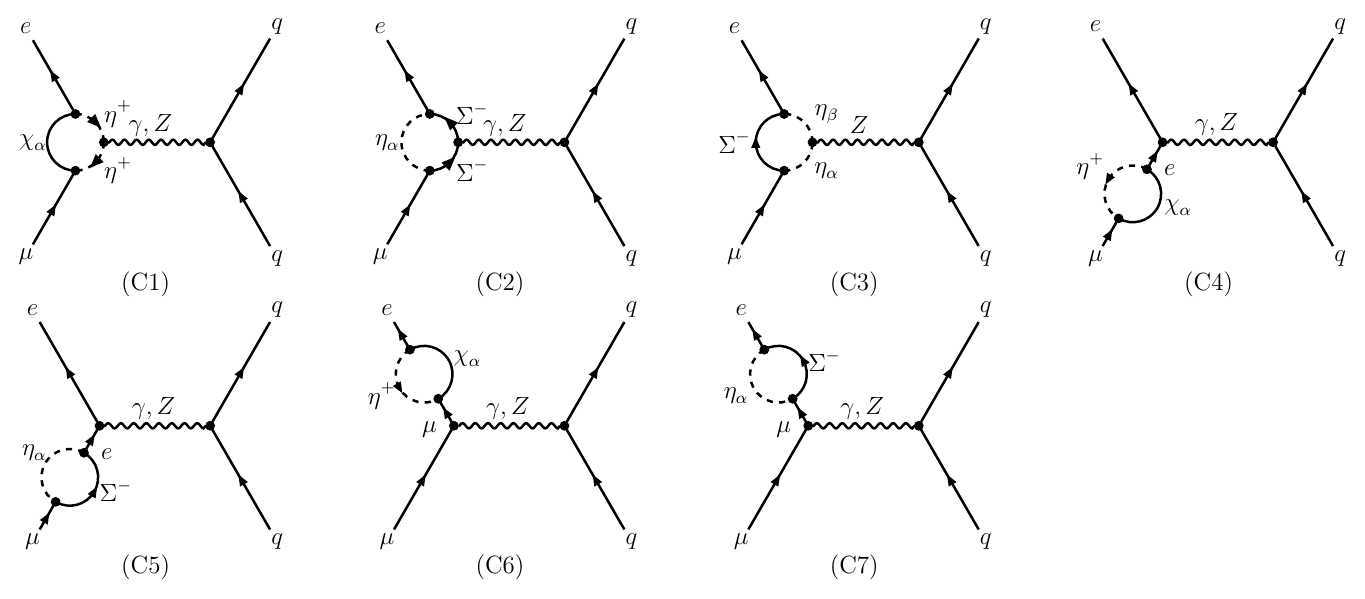}
\caption{Feynman diagrams for the $\mu$--$e$ conversion process $\mu^-+(A,Z)\rightarrow e^-+(A,Z)$, where $\chi_\alpha = \chi_1, \chi_2$ and $\eta_\alpha = \eta_R, \eta_I$.\label{fi:mue_conv}  }
\end{figure}

Besides the processes $e_j\to e_i\gamma$ and $e_j\to e_i\overline{e}_ie_i$, another important LFV channel is the $\mu$--$e$ conversion process,
\begin{eqnarray}
\mu^-+(A,Z)\rightarrow e^-+(A,Z),
\end{eqnarray}
which describes the capture of the $\mu^-$ by a nucleus with mass number $A$ and atomic number $Z$, followed by its conversion into an electron while leaving the nucleus unchanged. For the coherent $\mu$--$e$ conversion, the general effective interactions at the quark level are given by~\cite{Arganda:2007jw}: 
\begin{eqnarray}
 \mathcal{L}^{\text{eff}}_{\mu-e}&=&-\frac{G_F}{\sqrt{2}}\sum_q\big\{\left[g_{LS(q)}\overline{e_L}\mu_R+g_{RS(q)}\overline{e_R}\mu_L\right]\overline{q}q\nonumber\\
&&+\left[g_{LV(q)}\overline{e_L}\gamma^\mu\mu_L+g_{RV(q)}\overline{e_R}\gamma^\mu\mu_R\right]\overline{q}\gamma_\mu q\big\}\,,\label{eq:STM-muecon}
\end{eqnarray}
where $G_F$ denotes the Fermi coupling constant. Notice that only the scalar ($S$) and vector ($V$) effective operators are relevant here, with corresponding couplings $g_{LS(q)}$, $g_{RS(q)}$, and $g_{LV(q)}$, $g_{RV(q)}$ (for left- and right-handed components, respectively). 

Similar to the $e_j\to e_i\overline{e}_ie_i$ decay, the $\mu$--$e$ conversion process also receives contributions from the penguin diagrams shown in figure~\ref{fi:mue_conv}. No counterpart of the box diagrams can contribute to the $\mu$--$e$ conversion, because all the new fields are $SU(3)_C$ singlets.
In addition, Higgs-penguin diagrams are neglected because they are suppressed by the small Yukawa couplings. Consequently, the relevant effective couplings satisfy
\begin{eqnarray}
g_{LS(q)}&=&0\,,~~~     g_{RS(q)}=0\,,~~~~
g_{LV(q)}=g_{LV(q)}^\gamma+g_{LV(q)}^Z\,,
~~
g_{RV(q)}=g_{RV(q)}^\gamma+g_{RV(q)}^Z\,.
\end{eqnarray}
The photon- and $Z$-boson–induced vector couplings can be derived from the Feynman diagrams shown in figure~\ref{fi:mue_conv}. Specifically,
\begin{eqnarray}
g_{LV(q)}^\gamma&=&\frac{\sqrt{2}}{G_F}e^2Q_q\left(K_1^L-K_2^R\right)\,,~~~~\quad       g_{LV(q)}^Z=-\frac{\sqrt{2}}{G_F}\frac{Z_L^q+Z_R^q}{2}\frac{F_L}{m_Z^2}\,,\label{eq:gLVq-def}
\end{eqnarray}
where $Q_q$ is the electric charge of the corresponding quark and
\begin{eqnarray}
Z_L^q&=&\frac{g}{\cos\theta_W}\left(Q_q\sin^2\theta_W-T_3^q\right)\,,~~~~Z_R^q=\frac{g}{\cos\theta_W}Q_q\sin^2\theta_W\,,
\end{eqnarray}
represent the tree-level $Z$-boson couplings to quarks.

\section{Effective potential and phase transition gravitational waves\label{sec:effV_PTGW}}

In this section, we investigate whether the singlet--triplet scotogenic model can realize a strong electroweak FOPT in the early Universe, which may produce detectable GW signals. This is in contrast to the SM, where the observed Higgs boson mass of 125~GeV~\cite{ATLAS:2012yve,CMS:2012qbp} leads to a smooth crossover transition rather than a first-order one~\cite{Kajantie:1995kf,Kajantie:1996mn,Kajantie:1996qd}. In the present model, the additional scalars $\eta$ and $\Omega$ extend the SM scalar potential, whose vacuum structure is further modified by finite-temperature corrections. At sufficiently high temperatures, the effective potential develops a minimum at $(h_0,\Omega_0)=(0,0)$, corresponding to the symmetric phase with restored symmetry. As the temperature decreases during the evolution of the Universe, the symmetry is broken, and at $T=0$ the VEVs approach $(h_0,\Omega_0)=(v_\phi,v_\Omega)$, corresponding to the broken phase. During this process, a second minimum away from the origin may appear. As the temperature decreases further, the two phases become degenerate at the critical temperature $T_C$, and the field eventually tunnels from the false vacuum to a deeper true vacuum, which we take to lie along the $h_0$ direction. This tunneling proceeds through the nucleation of true-vacuum bubbles, which expand and collide with one another, gradually converting the false vacuum into the true vacuum throughout the Universe. These bubble collisions generate a stochastic GW background. The temperature at which approximately one bubble nucleates per Hubble volume is referred to as the nucleation temperature $T_N$. For a fast FOPT without substantial reheating, $T_N$ can be regarded as the temperature at which GWs are produced \cite{Guo:2021qcq}. Finally, the Higgs VEV smoothly evolves toward its tree-level value $v_\phi$ as the temperature approaches zero. In this work, we take $T_N$ to be the characteristic temperature of GW production.

\subsection{Potential for phase transitions\label{sec:FTV}}

To study the cosmological phase transitions in this model, we first construct the effective potential. Throughout the cosmological evolution, we assume that the $\mathbb{Z} _{2}$ symmetry remains intact, thereby forbidding $\eta$ from developing a VEV, $\langle \eta^0 \rangle = 0$. As a result, only the CP-even scalar neutral fields are permitted to acquire nonzero background field values, which are defined as $\varphi_1 \equiv v_\phi + h_0$ and $\varphi_2 \equiv v_\Omega + \Omega_0$. Accordingly, the total effective potential depends only on these two background fields, $V_{\rm tot}(\varphi_1,\varphi_2)$.

\subsubsection{Tree-level potential  }

The tree-level effective potential in terms of $\varphi_1$ and $\varphi_2$ derived from Eq.~\eqref{eq:scalarpot} is
given by
\begin{eqnarray}
  V_{\text{tree}} &=& -\frac{m_{\phi}^{2}}{2}\varphi_1^{2}
  -\frac{m_{\Omega}^{2}}{2}\varphi_2^{2}
  +\frac{\lambda_1}{8}\varphi_1^{4}
  -\frac{\mu_{1}}{2\sqrt{2}}\varphi_1^{2}\varphi_2
  +\frac{\lambda_{1}^{\Omega}}{4}\varphi_1^{2}\varphi_2^{2}
  +\frac{\lambda_{2}^{\Omega}}{4}\varphi_2^{4}\,,\label{eq:treeV-vev}
\end{eqnarray}
where the $m_{\phi}^{2}$ and $m_{\Omega}^{2}$ are given in Eq.~\eqref{eq:mphisq} and Eq.~\eqref{eq:mphimOmg} 
respectively.

\subsubsection{Quantum corrections to the potential}

The CW potential encodes the quantum corrections to the tree-level potential in Eq.~\eqref{eq:treeV-vev}. It is obtained from the sum of one-particle-irreducible diagrams with external legs corresponding to the classical background fields $\varphi_1$ and $\varphi_2$. In the Landau gauge, the one-loop CW potential is given by~\cite{Coleman:1973jx} 
\begin{equation}
V_{\text{CW}}= \sum_{i}(-1)^{2s_{i}}n_{i}\frac{m_{i}^{4}}{64\pi^{2}}\left[\ln \frac{m_{i}^{2}}{\bar{\mu}^{2}}-C_{i} \right]\,,\label{eq:CWV-vev}
\end{equation}
where the sum runs over the fields $i=S_{1,2},\ \phi_I,\ \kappa_{1,2}^{\pm}, \ \eta_R,\ \eta_I,\ \eta^{\pm},\ W^{\pm},\ Z,\ t,\,b$. 

Here $m_{i}^{2}\equiv m_{i}^{2}(\varphi_1,\varphi_2)$
is the field-dependent tree-level mass squared of the particle species $i$ in our model. For the top and bottom quarks, the field-dependent masses are given by $m_{t,b}^2=\frac{y_{t,b}^2}{2}\varphi_1^2$, where $y_{t,b}$ denotes the corresponding Yukawa coupling. We choose the renormalization scale to be $\bar{\mu}=246$ GeV. The constants $C_i$ depend on the renormalization scheme. In our analysis, we adopt the modified minimal subtraction ($\overline{\text{MS}}$) scheme, in which $C_i=\frac{5}{6}$ for the gauge bosons $W^\pm$ and $Z$, while $C_i=\frac{3}{2}$ for all scalar and fermionic degrees of freedom. 
The parameters $n_i$ and $s_i$ denote the number of degrees of freedom and the spin of each field, respectively.  The multiplicities $n_i$ are given by 
\begin{eqnarray}
\nonumber&&n_{S_{1,2}}=n_{\phi_{I}}=n_{\eta_{R}}=n_{\eta_{I}}=n_{Z_{L}}=1\,,\quad n_{W^{\pm}_{T}}=4\,,\\
&&n_{\kappa_{1,2}^+}=n_{\eta^{\pm}}=n_{Z_{T}}=n_{W^{\pm}_{L}}=2\,,\quad n_{t}=n_{b}=12\,, \label{eq:n_degree}
\end{eqnarray}
where the subscripts $T$ and $L$ refer to the transverse and longitudinal components of the gauge bosons. We neglect the lighter fermions because their Yukawa couplings are much smaller than the top and bottom Yukawa couplings. Besides, the ghost bosons decouple in the Landau gauge, thus we do not need to consider them in our calculation~\cite{Ford:1992mv}. 

The CW correction $V_{\text{CW}}$ will shift both the position of the minimum and the second-order partial derivatives of the
tree-level potential. Thus, in order to offset the modification, counter terms $V_{\text{CT}}$ need to be added to the Lagrangian. $V_{\text{CT}}$ can be written as
\begin{eqnarray}
V_{\text{CT}} = \delta m_{\phi}^2 \, \varphi_1^2 + \delta m_{\Omega}^2\, \varphi_2^2 + \delta\lambda_1\, \varphi_1^4 + \delta \mu_{1} \varphi_1^{2}\,\varphi_2 +\delta\lambda_{1}^{\Omega} \, \varphi_1^2 \varphi_2^2 + \delta\lambda_2^\Omega\, \varphi_2^4 \,.\label{eq:CTV-vev}
\end{eqnarray}
The coefficients \(\{\delta m_{\phi}^2,\delta m_{\Omega}^2,\delta\lambda_1,\delta \mu_{1},\delta\lambda_{1}^{\Omega},\delta\lambda_2^\Omega\}\) are fixed by imposing renormalization conditions at the tree-level vacuum, \(\langle \varphi_1\rangle = v_\phi\) and \(\langle\varphi_2\rangle = v_\Omega\):
\begin{eqnarray}
\left.\partial_{\varphi_i}\!\big( V_{\text{CW}} + V_{\text{CT}} \big)\right|_{\langle\varphi_i\rangle} = 0\,,~~~\quad  \left.\partial_{\varphi_i}\partial_{\varphi_j}\!\big( V_{\text{CW}} + V_{\text{CT}} \big)\right|_{\langle\varphi_i\rangle,\langle\varphi_j\rangle} = 0\,,\label{eq:renormalization_condition}
\end{eqnarray}
where $i,j\in\{1,2\}$.

In the Landau gauge, the Goldstone boson masses vanish at the physical minimum, which can induce infrared (IR) divergences in derivatives of the CW potential through logarithmic terms~\cite{Martin:2014bca,Elias-Miro:2014pca}. To regulate these divergences, we follow the prescription of~\cite{Chatterjee:2022pxf,Baum:2020vfl} and shift the Goldstone masses as
$m_G^2\to m_G^2 + \mu_{\rm IR}^2$
with \(\mu_{\rm IR}^2 = 1~\text{GeV}^2\) adopted in the numerical analysis.

\subsubsection{Thermal corrections to the potential}
At finite temperature, the effective potential receives additional one-loop corrections~\cite{Dolan:1973qd},
\begin{equation}
  V_{\text{th}}
  = \frac{T^{4}}{2\pi^{2}}
    \sum_{i}\Big[
      n_{B_{i}}\,J_{B}\!\left( \frac{m_{B_{i}}}{T}\right)
      + n_{F_{i}}\,J_{F}\!\left( \frac{m_{F_{i}}}{T}\right)
    \Big],\label{eq:thV-vev}
\end{equation}
where $n_{B_i}$ and $n_{F_i}$ are the degrees of freedom of bosons and fermions, respectively, as given in Eq.~\eqref{eq:n_degree}.  
The thermal functions $J_{B(F)}$ for bosons and fermions take the following form,
\begin{equation}
J_{B(F)}(y)=\pm \int_{0}^{\infty} \! dx\, x^{2}\,\ln\!\left(1\mp e^{-\sqrt{x^{2}+y^{2}}}\right)\,,
\end{equation}
with the upper (lower) sign corresponding to bosons (fermions).  

In evaluating the finite-temperature effective potential, we include the daisy (ring) resummation of the \(n = 0\) Matsubara modes for the longitudinal components of the gauge bosons \(W^{\pm}_{L}\), \(W^{3}_{L}\), and \(B_{L}\), as well as the scalar fields \(\phi\), \(\eta\), and \(\Omega\). These resummations account for the Debye mass corrections arising at nonzero temperature and prevent the IR divergences that occur when scalar masses become small compared to the thermal scale. The daisy resummation plays a crucial role in a FOPT, as it significantly modifies the cubic term in the finite-temperature potential. Only the longitudinal modes of vector bosons and scalar fields receive thermal mass corrections, whereas the transverse modes remain unaffected due to gauge invariance~\cite{Espinosa:1992kf}. 
 
The thermally corrected mass eigenvalues of particle \(s\) are denoted by \(\widetilde{m}_{s}\), and  the daisy contributions can be included through the replacement,
\begin{equation}
\widetilde{m}_{s}^{2} = \text{eigenvalues}\!\left[ M_{s}^{2} + \Pi_{s}(T) \right],
\end{equation}
where \(M_{s}^{2}\) represents the tree-level field-dependent mass-squared matrices, and \(\Pi_{s}(T)\) denotes their thermal corrections. The subleading off-diagonal elements of \(\Pi_{s}(T)\) can be safely neglected~\cite{Carrington:1991hz,Arnold:1992rz}. The diagonal components of the thermal self-energies \(\Pi_{s}(T)\) for scalar bosons are
\begin{eqnarray}
\nonumber \Pi_{S,11} &=& \Pi_{H^{\pm},11} = \frac{T^{2}}{48} \left( 9g^{2} + 3g'^{2} + 12y_{t}^{2} + 12y_{b}^{2} + 12\lambda_{1} + 8\lambda_{3} + 4\lambda_{4} + 6\lambda_{1}^{\Omega} \right), \\
\Pi_{S,22} &=& \Pi_{H^{\pm},22} = \frac{T^{2}}{12} \left( 6g^{2} + 2\lambda_{1}^{\Omega} + 5\lambda_{2}^{\Omega} + 2\lambda_{3}^{\Omega} \right)\,,
\end{eqnarray}
and for other scalar fields,
\begin{eqnarray}
\nonumber \Pi_{\phi_{I}} &=& \frac{T^{2}}{48} \left( 9g^{2} + 3g'^{2} + 12y_{t}^{2} + 12y_{b}^{2} + 12\lambda_{1} + 8\lambda_{3} + 4\lambda_{4} + 6\lambda_{1}^{\Omega} \right), \\
\Pi_{\eta_{R},\eta_{I},\eta_{\pm}} &=& \frac{T^{2}}{48} \left( 9g^{2} + 3g'^{2} + 8\lambda_{3} + 4\lambda_{4} + 12\lambda_{2} + 6\lambda_{3}^{\Omega} \right)\,.
\end{eqnarray}
The thermal corrections to the electroweak gauge bosons are
\begin{eqnarray}
  \Pi_{W_{L}^{\pm}} = \Pi_{W_{L}^{3}} = \frac{7g^{2}}{3}T^{2}, \quad
  \Pi_{B_{L}} = 2g'^{2}T^{2}, \quad
  \Pi_{W_{T}^{\pm}} = \Pi_{B_{T}} = \Pi_{W^{3}_{T}} = 0.
\end{eqnarray}
Here, \(\Pi_{W_{L}^{3}}\) and \(\Pi_{B_{L}}\) correspond to the thermal corrections to the diagonal elements of the neutral gauge boson mass matrix. The resulting daisy contribution to the effective potential is given by the resummation formula~\cite{Carrington:1991hz,Arnold:1992rz}:
\begin{equation}
V_{\text{Daisy}} = -\frac{T}{12\pi} \sum_{s=\text{bosons}} n_{s} \left[ \left(\widetilde{m}_{s}^{2}\right)^{3/2} - \left(m_{s}^{2}\right)^{3/2} \right],\label{eq:DSV-vev}
\end{equation}
where \(n_s\) denotes the number of degrees of freedom of species \(s\). Note that the Yukawa couplings between the scalars $\eta$, $\Omega$ and the Majorana fermions $\Sigma$, $N$ given by Eq.~\eqref{eq:laginteraction}, may give additional contributions to $\Pi_{S}$, $\Pi_{H^{\pm}}$ and $\Pi_{\eta_{R},\eta_{I},\eta_{\pm}}$. Besides, the kinetic terms of $\Sigma$ and $N$ can also induce contributions to $\Pi_{W}$. However, as shown below, the Majorana fermions $N$ and $\Sigma$ are much heavier than the nucleation temperature. Their thermal contributions are therefore strongly Boltzmann suppressed at temperatures of order $100~\mathrm{GeV}$. We therefore neglect those contributions in our analysis.

\subsubsection{Full thermally corrected effective potential}
The full thermally corrected effective potential required for the computation of the phase-transition parameters is obtained by adding to the tree-level potential in Eq.~\eqref{eq:treeV-vev} the quantum corrections summarized in Eqs.~\eqref{eq:CWV-vev} and \eqref{eq:CTV-vev}, together with the thermal corrections given in Eqs.~(\ref{eq:thV-vev}, \ref{eq:DSV-vev}):
\begin{equation}
V_{\text{tot}} = V_{\text{tree}} + V_{\text{CW}} + V_{\text{th}} + V_{\text{daisy}} + V_{\text{CT}}\,.\label{eq:Vtot}
\end{equation}

\subsection{Gravitational waves induced by a first-order phase transition\label{eq:GW_FOPT}}

With the effective potential and the corresponding VEVs, including the one-loop corrections, we can investigate the phenomenology of FOPT-induced PTGWs in the singlet--triplet scotogenic model. In this work, we employ the package \texttt{PhaseTracer2}~\cite{Athron:2024xrh} to compute the dynamics of the FOPT and the associated PTGW spectrum.

At sufficiently high temperatures, the electroweak symmetry is restored. As the Universe cools, a symmetry-breaking minimum develops in the finite-temperature effective potential. The critical temperature, $T_C$, is defined as the temperature at which the symmetric and broken vacua become degenerate in free energy, indicating that the phase transition becomes thermodynamically possible. However, the transition does not occur immediately at $T_C$, since the two phases are still separated by a potential barrier. The Universe therefore remains trapped in the metastable false vacuum and continues to supercool until thermal fluctuations become sufficiently efficient to nucleate bubbles of the true vacuum.

The nucleation temperature, $T_N$, is defined as the temperature at which the probability of nucleating one critical bubble per Hubble volume becomes of order unity~\cite{Affleck:1980ac,Linde:1980tt,Linde:1981zj}. Equivalently, it is determined by the condition
\begin{eqnarray}
\int_{T_C}^{T_N} dT\frac{\Gamma(T)}{H(T)T^4}=1\,,
\end{eqnarray}
where $H(T)$ is the Hubble expansion rate and $\Gamma(T)$ denotes the bubble nucleation rate per unit volume. Employing the semiclassical expression
\begin{eqnarray}
\Gamma(T)\propto T^4
\exp\left[-\frac{S_3(T)}{T}\right]\,.
\end{eqnarray}
The nucleation condition is well approximated by
\begin{eqnarray}
\frac{S_3(T_N)}{T_N}\simeq140,
\end{eqnarray}
which provides an accurate estimate of the nucleation temperature~\cite{Apreda:2001us}. Here, $S_3(T)$ denotes the three-dimensional Euclidean bounce action, which represents the free-energy cost of nucleating a critical bubble. It is evaluated on the $O(3)$-symmetric bounce solution and is given by
\begin{eqnarray}
S_3(T)=4\pi\int dr\,r^2\left[\frac12\left(\frac{d\varphi_1}{dr}\right)^2+\frac12\left(\frac{d\varphi_2}{dr}\right)^2
+V_{\rm tot}(\varphi_1,\varphi_2,T)\right],
\end{eqnarray}
where $\varphi_1(r)$ and $\varphi_2(r)$ denote the bounce profiles of the Higgs and triplet scalar fields, respectively. The finite-temperature effective potential $V_{\rm tot}$, defined in Eq.~\eqref{eq:Vtot}, is normalized such that $V_{\rm tot}(0,0,T)=0$~\cite{Kubo:2016kpb,Ellis:2020awk}. Since the bubble nucleation rate is exponentially suppressed by $\exp[-S_3(T)/T]$, a smaller value of $S_3(T)/T$ corresponds to a higher nucleation probability.

Below the nucleation temperature, the number density of true-vacuum bubbles rapidly increases. The nucleated bubbles expand and begin to collide with one another while continuing to convert the false vacuum into the true vacuum. As the bubble network grows, the system eventually reaches the percolation stage, at which the true-vacuum regions become sufficiently connected for the phase transition to effectively complete. The stochastic GW signal is generated during this stage through bubble collisions, long-lasting sound waves in the plasma, and the subsequent magnetohydrodynamic turbulence. Since the interval between nucleation and percolation is typically short for the transitions considered in this work, we adopt the nucleation temperature $T_N$ as the reference temperature for evaluating all phase-transition parameters relevant to the GW spectrum. Once the nucleation temperature $T_N$ has been determined, the key thermodynamic and dynamical parameters characterizing the phase transition can be evaluated. In particular, the strength of the phase transition is characterized by the parameter $\alpha$, while its characteristic time scale is described by the inverse duration parameter $\beta$.

The transition strength $\alpha$ is defined as the released vacuum energy normalized to the radiation energy density at the nucleation temperature~\cite{1004.4187},
\begin{eqnarray}
\alpha\simeq\frac{30}{\pi^{2} g_{*} T_{N}^{4}}\left(\Delta V(T)-\frac{T}{4}\frac{\partial \Delta V(T)}{\partial T} \right)\bigg|_{T=T_N}\,,\label{eq:alpha-def}
\end{eqnarray}
where $\Delta V(T)$ is the free-energy difference between the false and true vacua, and $g_*$ denotes  relativistic degrees of freedom of the thermal plasma at the temperature $T_N$. The characteristic duration of the phase transition is quantified by the parameter $\beta$, which is conventionally normalized to the Hubble expansion rate at the nucleation temperature, $H_*\equiv H(T_N)$~\cite{Grojean:2006bp}, yielding the dimensionless quantity
\begin{eqnarray}
\frac{\beta}{H_*}=T\frac{d}{dT}\left(\frac{S_3}{T}\right)\Big|_{T=T_N}\,.\label{eq:beta-def}
\end{eqnarray}
With these quantities at hand, we now discuss the sources of GW production in a FOPT. There are three primary sources of stochastic GWs associated with a FOPT: bubble collisions, sound waves, and magnetohydrodynamic  turbulence~\cite{Binetruy:2012ze,Caprini:2015zlo,Caprini:2019egz,Hindmarsh:2020hop}. Bubble collisions originate from the acceleration and subsequent collision of expanding bubble walls, which source tensor perturbations through the anisotropic stress stored in the scalar-field configuration \cite{Kosowsky:1992vn,Jinno:2016vai}. For a phase transition occurring in a thermal plasma, the bubble collision contribution is expected to be much smaller than the acoustic contribution and can therefore be neglected to a good approximation~\cite{Bodeker:2017cim}. We thus write the total GW energy density $\Omega_{\rm gw}h^2$ as
\begin{eqnarray}
    \Omega_{\rm gw} h^2 \simeq \Omega_{\rm sw} h^2 + \Omega_{\rm turb} h^2\,.
\end{eqnarray}
For non-runaway bubble walls, the acoustic contribution $\Omega_{\rm sw}h^2$ is often the dominant component of the stochastic GW background produced during a thermal FOPT~\cite{Hindmarsh:2013xza}. During the transition, the expanding bubble walls transfer part of the released vacuum energy to the surrounding plasma through friction, thereby sourcing bulk fluid motion in shells around the bubbles~\cite{Espinosa:2010hh}. After bubble collisions, this fluid motion can persist as acoustic waves whose anisotropic shear stresses source GWs, and the resulting spectrum is described by the following fitting formulae~\cite{Hindmarsh:2015qta,Hindmarsh:2017gnf,PhysRevD.101.089902,Caprini:2019egz}:
\begin{eqnarray}
\Omega_{\rm sw} h^2=2.061 \, F_{\rm gw,0}\, S_{\rm sw}(f)\,
\widetilde{\Omega}_{\rm gw}\,
\left(\frac{\kappa_{\rm sw}\alpha}{1+\alpha}\right)^2\, \min\!\left(\frac{H_* R_*}{\overline{U}_f},1\right)
(H_* R_*)\,h^2\,.\label{eq:Omg-sw}
\end{eqnarray}
The dimensionless coefficient $\widetilde{\Omega}_{\rm gw} \simeq 0.012$ is a numerical normalization calibrated by simulations, encoding the efficiency of GW production from the shear-stress unequal-time correlator of the acoustic source. The acoustic efficiency factor $\kappa_{\rm sw}$ quantifies the fraction of the released vacuum energy converted into bulk kinetic energy of the plasma~\cite{Espinosa:2010hh}, which subsequently sources the acoustic GW signal. Its value depends on the expansion mode of the bubble~\cite{Caprini:2015zlo}, with the wall velocity fixed at $v_w=0.3$ in this work. Accordingly, different fitting formulae are employed for deflagrations ($v_w<c_s$), supersonic deflagrations (hybrid solutions, $c_s<v_w<v_J$), and detonations ($v_J<v_w$), following Ref.~\cite{Espinosa:2010hh}. Here, $c_s=1/\sqrt{3}$ is the speed of sound in the relativistic plasma, while $v_J$ denotes the Jouguet velocity that separates the hybrid and detonation hydrodynamic regimes.  $\overline{U}_f$ denotes the enthalpy-weighted root-mean-square (RMS) of fluid velocity, which reduces to the usual RMS three-velocity in the non-relativistic limit. The parameter $R_*$ denotes the mean bubble separation around the completion of the phase transition, which sets the characteristic length scale of the acoustic source and is approximately given by $(8\pi)^{1/3}v_w/\beta$. The redshift factor $F_{\rm gw,0}$ and the dimensionless spectral shape function $S_{\rm sw}(f)$ are defined by
\begin{eqnarray}
F_{\rm gw,0} = 3.57\times10^{-5}\left(\frac{100}{g_*}\right)^{1/3},\quad 
S_{\rm sw}(f)= \left(\frac{f}{f_{\rm sw}}\right)^3
\left[\frac{7}{4 + 3(f/f_{\rm sw})^2}\right]^{7/2}\,,
\end{eqnarray}
where $f_{\rm sw}$ denotes the redshifted peak frequency,
\begin{eqnarray}
f_{\rm sw}&=& 26 \left(\frac{z_p}{10}\right)\left(\frac{T_*}{100\text{ GeV}}\right)\left(\frac{g_*}{100}\right)^{1/6}\left(\frac{1}{H_* R_*}\right)\,\mu\text{Hz}\,.
\end{eqnarray}
Here, the thermal parameter $z_p \simeq 10$ is preferred by numerical simulation~\cite{Hindmarsh:2017gnf,Caprini:2019egz}.

The magnetohydrodynamic turbulence may develop from nonlinear plasma motions during and after the acoustic stage of the phase transition, providing an additional, typically subdominant, contribution to the GW background. The corresponding energy density spectrum is commonly described by the phenomenological fitting formula~\cite{Caprini:2009yp,Binetruy:2012ze}:
\begin{eqnarray}
\Omega_{\rm turb} h^2 = 3.35\times10^{-4}\, v_w \left(\frac{H_*}{\beta}\right)
\left(\frac{\kappa_{\rm turb}\alpha}{1+\alpha}\right)^{3/2}
      \left(\frac{100}{g_*}\right)^{1/3}
      \frac{r^3}{(1+r)^{11/3}(1+8\pi f/H_0)}\,.\label{eq:Omg-tur}
\end{eqnarray}
Here the efficiency coefficient $\kappa_{\rm turb}$ characterizes the fraction of the released vacuum energy transferred to turbulent plasma motion. Following numerical simulations, it is commonly taken to be a fixed fraction of the sound-wave efficiency, $\kappa_{\rm turb} \simeq (0.05\text{--}0.10)\,\kappa_{\rm sw}$~\cite{Hindmarsh:2015qta}. The parameter $r \equiv f/f_{\rm turb}$ denotes the frequency normalized to the peak frequency of the turbulent spectrum, and the redshifted peak frequency observed today, $f_{\rm turb}$, is given by
\begin{eqnarray}
f_{\rm turb}= 27\,\frac{1}{v_w} \left(\frac{\beta}{H_*}\right)
\left(\frac{T_*}{100\text{ GeV}}\right) \left(\frac{g_*}{100}\right)^{1/6}\,
\mu\text{Hz}.
\end{eqnarray}
The redshifted Hubble parameter $H_0$ at GW production in Eq.~\eqref{eq:Omg-tur} is
\begin{eqnarray}
H_0= 16.5\left(\frac{g_*}{100}\right)^{1/6} \left(\frac{T_*}{100\text{ GeV}}\right)\, \mu\text{Hz}.
\end{eqnarray}

\section{Theoretical and experimental constraints\label{sec:constraint}}
Having set up the framework for our study, we discuss various theoretical and phenomenological constraints on the model parameters.

\subsection{Theoretical constraints}

In this subsection, we highlight some of the key theoretical constraints on the parameter space of our model.

\subsubsection{Vacuum stability}

To ensure vacuum stability, the scalar potential $V$ in 
Eq.~\eqref{eq:scalarpot} must be bounded from below. Applying the 
copositivity criteria derived in Ref.~\cite{Kannike:2012pe}, we 
obtain the following constraints on the parameters appearing in 
Eq.~\eqref{eq:scalarpot}. For $\lambda_4 - |\lambda_5| \ge 0$, they 
read
\begin{eqnarray} \nonumber
&&\lambda_1 \ge 0,\qquad  \lambda_2 \ge 0,\qquad  \lambda_2^\Omega \ge 0,\qquad \lambda_3 + \sqrt{\lambda_1 \lambda_2} \ge 0, \\ \nonumber
&&\lambda_1^\Omega + \sqrt{2\lambda_1 \lambda_2^\Omega} \ge 0,\qquad \lambda_{3}^{\Omega} + \sqrt{2\lambda_2 \lambda_2^\Omega} \ge 0,\\ \nonumber &&\sqrt{2\lambda_1\lambda_2\lambda_2^\Omega} + \lambda_3\sqrt{2\lambda_2^\Omega} + \lambda_1^\Omega \sqrt{\lambda_2} + \lambda_{3}^{\Omega} \sqrt{\lambda_1} \\
&&+\sqrt{\left(\lambda_3+\sqrt{\lambda_1\lambda_2}\right)\left( \lambda_1^\Omega + \sqrt{2\lambda_1 \lambda_2^\Omega}\right)\left(\lambda_{3}^{\Omega}+\sqrt{2\lambda_2\lambda_2^\Omega}\right)} \ge 0\,, \label{eq:vaccum_stability}
\end{eqnarray}
while for $\lambda_4 - |\lambda_5| < 0$, the replacement 
$\lambda_3 \to \lambda_3 + \lambda_4 - |\lambda_5|$ should be made 
in Eq.~\eqref{eq:vaccum_stability}.

\subsubsection{Perturbativity}

In order to ensure the validity of standard perturbation theory, the one-loop level
correction to coupling should be smaller than the tree-level couplings. The quartic couplings $\lambda_i$ and $\lambda^{\Omega}_i$ in the scalar potential and the Yukawa 
couplings $y_{ij}$, $Y_{N,i}$, $Y_{\Sigma,i}$ and 
$Y_\Omega$ should satisfy~\cite{Karan:2023adm}
\begin{eqnarray}\nonumber
&&|\lambda_i|<4\pi\,,\quad |\lambda^{\Omega}_{1,2,3}|<4\pi\,,\\
&&|y_{ij}|<\sqrt{4\pi}\,,\quad |Y_{N,i}|<\sqrt{4\pi}\,,\quad |Y_{\Sigma,i}|<\sqrt{4\pi}\,,\quad |Y_{\Omega}|<\sqrt{4\pi}\,.\label{eq:perturbativity}
\end{eqnarray}

\subsection{Phenomenological constraints}
\subsubsection{Electroweak precision observables}
 
 It is well-known that the presence of a Higgs triplet affects the $\rho$-parameter at tree-level~\cite{Schechter:1980gr}. Indeed, our triplet $\Omega$ contributes to the mass of the $W$ boson (but not to the $Z$ boson mass), so that the $\rho$-parameter at tree-level becomes~\cite{Diaz-Cruz:2003kcx} 

\begin{equation}
\rho=\frac{\sum_{i}c_i\langle\phi_i^0\rangle^2[T_i(T_i+1)-Y_i^2]}{2\sum_{i}^{}\langle\phi_i^0\rangle^2 Y_i^2}\Bigg|_{\phi_{i}\in\{\phi,\Omega\}}=1+\frac{4\,v_\Omega^2}{v_\phi^2}\,.\label{eq:rho_def}
\end{equation}
Here $\langle\phi_i^0\rangle$ is the VEV of the neutral component of  $\phi_i$, $T_i$ and $Y_i$ are the weak isospin and the hypercharge of $\phi_i$, and $c_i$ is a constant equal to $1/2$ and $1$ for real and complex scalar fields, respectively. Hence the precise measurement of the $\rho$-parameter puts a severe restriction on $v_\Omega$. The current global fit for $\rho$-parameter is $\rho=1.00038\pm 0.00020$ \cite{ParticleDataGroup:2024cfk}, nearly $2\sigma$ away from the tree-level SM expectation.  Taking the $3\sigma$ range of the fit, one can restrict the triplet VEV as $v_\Omega\lsim 4 \text{GeV}$.

\subsubsection{Neutrino oscillation constraints}

Explaining the smallness of neutrino masses is one of the key motivations of the model under
consideration. Hence, we briefly mention the constraints imposed on the model parameter space coming from neutrino oscillation data. Neutrino oscillation experiments provide bounds on  neutrino mass-squared differences and the mixing angles~\cite{deSalas:2020pgw,Esteban:2024eli}. In the following analysis, we use the $3\sigma$ ranges of the neutrino oscillation parameters reported by the global fit of NuFIT v6.1~\cite{Esteban:2024eli},  
\begin{eqnarray}\nonumber
  &&\sin^{2}\theta_{12}\in [0.2893\,,0.3295]\,,\\\nonumber
  &&\sin^{2}\theta_{13}\in [0.02064\,,0.02418]~\left( [0.02093\,,0.02441]\right)\,,\\\nonumber
  &&\sin^{2}\theta_{23}\in [0.435\,,0.584]~\left( [0.439\,,0.584]\right)\,,\\\nonumber
  &&\delta_{CP} \in [125^{\circ}\,,365^{\circ}]~\left( [203^{\circ}\,,335^{\circ}] \right)\,,  \\\nonumber
  &&\Delta m^2_{21}/(10^{-5}\text{eV}^2)\in [7.236\,,7.823]~\left( [7.236\,,7.822]\right)\,,\\
  &&\Delta m^2_{3\ell}/(10^{-3}\text{eV}^2)\in [2.450\,,2.576]~\left([-2.547\,,-2.421]\right)\,,\label{eq:nufit}
\end{eqnarray}
for the NO (IO) neutrino mass spectrum. Here, $\ell=1$ for NO and
$\ell=2$ for IO. From Eq.~\eqref{eq:CI}, we know that the Yukawa couplings $Y_{N,i}$ and $Y_{\Sigma,i}$ can be expressed in terms of the neutrino oscillation parameters and the free complex angle $\varrho$. In our analysis we vary the values of the neutrino oscillation parameters randomly in the corresponding $3\sigma$ regions.

\subsubsection{Collider constraints}

Collider searches for the inert doublet model (IDM), which extends the SM by introducing the inert scalar doublet $\eta$, provide the most relevant experimental constraints on the inert scalar masses $m_{\eta_R}$, $m_{\eta_I}$, and $m_{\eta^\pm}$. Approximate limits from the second phase of the LEP on the neutral inert scalars were obtained in Ref.~\cite{Lundstrom:2008ai} by simulating the process $e^+e^-\rightarrow\eta_R\eta_I$ under the assumption $m_{\eta_I}>m_{\eta_R}$. For mass splittings satisfying $m_{\eta_I}-m_{\eta_R}>8~\mathrm{GeV}$, the region of $m_{\eta_R}<80~\mathrm{GeV}$ and $m_{\eta_I}<100~\mathrm{GeV}$ is excluded.  In the present work, $|\lambda_5|$ is very small to accommodate the tiny neutrino mass, implying $|m_{\eta_I}-m_{\eta_R}|\sim\mathcal{O}(\mathrm{keV})$. Such an almost degenerate neutral spectrum is expected to substantially weaken the LEP II sensitivity, so that the above limits are not directly applicable. For the charged inert scalar $\eta^\pm$, reinterpretations of the OPAL searches~\cite{OPAL:2003nhx,OPAL:2003wxm} in Refs.~\cite{Pierce:2007ut,Blinov:2015qva} yield an approximate lower bound $m_{\eta^\pm}\gtrsim70\text{--}90~\mathrm{GeV}$ for $m_{\eta^\pm}-m_{\eta_R}\gtrsim5~\mathrm{GeV}$. However, for $m_{\eta^\pm}\gtrsim m_Z/2$ and a mass splitting $\text{a few hundred MeV}\lesssim m_{\eta^\pm}-m_{\eta_R}\lesssim5~\mathrm{GeV}$, the LEP searches don't constrain this model~\cite{Blinov:2015qva}.

The scalar triplet $\Omega$ is also subject to constraints from collider experiments. 
 In particular, the pair production of the charged scalar component, $pp \to \kappa_1^+ \kappa_1^- \to \tau^+\tau^-\nu\bar{\nu}$, gives rise to a di-tau plus missing transverse momentum signature. By reinterpreting recent searches at the LHC by the ATLAS and the CMS collaborations for supersymmetric tau partners, Ref.~\cite{Ashanujjaman:2024lnr} derived a lower bound of $m_{\kappa_1^\pm}>110~\mathrm{GeV}$ at the $95\%$ CL.
For the fermion triplet $\Sigma$, the type-III scotogenic model studied in Ref.~\cite{Singh:2025jtn} showed that the ATLAS search for electroweak production of triplet fermions using the $13$ TeV data with an integrated luminosity of $136~\mathrm{fb}^{-1}$ excludes triplet fermion masses below approximately $M_{\Sigma}\lesssim580~\mathrm{GeV}$ at $95\%$ CL. In the scenario of scalar DM, the singlet Majorana fermion $N$ is not directly constrained by current LHC searches due to its electroweak singlet nature, and no model-independent lower bound on $M_N$ is available. 

The most comprehensive LHC study of the $\Sigma$ and $N$ sector has recently been performed in Ref.~\cite{Lozano:2025tst}, where the parameter space was confronted with the current ATLAS and CMS searches. The dominant production channels are $pp\to\Sigma^+\Sigma^-$, $pp\to\Sigma^\pm\chi_i$, and $pp\to\chi_i\chi_j$ with $i,j=1,2$. The production cross section of the neutral-pair channel is typically too small compared with the other channels. The charged states $\Sigma^\pm$ mainly decay through $\Sigma^\pm\to W^{(*)}\chi_1\to\ell^\pm\nu\chi_1$. Using the latest ATLAS and CMS searches, Ref.~\cite{Lozano:2025tst} derived lower bounds on the mass of $\Sigma$ that depend on the scalar-sector parameters. For $v_\Omega=1.5~\mathrm{GeV}$, the excluded region approximately corresponds to $M_\Sigma\lesssim100~\mathrm{GeV}$ for $\mu_1=1.1~\mathrm{GeV}$ and $M_\Sigma\lesssim 150~\mathrm{GeV}$ for $\mu_1=137.6~\mathrm{GeV}$. For $v_\Omega=4~\mathrm{GeV}$ and $\mu_1=54~\mathrm{GeV}$, the corresponding limit is relaxed to approximately $M_\Sigma\gtrsim130~\mathrm{GeV}$. In contrast, no significant  constraint can be obtained on the mass of $N$ from the current LHC searches.

\section{Phenomenological implications\label{sec:ph}}

In the preceding sections, we have developed the theoretical formulation of LFV processes and the electroweak phase transition within the singlet--triplet scotogenic model and discussed the relevant theoretical and experimental constraints. We now proceed to a numerical investigation of its phenomenological predictions for LFV, GWs associated with an electroweak FOPT, and DM. A central objective is to determine whether a common and consistent choice of model parameters can simultaneously account for all the observables considered in this work. The corresponding numerical results and experimental prospects are presented in the following subsections.

\subsection{Lepton flavor violation\label{subsec:ph-lfv}}

Using the analytical expressions derived in section~\ref{sec:LFV}, we now investigate the numerical predictions for $\mu\to e\gamma$, $\mu\to e\overline{e}e$, and $\mu$--$e$ conversion in Ti. The LFV amplitudes mainly depend on the Yukawa couplings $Y_\Sigma$ and $Y_N$ and the masses of the particles entering the loops. Through the Casas-Ibarra parametrization in Eq.~\eqref{eq:CI}, these Yukawa couplings are expressed in terms of the neutrino mixing matrix, the light neutrino masses, the angular parameter $\varrho$, and the loop factors $\Lambda_{\alpha}$ governing radiative neutrino mass generation. The latter depend on the new-particle masses and the scalar coupling $\lambda_5$, which therefore also affect the Yukawa couplings required to reproduce the light neutrino masses. We thus examine the dependence of the LFV observables on the new-particle masses, $\lambda_5$, the neutrino mixing parameters, and $\varrho$, considering both NO and IO neutrino mass spectra. The mixing angles $\theta_{12}$, $\theta_{23}$, and $\theta_{13}$, the Dirac CP phase $\delta_{\mathrm{CP}}$, and the mass-squared differences $\Delta m^2_{21}$ and $\Delta m^2_{3\ell}$ are varied within their respective $3\sigma$ ranges given in Eq.~\eqref{eq:nufit}. Since the lightest neutrino is massless in this model, the light neutrino masses are fixed by the mass-squared differences:
\begin{align}
\text{NO}: \quad&
m_1=0,\qquad
m_2=\sqrt{\Delta m^2_{21}},\qquad
m_3=\sqrt{\Delta m^2_{31}},
\nonumber\\
\text{IO}: \quad&
m_1=\sqrt{|\Delta m^2_{32}|-\Delta m^2_{21}},\qquad
m_2=\sqrt{|\Delta m^2_{32}|},\qquad
m_3=0.
\end{align}

We first study the effects of the mass parameters $m_\eta$, $M_N$, and $M_\Sigma$ and the coupling $\lambda_5$, setting the Majorana phase $\alpha_{21}$ and the angular parameter $\varrho$ to zero. The effects of nonzero $\alpha_{21}$ and $\varrho$ are examined subsequently. The remaining parameters are chosen as
\begin{align}
Y_{\Omega}&=7\times10^{-3},
&
v_\Omega&=0.05~\mathrm{GeV},
&
\lambda_3&=7\times10^{-2},
\nonumber\\
\lambda_4&=5\times10^{-4},
&
\lambda^{\Omega}_{3}&=0.1,
&
\mu_2&=300~\mathrm{GeV}.
&&
\label{eq:LFV-numvalue}
\end{align}
These parameters affect the LFV observables only indirectly through their contributions to the mass spectra and therefore do not induce sizable variations. In particular, the parameter $\mu_2$ is crucial for ensuring that the $\mathbb{Z}_2$ symmetry remains intact up to high energy scales~\cite{Merle:2016scw,Rocha-Moran:2016enp}. Meanwhile, the small value of $v_\Omega$ guarantees that the deviation of the electroweak precision parameter $\rho$ from unity remains negligible, maintaining consistency with SM constraints.

\subsubsection{Predictions for $\mu\rightarrow e\gamma$}

\begin{figure}[t!]
\centering
\includegraphics[scale=0.29,valign=c]{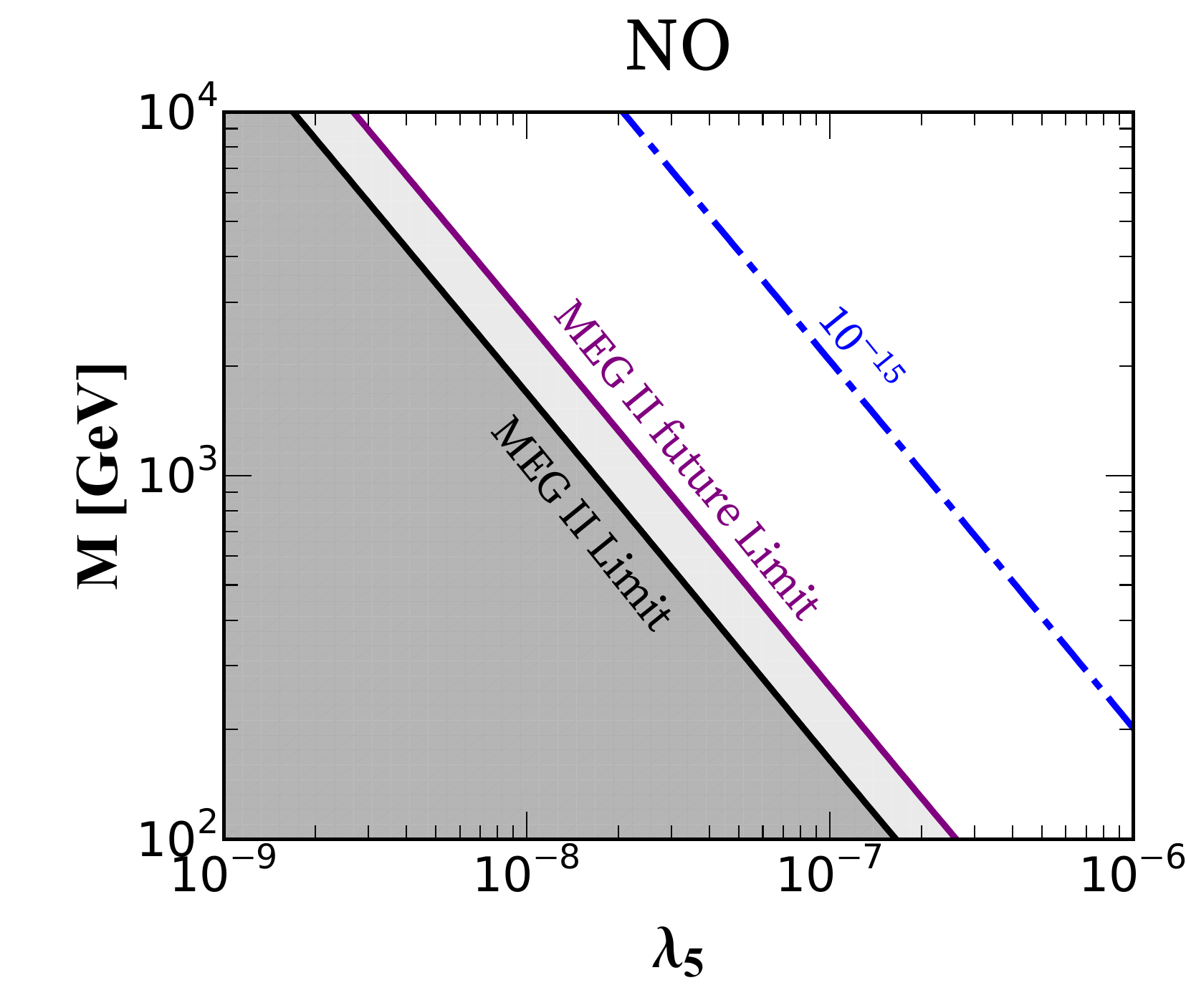}
\includegraphics[scale=0.29,valign=c]{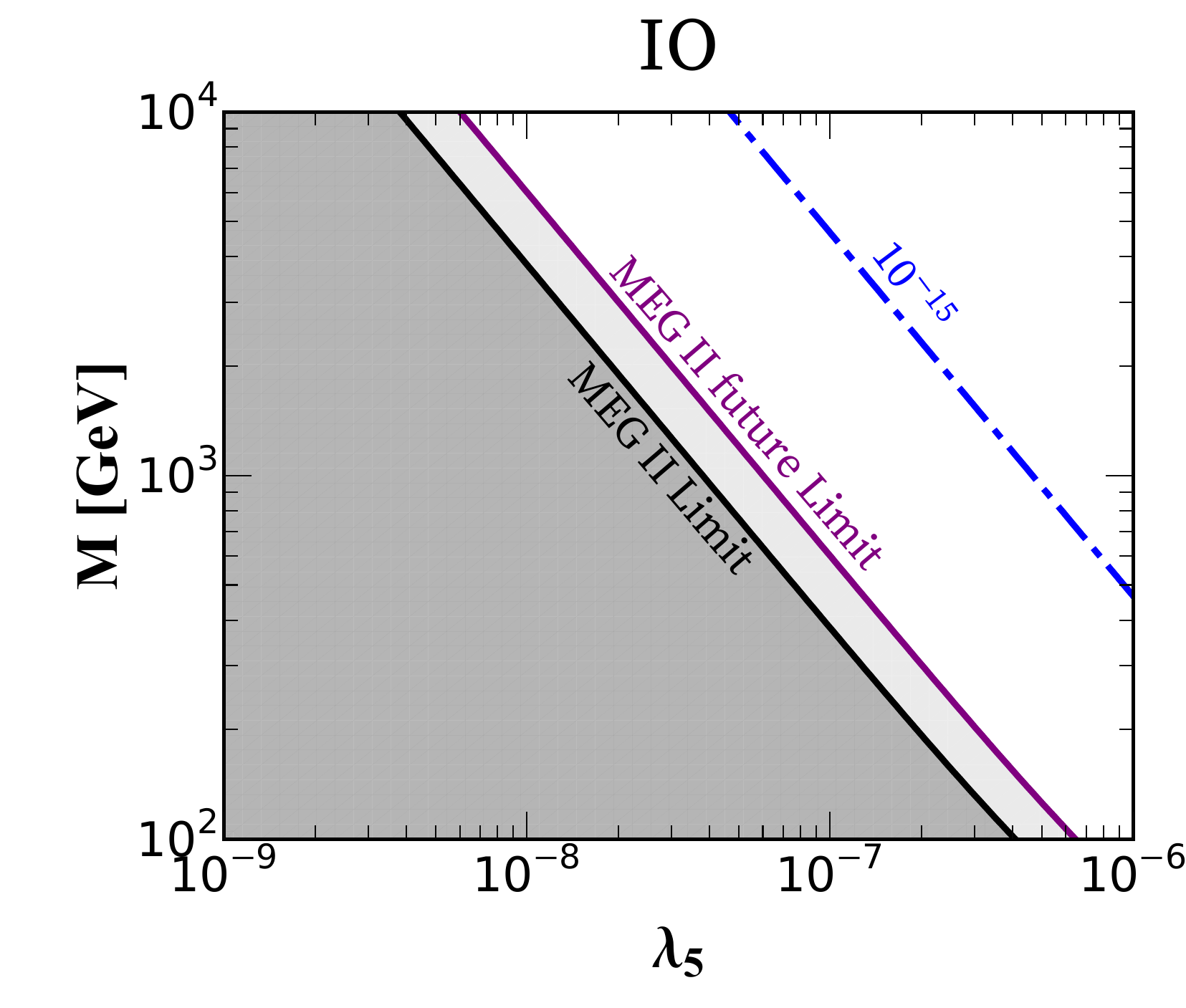}
\includegraphics[scale=0.29,valign=c]{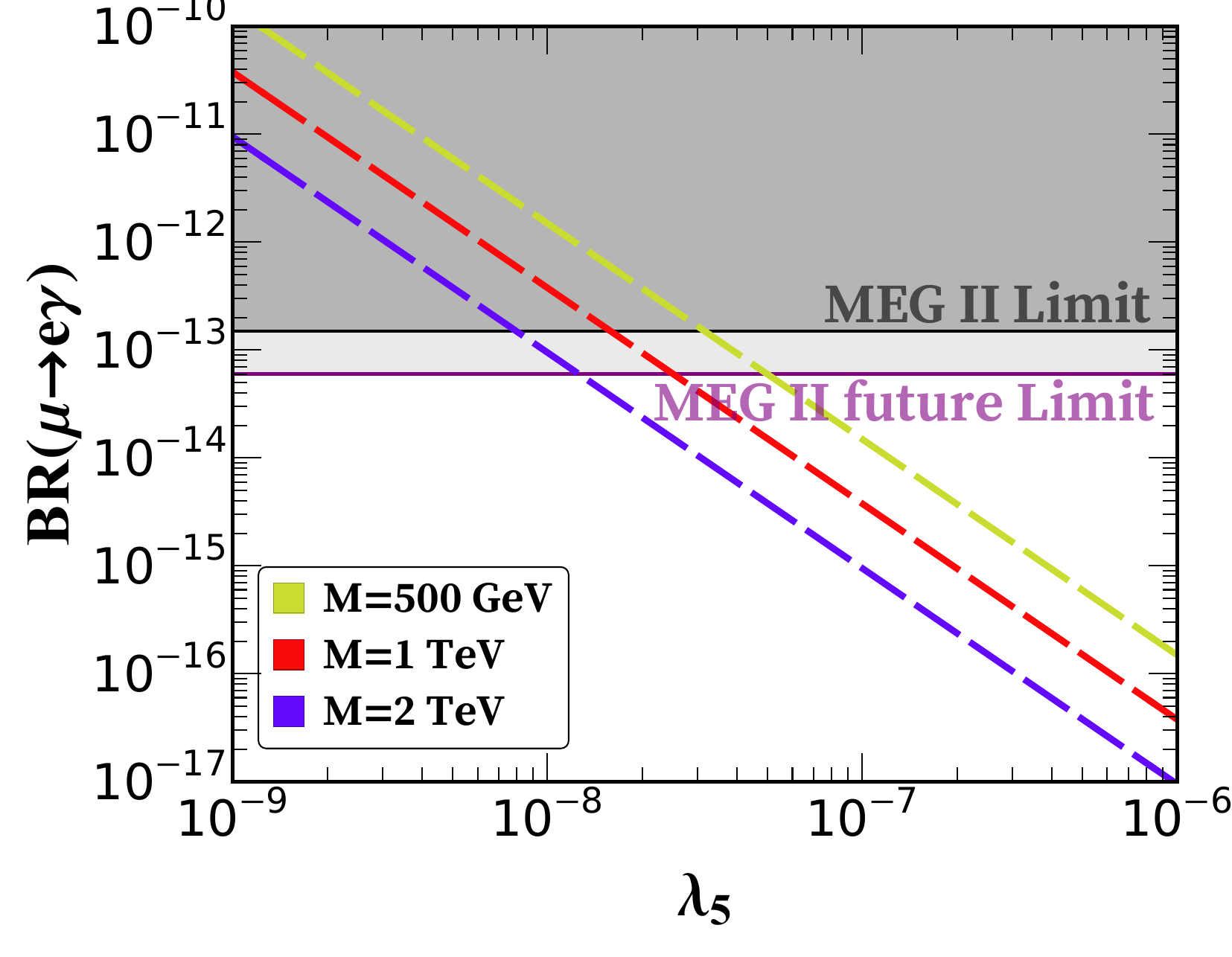}
\includegraphics[scale=0.29,valign=c]{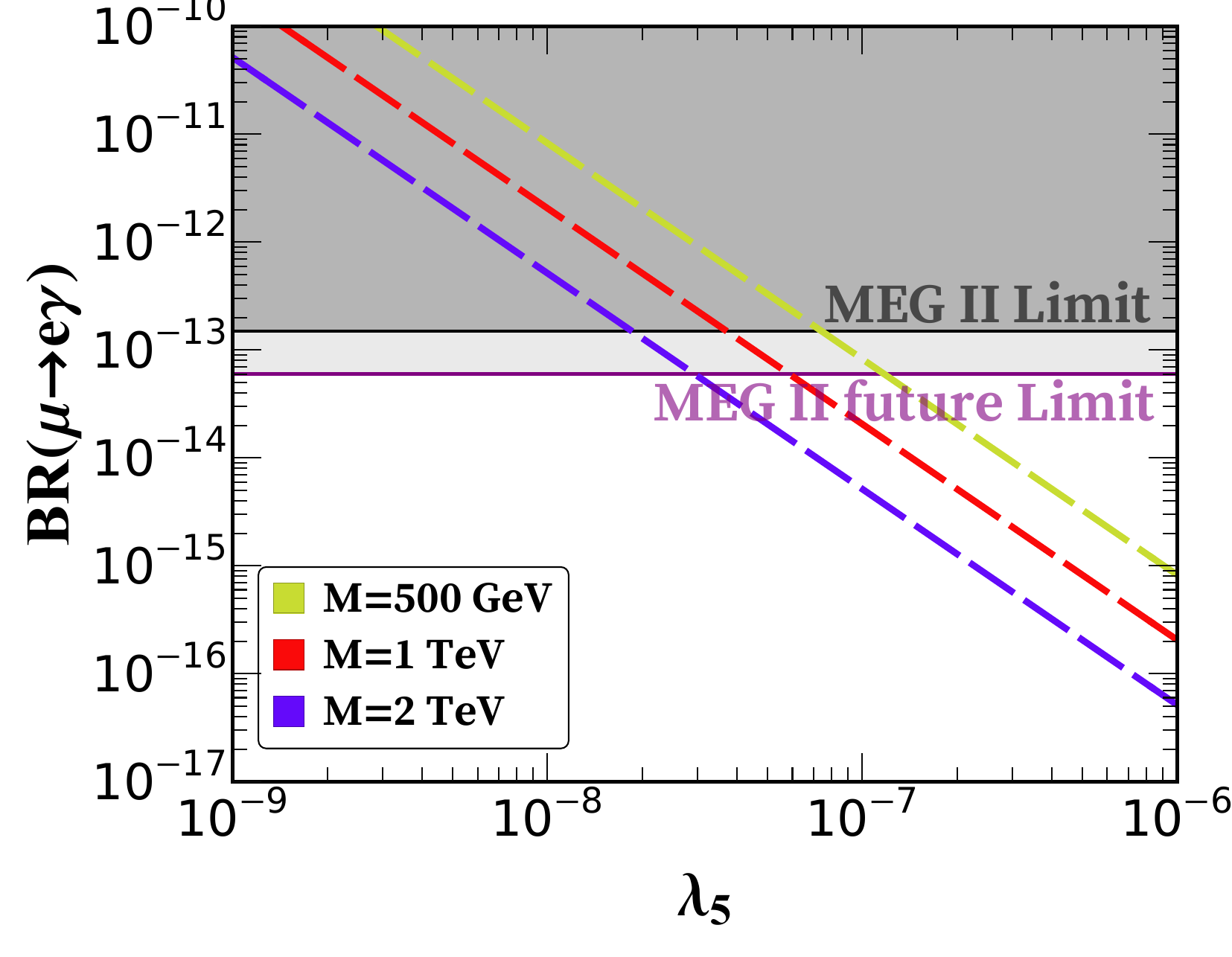}
\caption{The first-row panels show contour plots of the $\mathrm{BR}(\mu \to e \gamma)$ in the $(M,\lambda_5)$ plane, where $M\equiv m_\eta=M_\Sigma$ and $M_N=M+\Delta M$ with $\Delta M=10~\mathrm{GeV}$. The second-row panels display the dependence of $\mathrm{BR}(\mu \to e \gamma)$ on $\lambda_5$ for the representative mass values $M=500~\mathrm{GeV}$, $1~\mathrm{TeV}$, and $2~\mathrm{TeV}$. The two left panels correspond to NO, while the two right panels correspond to IO. 
  }
        \label{fi:RP-meg-lam5}
\end{figure}

In figure~\ref{fi:RP-meg-lam5}, we show the dependence of $\mathrm{BR}(\mu \to e \gamma)$ on the mass scale $M$ and the dimensionless coupling $\lambda_5$, where $M=m_\eta=M_\Sigma$ and $M_N=M+\Delta M$ with $\Delta M=10~\mathrm{GeV}$. Here, the neutrino mixing parameters are set to their best-fit values from the latest global analysis~\cite{Esteban:2024eli}, while $\alpha_{21}$ and $\varrho$ are set to zero. The remaining parameters are fixed as specified in Eq.~\eqref{eq:LFV-numvalue}. The first-row panels show that the $\mathrm{BR}(\mu\to e\gamma)$ decreases as $M$ and $\lambda_5$ increase.  The second-row panels further demonstrate that $\ln \mathrm{BR}(\mu\to e\gamma)$ exhibits an approximately linear dependence on $\ln \lambda_5$ for fixed $M$. This linear behavior can be understood as follows. In the range $100~\mathrm{GeV}\leq M\leq 100~\mathrm{TeV}$ shown in figure~\ref{fi:RP-meg-lam5}, we have $m_\eta=M_\Sigma\simeq M_N$, and hence $m_{\eta_R}\simeq m_{\eta_I}\simeq m_{\chi_1}\simeq m_{\chi_2}\simeq M$. The small mass splitting $\Delta M$ drives the mixing angle $\theta\simeq0$. From Eqs.~(\ref{eq:Lam-chi-eq-eta}, \ref{eq:CI}), we find that $h$ at leading order in $\lambda_5$ can be approximated as
\begin{eqnarray}
h\propto\sqrt{\Lambda}^{-1}\propto\sqrt{\frac{M}{\lambda_5}}\,.
\end{eqnarray}
Therefore the Yukawa couplings $Y_\Sigma$ and $Y_N$ scale as
\begin{eqnarray}
Y_{\Sigma, N}\propto\sqrt{\frac{M}{\lambda_5}}\,.
\end{eqnarray}
From Eq.~\eqref{eq:K2R-exp}, we know that the dipole coefficient $K^{R}_2$ scales as $K_{2}^{R}\propto M^{-2}Y^2$. As a consequence, we have
\begin{equation}\label{eq:BR_lambda5}
\mathrm{BR}(\mu\to e\gamma)\propto |K_{2}^{R}|^{2}
\propto\left(\frac{1}{M\lambda_{5}}\right)^{2}\,.
\end{equation}
This analytic scaling is consistent with the numerical results shown in figure~\ref{fi:RP-meg-lam5}, confirming that $\mathrm{BR}(\mu\to e\gamma)$ decreases with increasing $\lambda_5$. The values of $\lambda_5$ allowed by the adopted MEG~II bound also depend on the mass scale $M$. For $M\simeq2~\text{TeV}$, the lower limit on $\lambda_5$ is of order $10^{-8}$, with a stronger restriction for IO than for NO. 

\begin{figure}[t!]
\centering
\includegraphics[scale=0.38,valign=c]{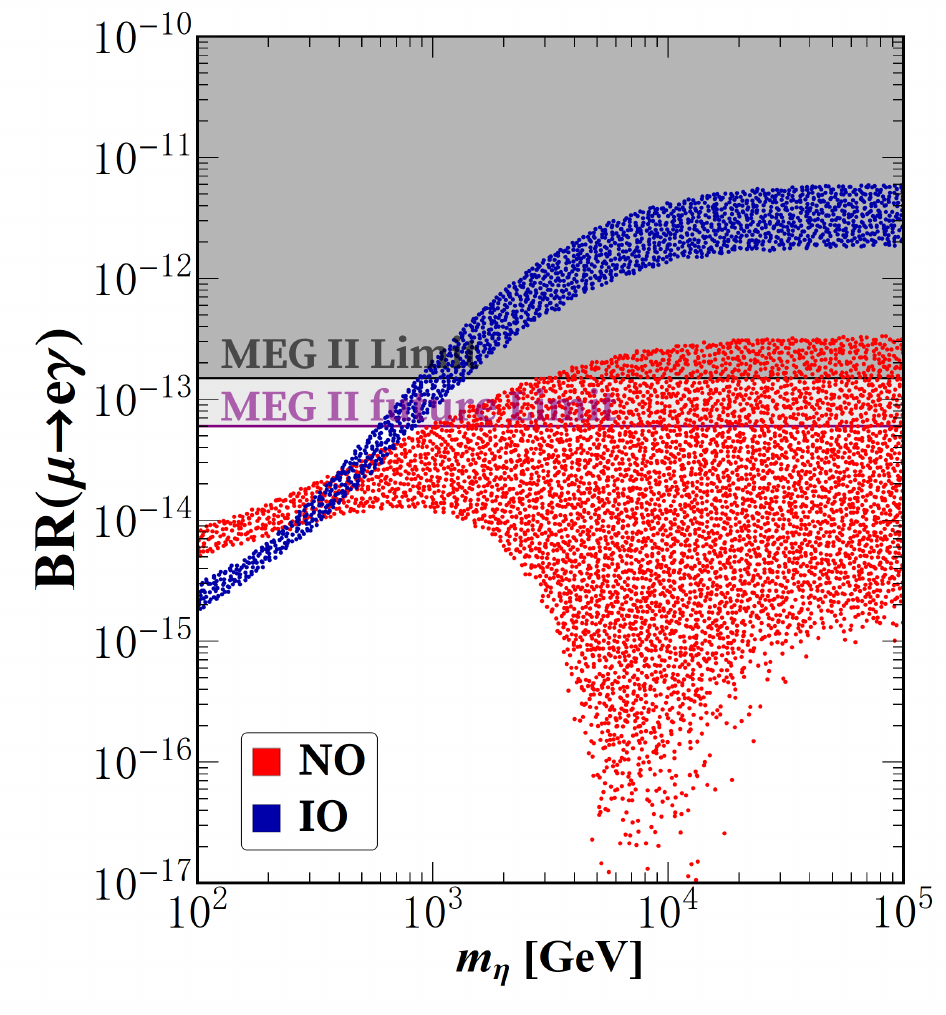}
\includegraphics[scale=0.317,valign=c]{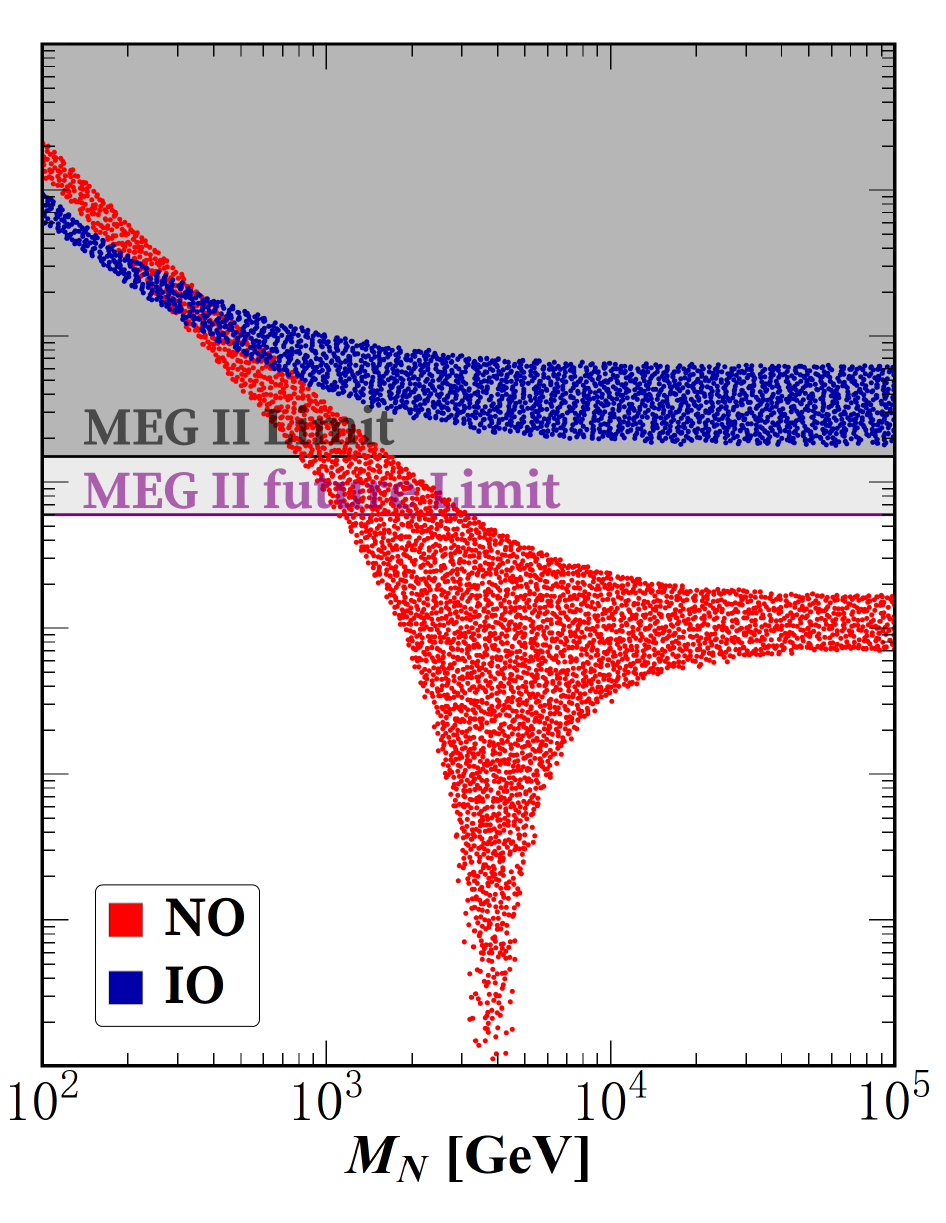}
\includegraphics[scale=0.317,valign=c]{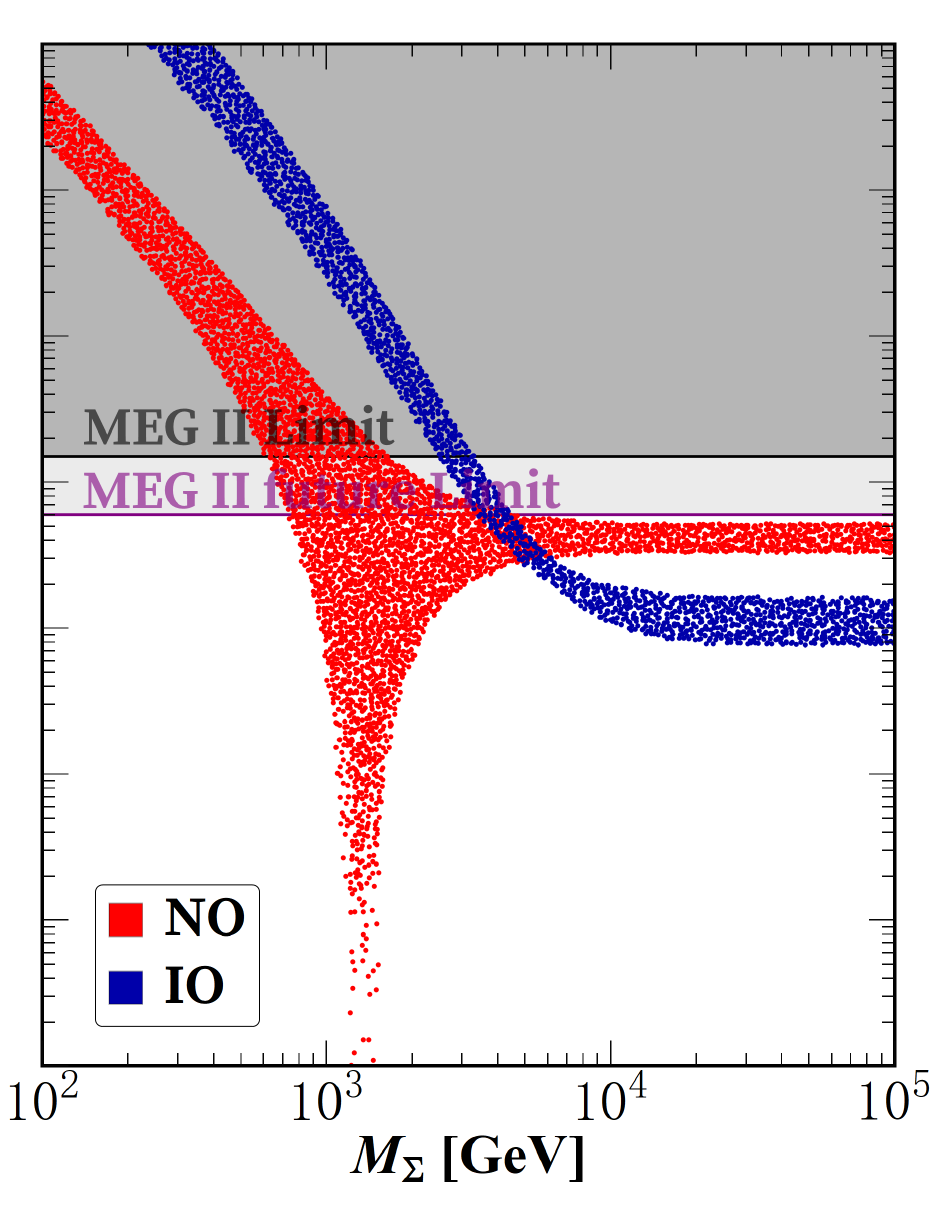}
\caption{The BR$(\mu \to e \gamma)$ as a function of $m_\eta$ (GeV), $M_N$ (GeV), and $M_\Sigma$ (GeV). The red points correspond to the NO, while the blue points represent the IO. The black horizontal lines indicate the  current  upper limit $1.5\times10^{-13}$~\cite{MEGII:2025gzr} from MEG~II, and the purple horizontal lines denote its projected sensitivity, $6\times10^{-14}$~\cite{Venturini:2024keu}.\label{fi:RP-meg}  }
\end{figure}

In figure~\ref{fi:RP-meg}, we show the dependence of $\mathrm{BR}(\mu\to e\gamma)$ on the masses $m_{\eta}$, $M_N$, and $M_{\Sigma}$ by varying one mass parameter within the range $[10^{2},10^{5}]$ GeV at a time, while fixing the other two masses to their benchmark values, $m_\eta=M_\Sigma=2~\text{TeV}$ and $M_N=2.01~\text{TeV}$. The neutrino parameters are varied within the ranges shown in Eq.~\eqref{eq:nufit}, while the other model parameters are fixed as in Eq.~\eqref{eq:LFV-numvalue}. In addition, we take $\alpha_{21}=\varrho=0$ and $\lambda_5 = 10^{-8}$ for illustration. 

As can be seen from figure~\ref{fi:RP-meg}, the MEG~II limit gives a weaker constraint on $m_{\eta}$ for NO, whereas $m_{\eta}$ is required to be less than $1\,\mathrm{TeV}$ for IO neutrino masses.  When varying $M_N$, the MEG II limit can be saturated for $M_N\gtrsim1~\text{TeV}$ in the case of NO, whereas the IO case is already excluded by the current data of MEG II. Similarly, for $M_\Sigma$, larger values of $M_{\Sigma}$ are preferred by both the present and future bounds of the MEG II experiment.  In addition, we see pole-like structure in figure~\ref{fi:RP-meg} in the regions $m_{\eta}\sim 10~\text{TeV}$, $M_N\sim 4~\text{TeV}$ and $M_{\Sigma}\sim 1~\text{TeV}$ in the case of NO. This feature originates from occasional cancellation between $D^0$ and $D^-$ in Eq.~\eqref{eq:K2R-exp} at the mass values specified above. When the neutrino parameters are varied within the ranges given in Eq.~\eqref{eq:nufit}, $\mathrm{BR}(\mu\to e\gamma)$ develops a width as the mass parameters $m_\eta$, $M_N$, and $M_\Sigma$ are varied. The width of the band is determined mainly by $\delta_{\mathrm{CP}}$, owing to its relatively large experimental uncertainty. Therefore, a more precise measurement of $\delta_{\mathrm{CP}}$ at future neutrino oscillation experiments would help to sharpen the prediction for $\mathrm{BR}(\mu\to e\gamma)$.

It is also worth noting that when the masses $m_\eta$, $M_N$, and $M_\Sigma$ increase beyond a certain scale, $\mathrm{BR}(\mu\to e\gamma)$ approaches a plateau. For larger $m_\eta$, the variables $\zeta_{1,2}$ and $\rho_{R,I}$ tend to zero, as follows from their definitions in Eq.~\eqref{eq:zeta-rho-def}. As a result, the loop functions $F_2(\zeta_{1,2})$ and $G_2(\rho_{R,I})$ approach $1/6$ and $1/3$, respectively. From Eq.~\eqref{eq:Lam-chi-ll-eta}, we have
\begin{eqnarray}
  \Lambda_\alpha\propto\frac{1}{m_\eta^2}\,.
\end{eqnarray}
Then, from Eq.~\eqref{eq:CI}, we find that $h$ can be approximated as
\begin{eqnarray}
    h\propto\sqrt{\Lambda_\alpha}^{-1}\propto m_\eta\,.
\end{eqnarray}
Since $m_\eta$ does not affect the fermion mixing angle $\theta$, the Yukawa couplings $Y_\Sigma$ and $Y_N$ scale as
\begin{eqnarray}
    Y_{\Sigma, N}\propto m_\eta\,.\label{eq:Y-meta}
\end{eqnarray}
From Eq.~\eqref{eq:K2R-exp}, the dipole coefficient scales as $K_{2}^{R}\propto m_\eta^{-2}Y^2$. Consequently, we have
\begin{equation}
\mathrm{BR}(\mu\to e\gamma)\propto |K_{2}^{R}|^{2}
\propto m_\eta^0\,,
\end{equation}
for large $m_\eta$. This is in agreement with the behavior shown in the left panel of figure~\ref{fi:RP-meg}. 

In the regime $M_{N}\gg 2~\text{TeV}$ with $M_\Sigma=m_\eta = 2~\text{TeV}$, Eq.~\eqref{eq:mchi2} implies $m_{\chi_1}\simeq m_\eta = 2~\text{TeV}$, $m_{\chi_2}\simeq M_{N}$, and $\theta \simeq 0$. In this limit, the parameters $\Lambda_{1,2}$ scale as $\Lambda_{1}\propto m_{\chi_1}^{-1}$ and $\Lambda_{2}\propto m_{\chi_2}^{-1}$, as shown in Eqs.~\eqref{eq:Lam-chi-eq-eta} and \eqref{eq:Lam-chi-gg-eta}. From the relation given in Eq.~\eqref{eq:CI}, one finds $Y_{N,i}\propto m_{\chi_2}^{1/2}$, while $Y_{\Sigma,i}\propto m_{\chi_1}^{1/2}$ remains approximately unchanged. Moreover, the mass-squared ratios defined in Eq.~\eqref{eq:zeta-rho-def} give $\zeta_1\simeq \rho_{R,I}\simeq 1$ and $\zeta_2\gg 1$. Consequently, the loop functions satisfy $F_2(\zeta_1)\simeq G_2(\rho_{R,I})\simeq 1/12$ and $F_2(\zeta_2)\simeq 0$. Therefore, from Eq.~\eqref{eq:D0Dm}, both $D^0$ and $D^-$ are proportional to $m_{\chi_1}$, implying that $\mathrm{BR}(\mu\to e\gamma)$ becomes independent of the large value of $M_N$. Similarly, in the limit $M_\Sigma\gg 2~\text{TeV}$ with $M_{N}\simeq m_{\eta}=2~\text{TeV}$, one has $m_{\chi_1}\simeq 2~\text{TeV}$, $m_{\chi_2}\simeq M_\Sigma$, and $\theta \to \pi/2$. In this case, $Y_{\Sigma,i}\propto m_{\chi_2}^{1/2}$, whereas $Y_{N,i}\propto m_{\chi_1}^{1/2}$ remains unchanged. Furthermore, $\zeta_1\simeq 1$ and $\zeta_2\simeq \rho_{R,I}\gg 1$, leading to $F_2(\zeta_1)\simeq 1/12$ and $F_2(\zeta_2)\simeq G_2(\rho_{R,I})\simeq 0$. Thus, Eq.~\eqref{eq:D0Dm} reduces to $D^0\propto m_{\chi_{1}}$ and $D^-\simeq 0$, implying that $\mathrm{BR}(\mu\to e\gamma)$ becomes independent of the large value of $M_\Sigma$. In summary, $\mathrm{BR}(\mu\to e\gamma)$ approaches a constant in either the large-$M_N$ or large-$M_\Sigma$ limit.

\subsubsection{Predictions for $\mu\rightarrow e\bar{e}e$}

\begin{figure}[htbp]
\centering
\includegraphics[scale=0.38,valign=c]{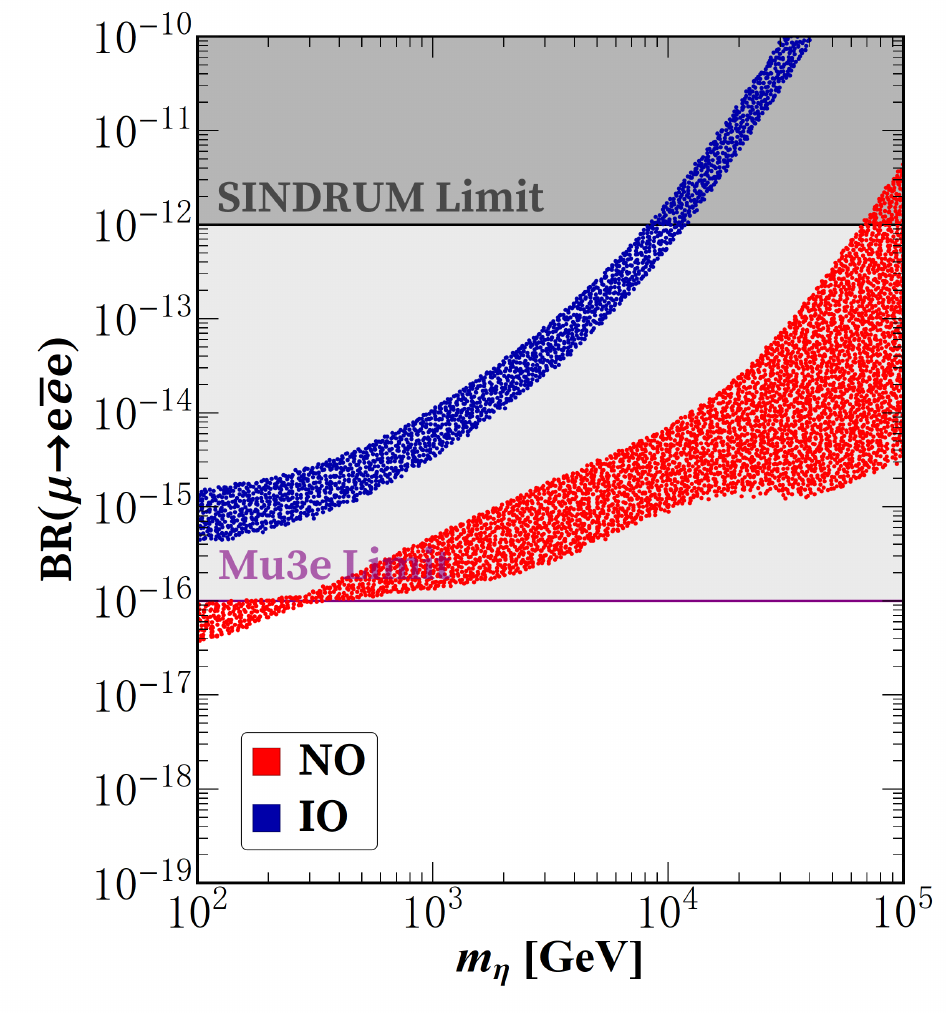}
\includegraphics[scale=0.317,valign=c]{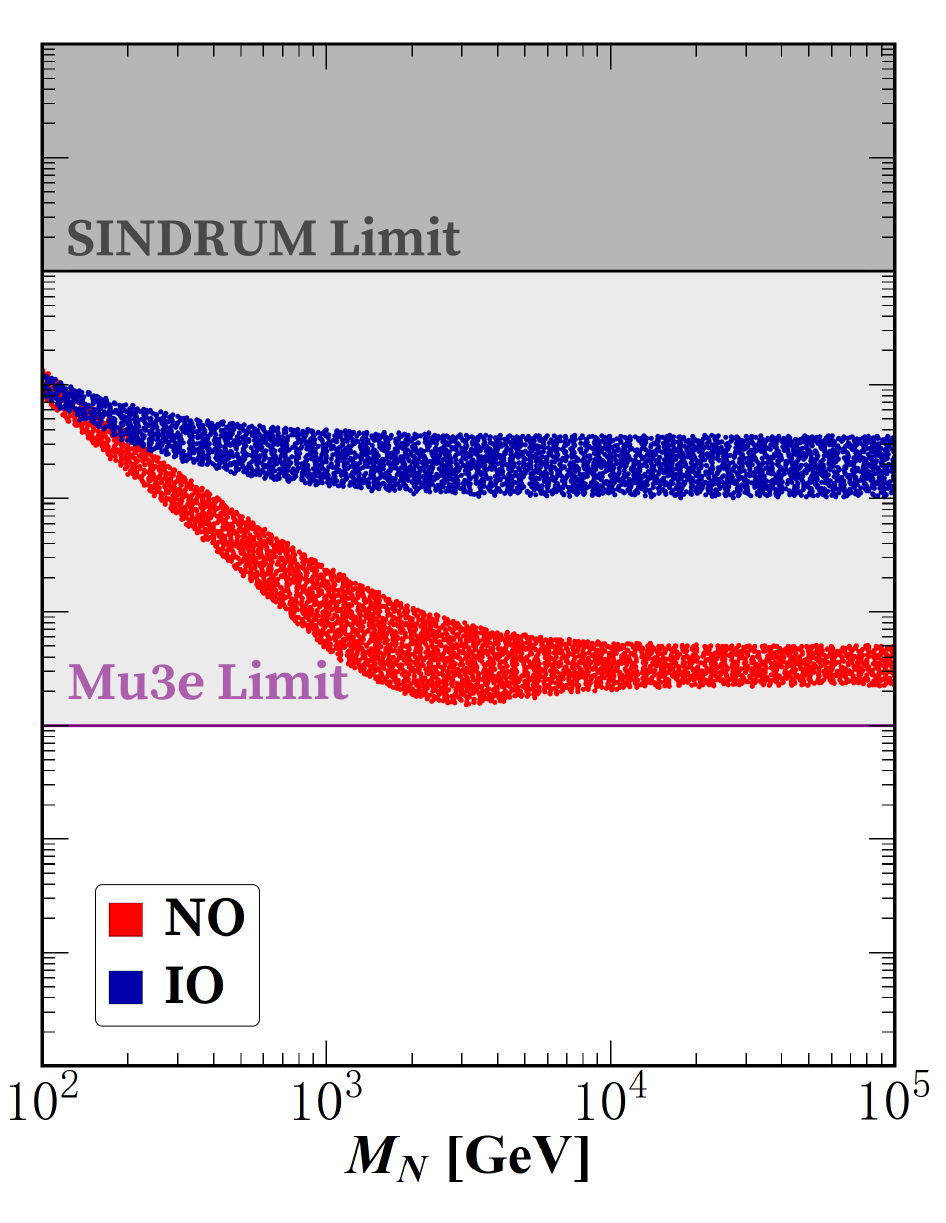}
\includegraphics[scale=0.317,valign=c]{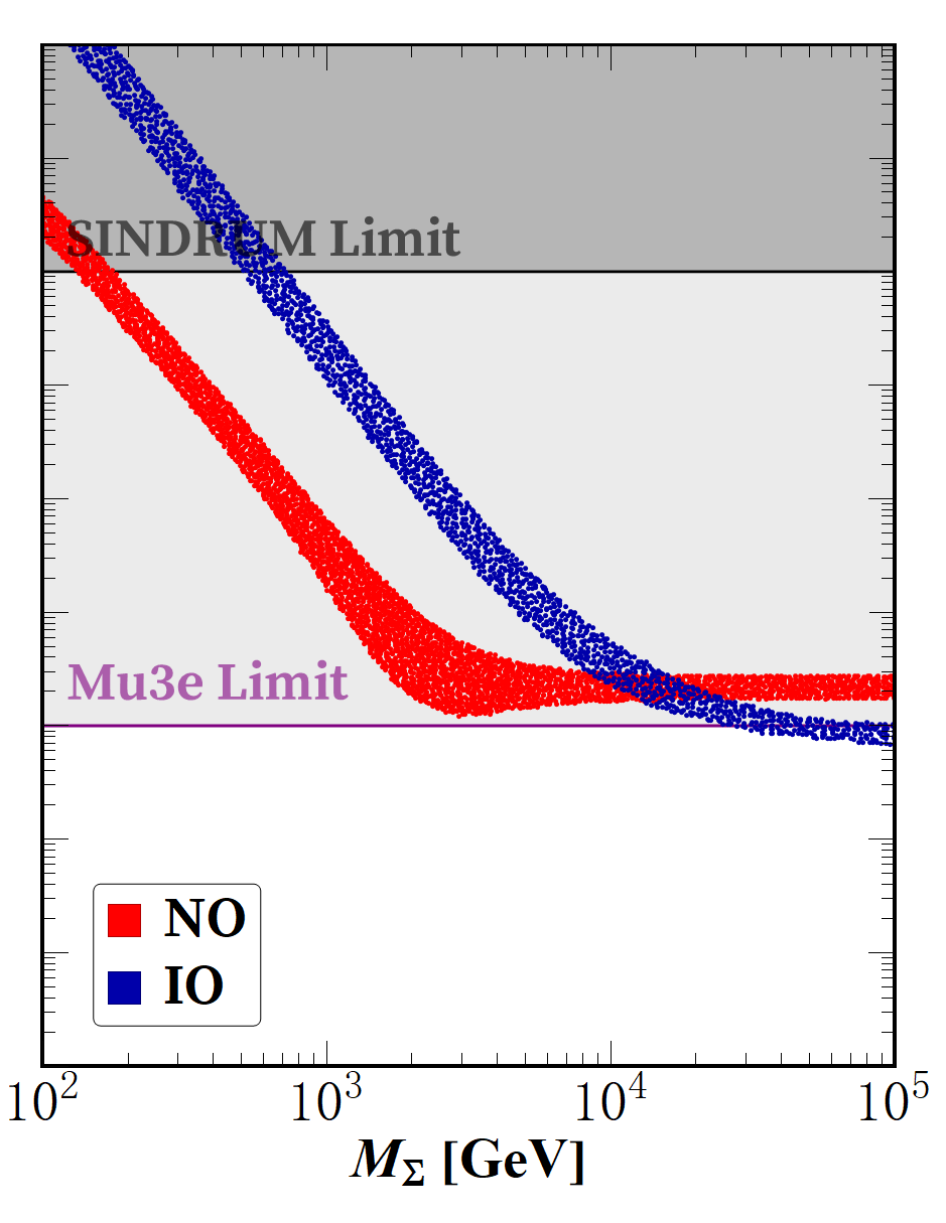}
\caption{The BR$(\mu\to e\overline{e}e)$ as a function of $m_\eta$ (GeV), $M_N$ (GeV), and $M_\Sigma$ (GeV). The red points correspond to the NO, while the blue points represent the IO. The black horizontal lines show the SINDRUM bounds~\cite{SINDRUM:1987nra}. The purple horizontal lines show the projected Mu3e sensitivity~\cite{MartinPerez:2023vfv}. \label{fi:RP-m3e}}
\end{figure}

For the $\mu\to e\overline{e}e$ process, figure~\ref{fi:RP-m3e} shows the dependence of $\mathrm{BR}(\mu\to e\overline{e}e)$ on the masses $m_{\eta}$, $M_N$, and $M_{\Sigma}$, where one mass parameter is varied at a time while the other two are fixed to the same benchmark values as in figure~\ref{fi:RP-meg}. The remaining parameters are also taken to be the same as in figure~\ref{fi:RP-meg}. The predictions for $\mathrm{BR}(\mu \rightarrow e\bar{e}e)$ form a band, whose thickness is predominantly induced by the large uncertainty in $\delta_{\text{CP}}$.

From the first panel of figure~\ref{fi:RP-m3e}, we find that in the NO case the $\mathrm{BR}(\mu\to e\overline{e}e)$ remains below the SINDRUM limit up to nearly $100~\text{TeV}$, whereas in the IO case it is compatible with the SINDRUM bound only for $m_\eta < 10~\text{TeV}$. The BR increases with $m_\eta$ and does not exhibit a plateau behavior. This behavior can be understood as follows. In the large-$m_\eta$ regime, the box diagrams provide the dominant contribution, while the $\gamma$-penguin contribution is subdominant and the $Z$-penguin contribution is negligible. The box Wilson coefficient $B_1^L$, shown in Eq.~\eqref{eq:B1L-exp}, depends quartically on the Yukawa couplings $Y_{\chi}$, which are linear combinations of $Y_\Sigma$ and $Y_N$. In the large-$m_\eta$ limit, $Y_{\Sigma,N} \propto m_\eta$ as given in Eq.~\eqref{eq:Y-meta}, and the loop functions satisfy $G_{13}(\zeta_\alpha,\zeta_\beta)\simeq 1/4$ and $G_{12}(\zeta_\alpha,\zeta_\beta)\simeq G_{14}(\rho_\alpha,\rho_\beta)\simeq 0$. Consequently, $B_1^L$ scales as $m_\eta^2$. For the $\gamma$-penguin contribution, the Wilson coefficient $K_2^R$ remains approximately constant, as in the $e_j \to e_i \gamma$ case. The coefficient $K_1^L$ receives two contributions, $M^0$ and $M^-$, as defined in Eq.~\eqref{eq:K1L-exp}. The $M^0$ term in Eq.~\eqref{eq:M0Mm} tends to a constant for large $m_\eta$, since $F_1(\zeta_{1,2})=1/3$. In contrast, the $M^-$ term depends on the loop function $G_1(\rho_{R,I})$, which scales logarithmically with $m_\eta$. As a result, $K_1^L$ increases as $\ln m_\eta$. Thus the $\gamma$-penguin contribution is subdominant compared to the box diagrams. The $Z$-penguin contribution to $\mathrm{BR}(\mu\to e\overline{e}e)$ is negligible over the range $100~\text{GeV} \lesssim m_\eta \lesssim 100~\text{TeV}$. This is because the corresponding Wilson coefficients $F_L$ and $F_R$, defined in Eqs.~\eqref{eq:FL} and~\eqref{eq:FR}, are strongly suppressed by the small value of $\lambda_5$ and by the ratio of charged-lepton masses to the $Z$-boson mass, respectively. For $m_\eta$ below the TeV scale, the Yukawa couplings $Y_N$ and $Y_\Sigma$ are of order $\mathcal{O}(0.1)$ or smaller. Consequently, the $\gamma$-penguin contribution, which scales with the square of these couplings, dominates over the box diagrams. As $m_\eta$ increases, the dominant contribution transitions from the $\gamma$-penguin to the box diagrams, while the $Z$-penguin contribution remains negligible throughout.

Over the range $100~\text{GeV}$--$100~\text{TeV}$ for both $M_N$ and $M_\Sigma$, the corresponding predictions for $\text{BR}(\mu\rightarrow e\bar{e}e)$ are below the SINDRUM limit, except for the very low-$M_\Sigma$ region below a few hundred GeV. The BR decreases as $M_N$ or $M_\Sigma$ increases and eventually reaches a plateau.  The $\gamma$-penguin contribution dominates over the other contributions in the concerned mass range. The origin of the plateau is the same as that in the $e_j\to e_i\gamma$ case.  In summary, the current SINDRUM result does not impose stringent bounds on the mass ranges of $m_\eta$, $M_N$, and $M_\Sigma$. However, the next-generation Mu3e experiment is expected to significantly tighten the allowed parameter space.

\subsubsection{Predictions for $\mu$--$e$ conversion }
\begin{figure}[htbp]
\centering
\includegraphics[scale=0.38,valign=c]{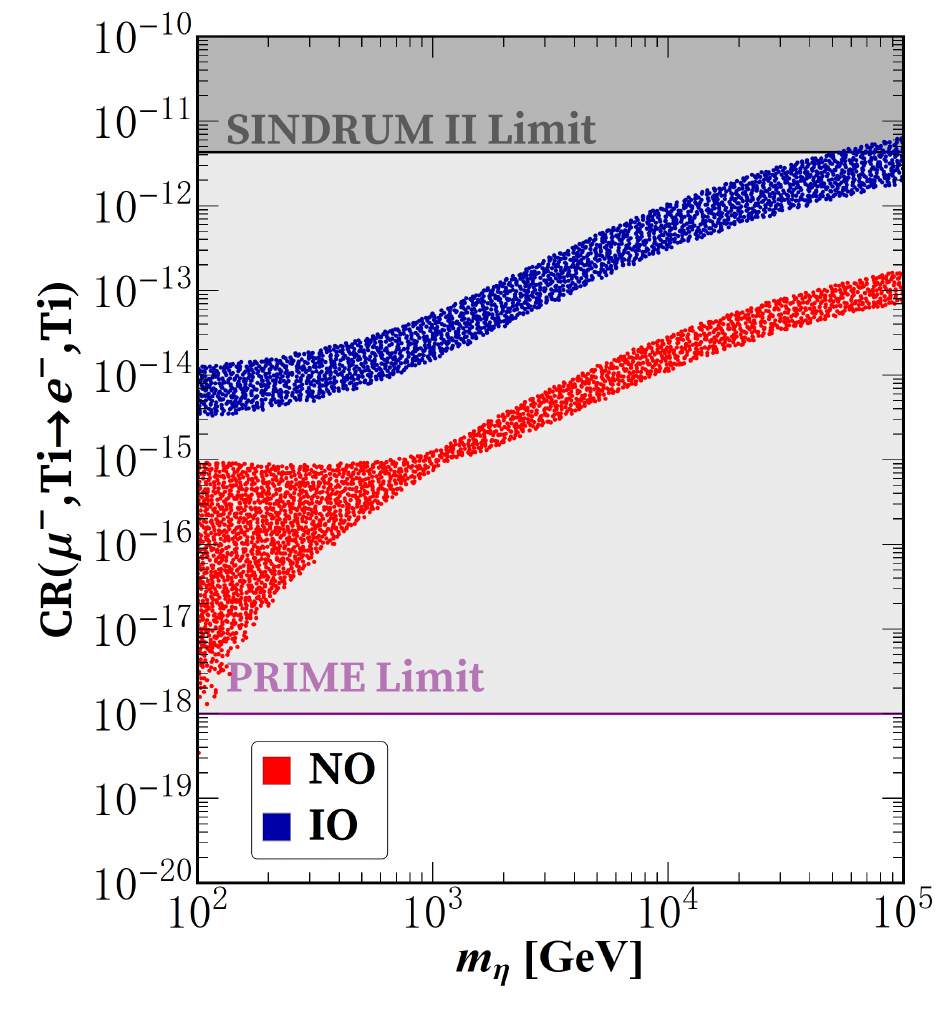}
\includegraphics[scale=0.317,valign=c]{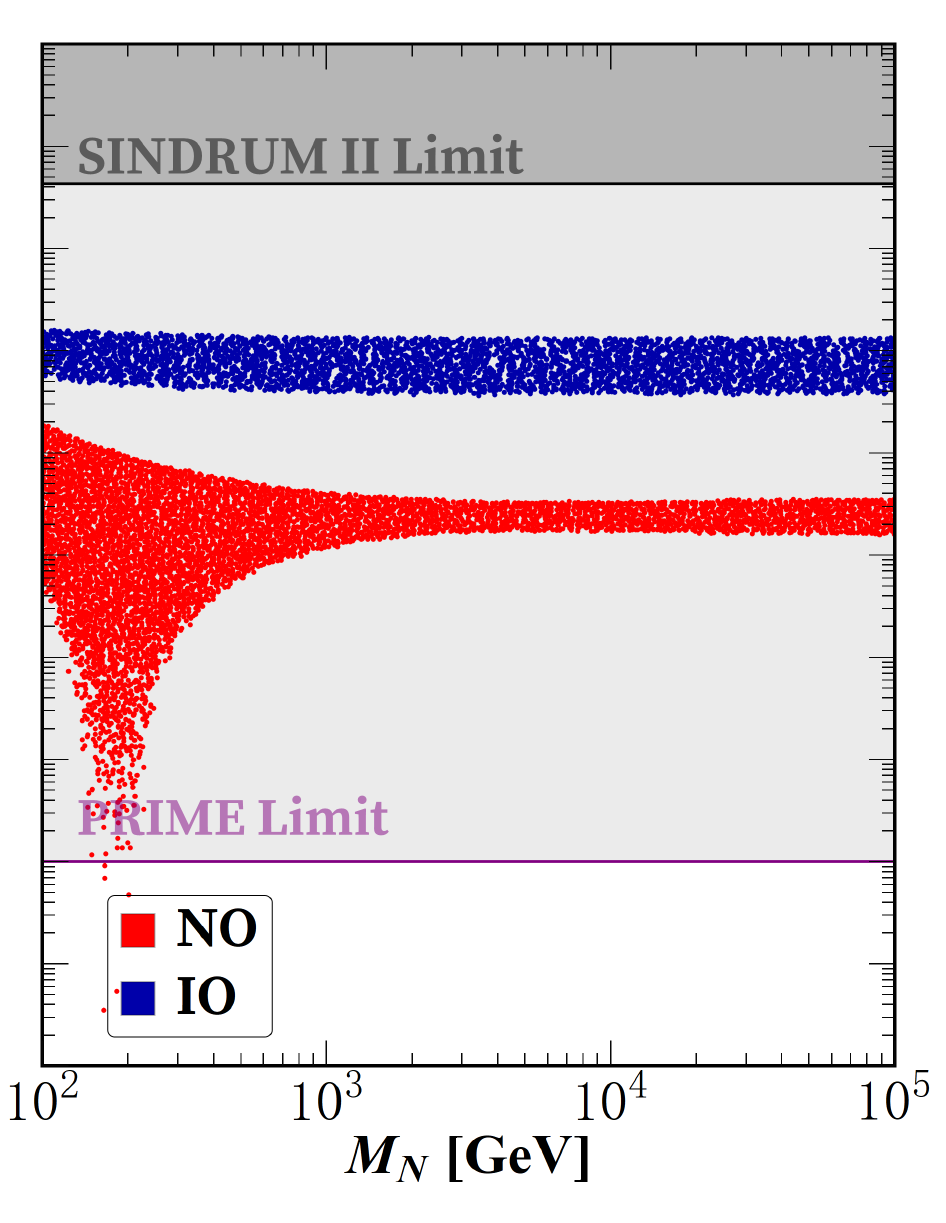}
\includegraphics[scale=0.317,valign=c]{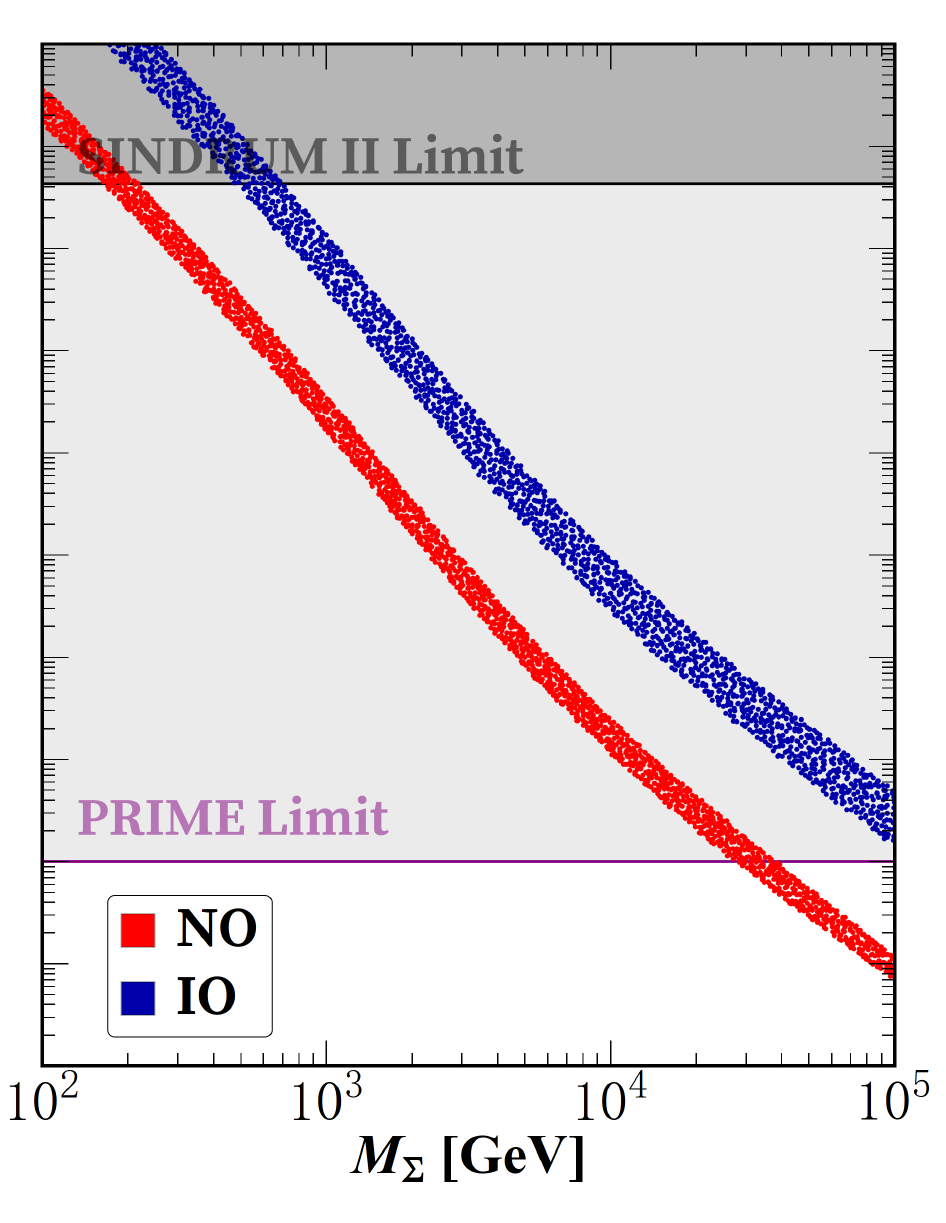}
\caption{The CR$(\mu^-,\text{Ti}\to e^-,\text{Ti})$ as a function of $m_\eta$ (GeV), $M_N$ (GeV), and $M_\Sigma$ (GeV). The red points correspond to the NO, while the blue points represent the IO. The black horizontal lines show the SINDRUM II bound~\cite{SINDRUMII:1993gxf}. The purple horizontal lines show the sensitivity target envisaged in the PRIME~\cite{prime}.\label{fi:RP-mec} }
\end{figure}

In this section, we analyze the process $\mathrm{CR}(\mu^-, \mathrm{Ti} \to e^-, \mathrm{Ti})$ as a representative channel for $\mu$--$e$ conversion. Figure~\ref{fi:RP-mec} shows the dependence of $\mathrm{CR}(\mu^-, \mathrm{Ti} \to e^-, \mathrm{Ti})$ on the masses $m_{\eta}$, $M_N$, and $M_{\Sigma}$. We find that $\mathrm{CR}(\mu^-, \mathrm{Ti} \to e^-, \mathrm{Ti})$ satisfies the SINDRUM~II bound~\cite{SINDRUMII:1993gxf}, except in the region where $M_\Sigma$ is of the order of a few hundred GeV. Over the range from $100~\mathrm{GeV}$ to $100~\mathrm{TeV}$, $\mathrm{CR}(\mu^-, \mathrm{Ti} \to e^-, \mathrm{Ti})$ increases with $m_\eta$ and decreases with $M_\Sigma$. In contrast, its dependence on $M_N$ is rather mild, except in the NO case when $M_N$ is around a few hundred GeV.

In all three panels of figure~\ref{fi:RP-mec}, the predictions for $\mathrm{CR}(\mu^-, \mathrm{Ti} \to e^-, \mathrm{Ti})$ appear as bands whose widths are mainly governed by the uncertainty in $\delta_{\text{CP}}$ as well. For NO, the dependence on $m_{\eta}$ shows that the band is wide at low $m_{\eta}$ (around a few hundred GeV), where the CR can span several orders of magnitude as $\delta_{\text{CP}}$ varies across its allowed range. This sensitivity decreases significantly at larger $m_{\eta}$. A similar behavior is found for the dependence on $M_N$, whereas no significant interplay between $M_\Sigma$ and $\delta_{\text{CP}}$ is observed in the last panel of figure~\ref{fi:RP-mec}. Overall, while the current SINDRUM~II limit can be satisfied over a wide region of parameter space, the upcoming $\mu$--$e$ conversion experiments such as PRIME are expected to provide significantly stronger constraints.

\subsubsection{Effects of $\varrho$ and $\alpha_{21}$}
\begin{figure}[htbp]
\centering
\includegraphics[scale=0.19,valign=c]{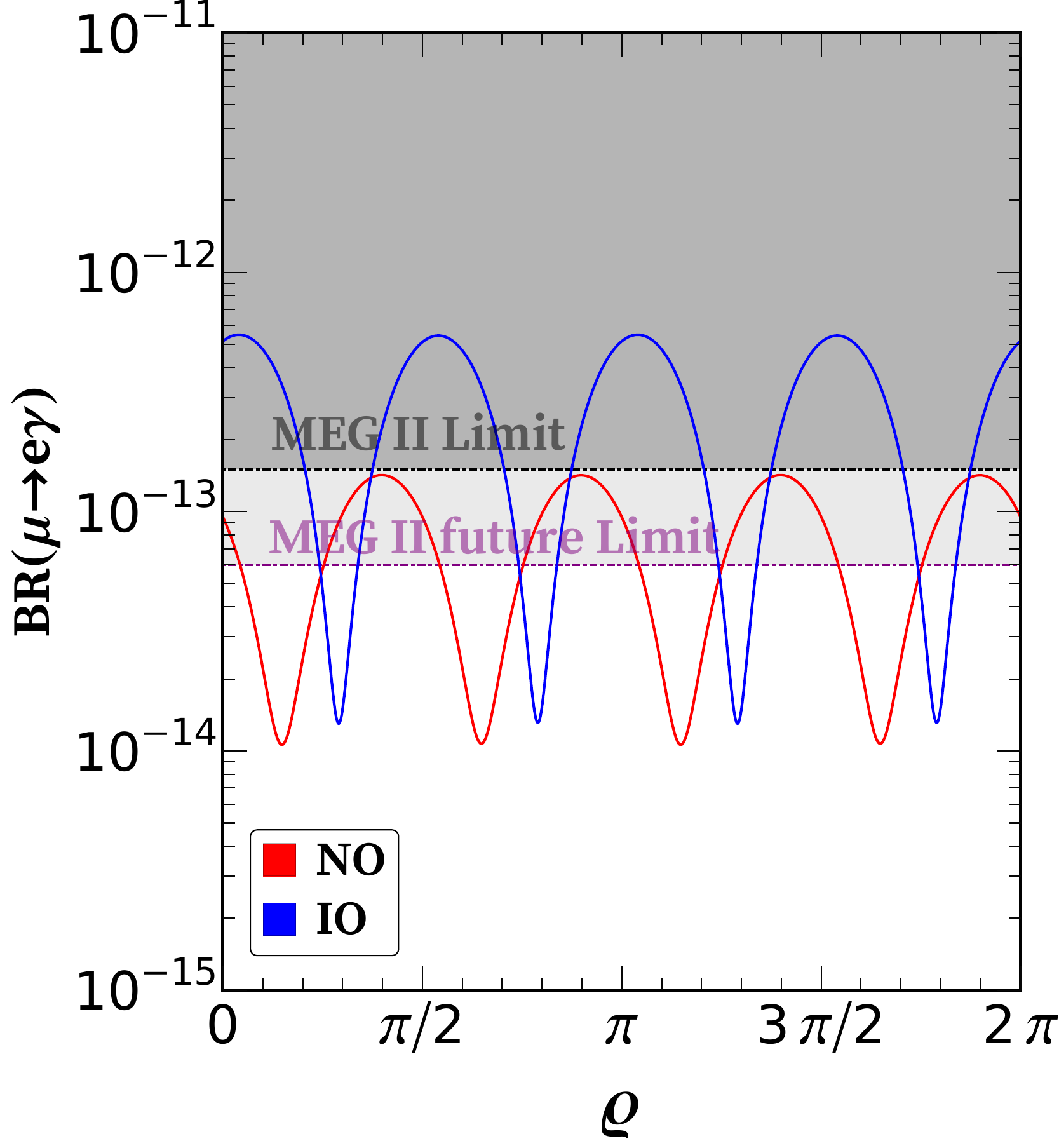}
\includegraphics[scale=0.19,valign=c]{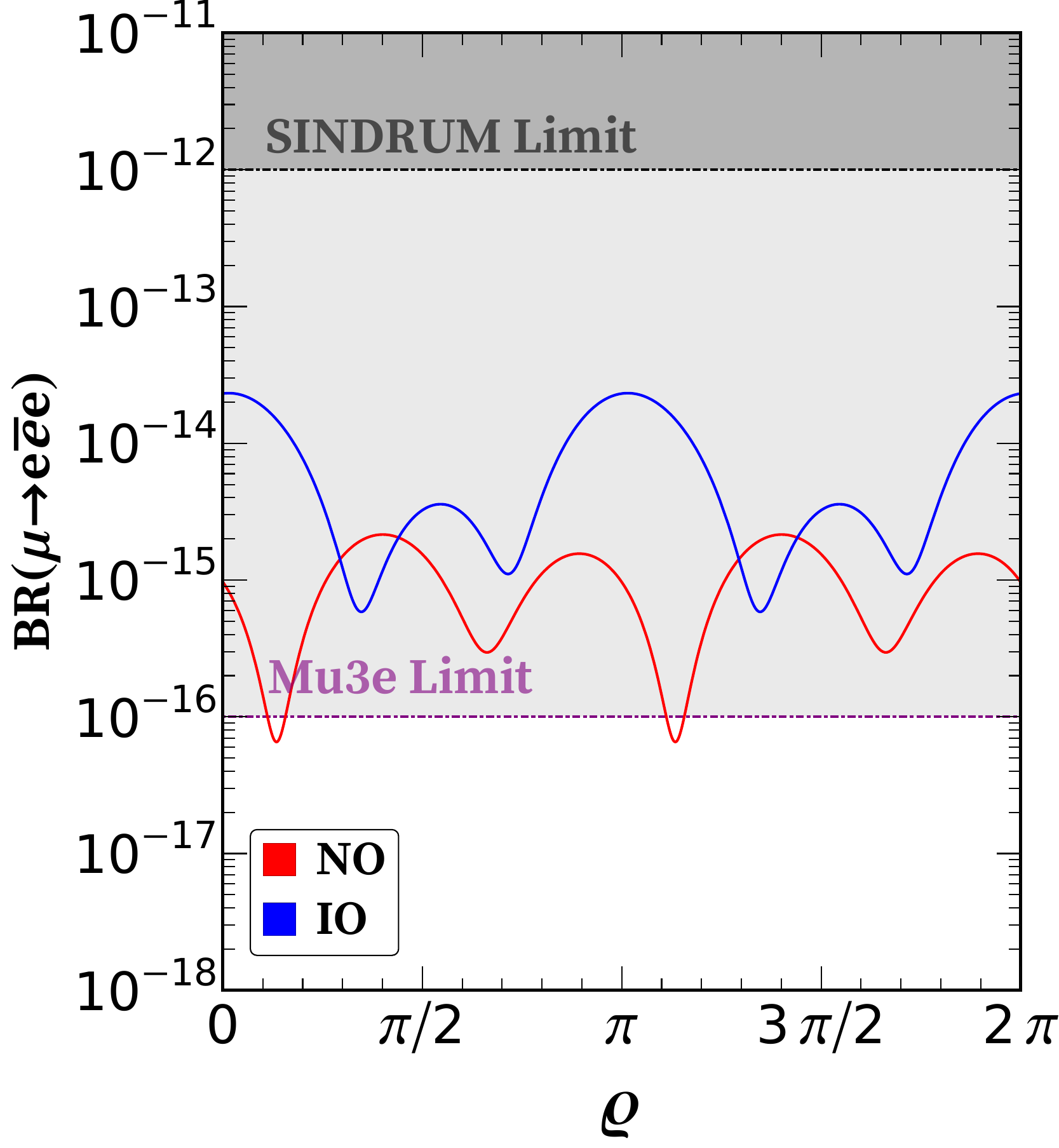}
\includegraphics[scale=0.19,valign=c]{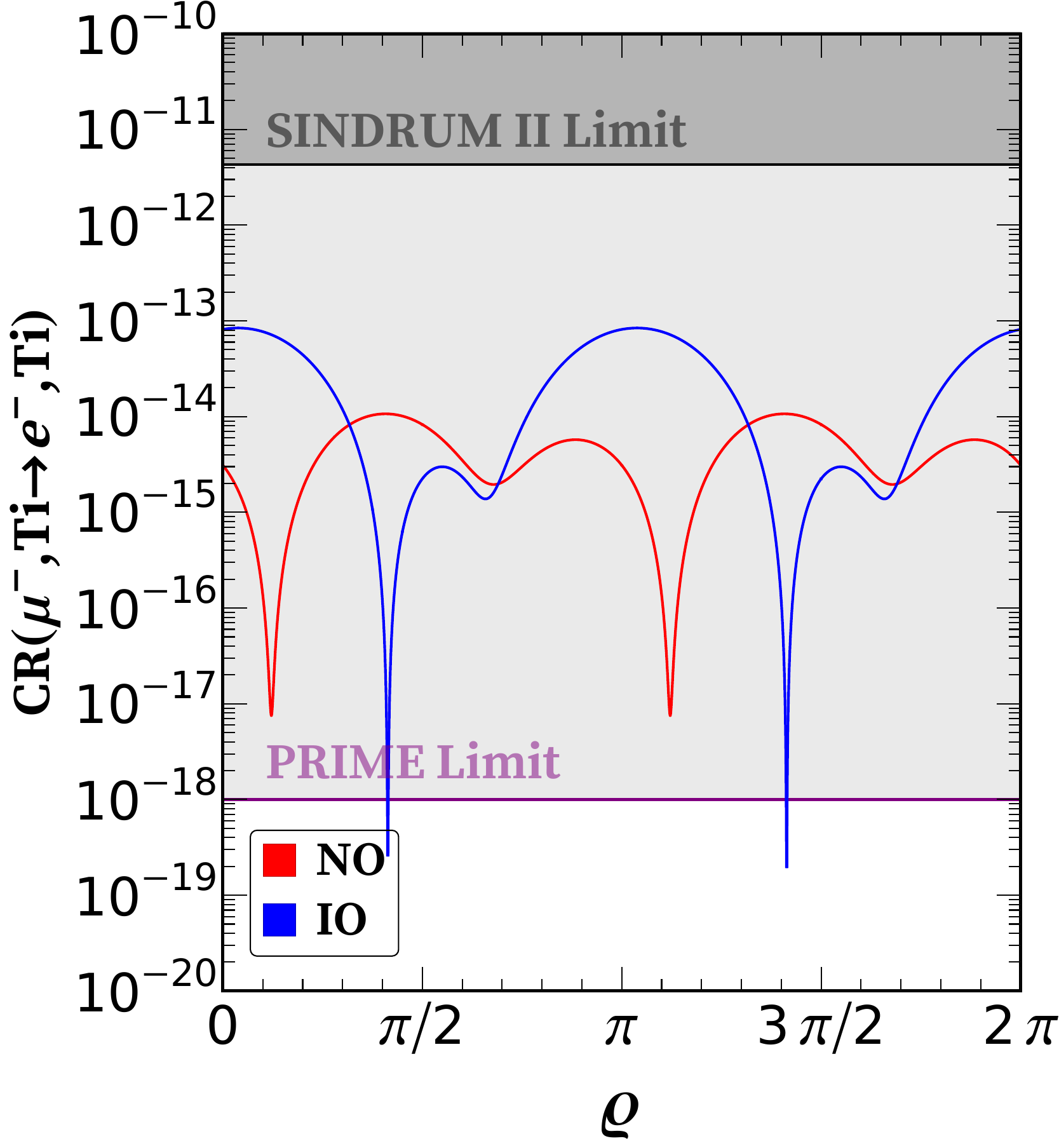}
\caption{BR$(\mu\to e\gamma)$, BR$(\mu\to e\overline{e}e)$ and CR$(\mu^-,\text{Ti}\to e^-,\text{Ti})$ as a function of $\varrho$. \label{fi:lfv-gamma}}
\end{figure}

In the previous analysis, the angle $\rho$ in the matrix $R$ and the Majorana CP-violating phase $\alpha_{21}$ were set to zero for illustration. However, as shown in Eq.~\eqref{eq:CI}, these two parameters enter the Yukawa couplings $Y_{\Sigma, N}$ and can therefore affect the amplitudes of LFV processes. In the following, we will discuss the impact of $\rho$ and $\alpha_{21}$ on LFV observables. 

In figure~\ref{fi:lfv-gamma}, we investigate the dependence of the three LFV quantities, BR$(\mu\to e\gamma)$, BR$(\mu\to e\overline{e}e)$, and CR$(\mu^-,\text{Ti}\to e^-,\text{Ti})$ on $\varrho$ for the benchmark value $\alpha_{21}=0$. From the left panel, we can read off that $\text{BR}(\mu\to e\gamma)$ can vary by one and two orders of magnitude for NO and IO, respectively. These results further demonstrate that the MEG~II experiment could impose significant constraints on the allowed values of $\varrho$ in the IO case, and the predicted values are below the present bound in the NO case.
Furthermore, for both NO and IO, there are regions of $\varrho$ for which $\text{BR}(\mu\to e\gamma)$ falls below the projected target sensitivity of MEG~II. The middle and right panels display $\text{BR}(\mu\to e\overline{e}e)$ and $\text{CR}(\mu^-,\text{Ti}\to e^-,\text{Ti})$, respectively. The predicted $\text{BR}(\mu\to e\overline{e}e)$ remains below the SINDRUM bound, varying by slightly more than one order of magnitude. By contrast, $\text{CR}(\mu^-,\text{Ti}\to e^-,\text{Ti})$ exhibits a significantly stronger dependence on $\varrho$, spanning several orders of magnitude, while still satisfying the SINDRUM~II constraint. All the three LFV quantities exhibit a $\pi$ periodicity in $\varrho$. This periodic behavior can be traced to the parametrization in Eq.~\eqref{eq:CI}, where $Y_N$ and $Y_\Sigma$ are linear combinations of $\cos\varrho$ and $\sin\varrho$. Because the Wilson coefficients $K_1^L$, $K_2^R$, $F_{L,R}$, and $B_1^L$ depend on $Y_N$ and $Y_\Sigma$ only through even powers, the corresponding LFV quantities exhibit the observed $\pi$ periodicity. By contrast, the Majorana phase $\alpha_{21}$ appears through phase factors $e^{\pm i\alpha_{21}}$ in $Y_N$ and $Y_\Sigma$, leading to a $2\pi$ periodicity in these LFV quantities.

\begin{figure}[htbp]
\centering
\includegraphics[scale=0.25,valign=c]{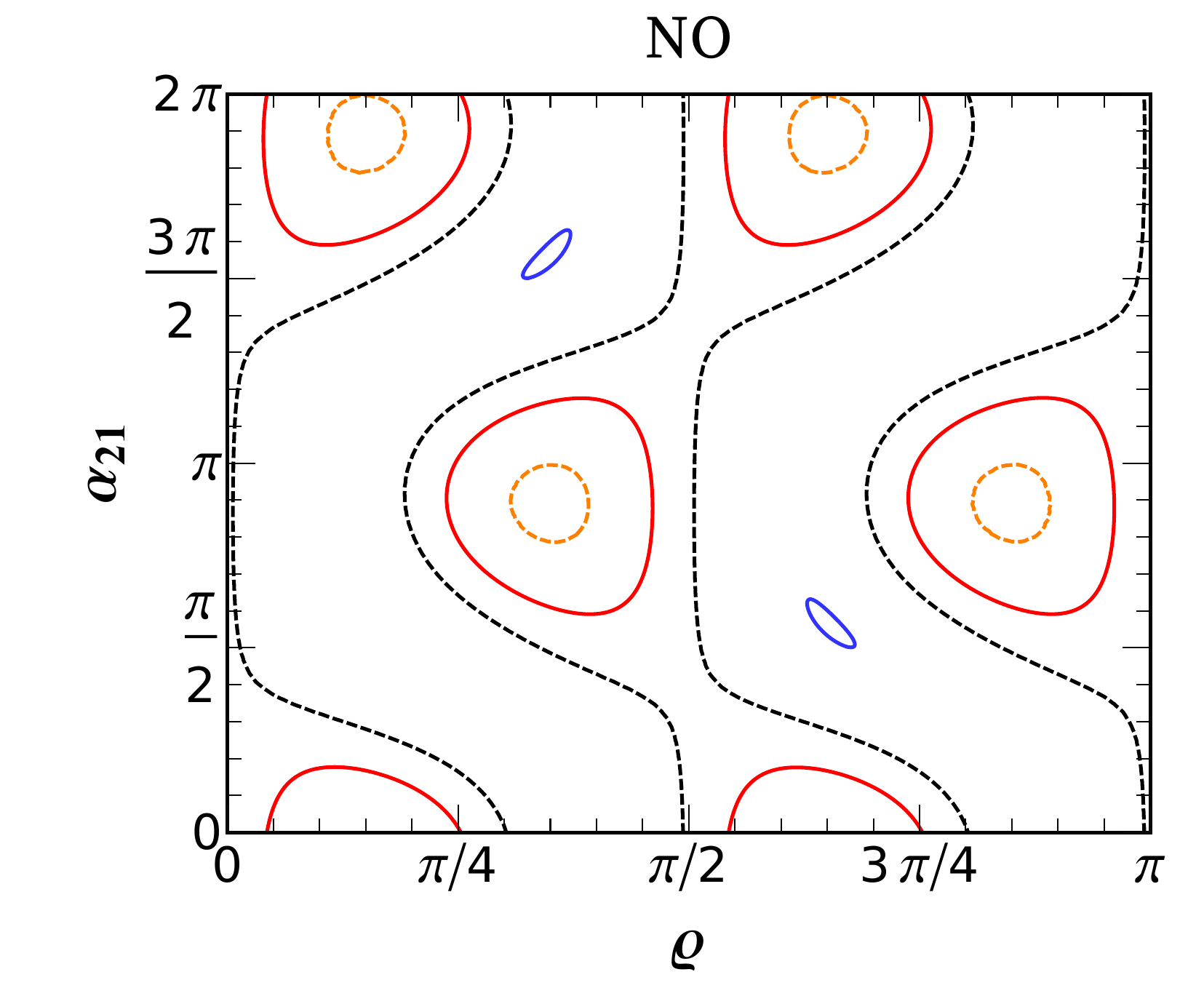}
\includegraphics[scale=0.25,valign=c]{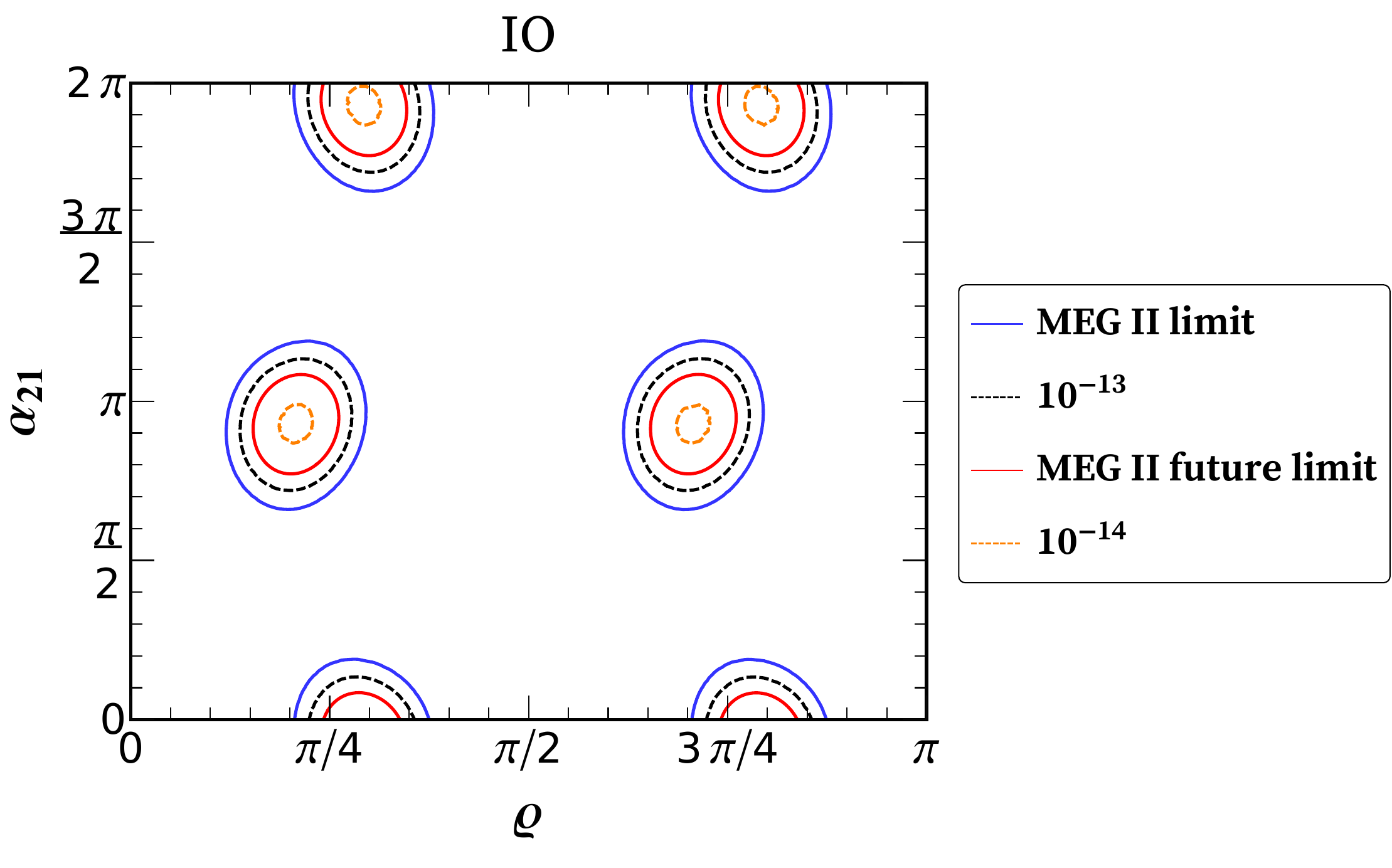}
\caption{Contour plot of BR$(\mu\to e\gamma)$ in the $(\varrho,\alpha_{21})$ plane. The left and right panels are for the NO and IO cases, respectively.\label{fi:Majvr-meg} } 
\end{figure}

In figure~\ref{fi:Majvr-meg}, we vary the Majorana phase $\alpha_{21}$ in the range $0\leq \alpha_{21}\leq 2\pi$ and the angular parameter $\varrho$ in the range $0\leq \varrho\leq \pi$. The resulting contours show that the interplay between $\alpha_{21}$ and $\varrho$ can have a sizable impact on BR$(\mu\to e\gamma)$. In the NO case, as shown in the left panel, most of the parameter space is consistent with the current MEG~II bound, except for the small regions enclosed by the blue contours. In contrast, the IO case shown in the right panel exhibits a much more constrained parameter space, with allowed regions inside the blue contours. After imposing the MEG~II future sensitivity, the viable parameter space is further reduced to several smaller isolated regions in the $(\varrho,\alpha_{21})$ plane for both NO and IO. These regions indicate the possibility of a strong cancellation between the $D^0$ and $D^-$ contributions to $K_2^R$, which allows BR$(\mu\to e\gamma)$ to fall below the projected MEG~II sensitivity. Therefore, the present MEG~II bound already constrains the viable choices of $\varrho$ and $\alpha_{21}$, especially in the IO case, while the projected MEG~II sensitivity is expected to impose substantially stronger restrictions for both mass orderings.

\subsection{Phase transition and GW \label{sec:PH_GW}}

In this section, we turn to another aspect of the phenomenology of the singlet--triplet scotogenic model. Based on the effective potential $V_{\text{tot}}$ constructed in section~\ref{sec:FTV} (see Eq.~\eqref{eq:Vtot}), we analyze its thermal evolution and the associated stochastic GW background. In this framework, only the neutral scalar fields $h_0$ and $\Omega_0$ develop nonzero VEVs. By contrast, the inert scalar doublet $\eta$ does not acquire a VEV, as it is odd under the $\mathbb{Z}_2$ symmetry, which is assumed to remain unbroken throughout the cosmological evolution.

We employ the package \texttt{PhaseTracer}~\cite{Athron:2020sbe,Athron:2024xrh}, with the finite-temperature effective potential as input, to determine the phase-transition dynamics and the corresponding GW signal. After scanning the free parameters introduced in section~\ref{sec:FTV}, we find that $\mu_1$ and $\lambda_1^\Omega$ have the most pronounced impact on the evolution of the scalar potential. This feature can be understood from the tree-level structure of the potential: the terms directly coupling $\varphi_1$ and $\varphi_2$ play a leading role in shaping the phase-transition pattern and the resulting GW phenomenology. More explicitly, the relevant portal interactions are schematically given by $\mu_1 \varphi_{1}^2\varphi_{2}$ and $\lambda_1^\Omega \varphi_1^2\varphi_2^2$, which provide the dominant communication between the two background-field directions. Consequently, larger values of $\mu_1$ and $\lambda_1^\Omega$ generally enhance the strength of the GW signal. We therefore characterize their effects in terms of the dimensionless ratio $\mu_1/v_\Omega$ and the coupling $\lambda_1^\Omega$, which are scanned over the following regions:
\begin{equation}\label{eq:region_mu_lambda}
1\leq \frac{\mu_1}{v_\Omega}\leq 5\,,\qquad 1\leq \lambda_1^\Omega\leq 6\,.
\end{equation}

The remaining parameters entering the effective potential, namely $\lambda_{1}$, $\lambda_2$, $\lambda_3$, $\lambda_4$, $\lambda_5$, $\lambda_2^\Omega$, $\lambda_3^\Omega$, $\mu_2$, $v_\Omega$, $v_\phi$, $m_\eta$, $m_{\phi}$, and $m_{\Omega}$, mainly affect the scalar mass spectra and finite-temperature corrections. However, by scanning the parameter space, we find that their impact on the strength of the electroweak phase transition is relatively mild. In the phenomenological analysis of LFV presented in section~\ref{subsec:ph-lfv}, the parameters $\lambda_3$, $\lambda_4$, $\lambda_3^\Omega$, $\mu_2$, $v_\Omega$, and $v_{\phi}$ are fixed to the benchmark values given in Eq.~\eqref{eq:LFV-numvalue}, and we adopt the same set of parameter values here to ensure the self-consistency of the analysis. Furthermore, taking into account the constraints from LFV processes discussed in section~\ref{subsec:ph-lfv}, we restrict the scalar mass to the range $m_\eta \in [100~\mathrm{GeV},\,1~\mathrm{TeV}]$. In addition, we fix $\lambda_5 = 10^{-8}$, consistent with the choice adopted in figures~\ref{fi:RP-meg}--\ref{fi:Majvr-meg}. The remaining independent parameters in the effective potential are therefore $\lambda_2$ and $\lambda_2^\Omega$. After imposing the theoretical constraints summarized in section~\ref{sec:constraint}, we take the representative values $\lambda_2=\lambda_{2}^{\Omega}=0.1$. Finally, the quartic coupling $\lambda_1$ is fixed by the Higgs mass, yielding $\lambda_1 \simeq 0.26$, while the mass parameters $m_\phi$ and $m_\Omega$ are determined from the tadpole equations in Eq.~\eqref{eq:mphimOmg}.

After specifying the parameter prescription, we quantify the strength of the phase transition at the nucleation temperature by the phase transition strength parameter
\begin{eqnarray}
\xi_N \equiv \frac{v_N}{T_N}\,,\label{eq:xiN}
\end{eqnarray}
where $v_N$ denotes the field-space separation between the false and true vacua at the nucleation temperature $T_N$, defined as $v_N=\left.\sqrt{\left(\Delta v_{\varphi_1}\right)^2+\left(\Delta v_{\varphi_2}\right)^2}\,\right|_{T=T_N}$. Here, $\Delta v_{\varphi_1}$ and $\Delta v_{\varphi_2}$ denote the variations of the VEVs along the $\varphi_{1}$ and $\varphi_{2}$ scalar field components, respectively, between the false and true vacua. A transition satisfying $\xi_N>1$ is classified as a strong FOPT. A sufficiently strong FOPT is required to generate a potentially observable stochastic GW background.

\begin{figure}[htbp]
\centering
\includegraphics[scale=0.60]{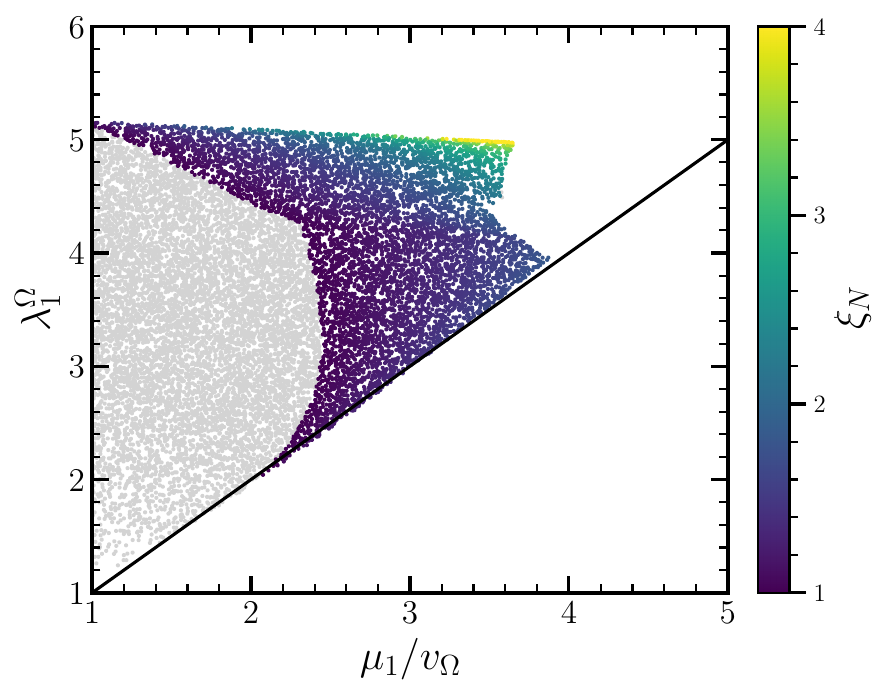}
\caption{Dependence of $\xi_N$ on $\lambda_1^{\Omega}$ and $\mu_1/v_\Omega$. The gray points indicate parameter points with $\xi_N<1$.   }
\label{fig:lamfr-xcxn}
\end{figure}

We then perform a numerical scan over the parameter space defined in Eq.~\eqref{eq:region_mu_lambda} to identify the regions capable of realizing a strong FOPT and generating a potentially observable GW signal. The viable points are shown in figure~\ref{fig:lamfr-xcxn} in the plane of $\lambda_{1}^{\Omega}$ versus $\mu_1/v_\Omega$.  A clear diagonal pattern is observed, indicating an approximate boundary between the viable and non-viable regions. Most of the viable points are located in the region with $\lambda_{1}^{\Omega} \gtrsim \mu_1/v_\Omega$. The absence of viable phase transitions in the opposite region, $\mu_1/v_\Omega \gtrsim \lambda_{1}^{\Omega}$, is mainly related to the stability of the zero-temperature effective potential. In this region, imposing the electroweak vacuum condition at
$v_{\varphi_1}=246~\mathrm{GeV}$ requires $\delta\lambda_2^\Omega$ to take a large negative value, leading to a sizable negative quartic CT contribution proportional to $\delta\lambda_2^\Omega\varphi_2^4$ given in Eq.~\eqref{eq:CTV-vev}, in order to satisfy the renormalization conditions in Eq.~\eqref{eq:renormalization_condition}. As a consequence, the renormalized zero-temperature effective potential,
$V_{\rm tot}|_{T=0}=V_{\rm tree}+V_{\rm CW}+V_{\rm CT}$,
develops an instability along the $\varphi_2$ ($\Omega_0$) direction. Thus the electroweak stationary point cannot be identified as a stable minimum of the effective potential, and the corresponding parameter points are excluded. After imposing this vacuum stability requirement, the viable parameter region is approximately restricted to $\lambda_1^\Omega \lesssim 5.2$ and $\mu_1/v_\Omega \lesssim 4$. Within this region, we find the parameter $\xi_N<1$ for the gray points and the phase transition is not strong enough, while the colored points satisfying $\xi_N>1$ correspond to a sizable domain of strong FOPTs. These points are mainly concentrated at relatively large values of both $\lambda_1^\Omega$ and $\mu_1/v_\Omega$, indicating that sufficiently strong portal interactions are preferred to enhance the phase-transition strength. The maximal value of the transition strength parameter reaches $\xi_N\simeq 4.55$, which leads to a potentially detectable stochastic GW signal. 
In the upper right corner of figure~\ref{fig:lamfr-xcxn} for larger values of $\mu_1/v_\Omega$ and $\lambda_1^\Omega$, although a non-zero true vacuum emerges below $T_C$, the bubble nucleation condition cannot be satisfied. Thus the transition to the desired electroweak-breaking vacuum is highly suppressed or even prevented, and the Universe is trapped in the false vacuum in this parameter space~\cite{Balazs:2023kuk}. As a consequence, the FOPT fails to complete, leading to no GW signal.

\begin{figure}[thbp!]
\centering
\includegraphics[scale=0.56]{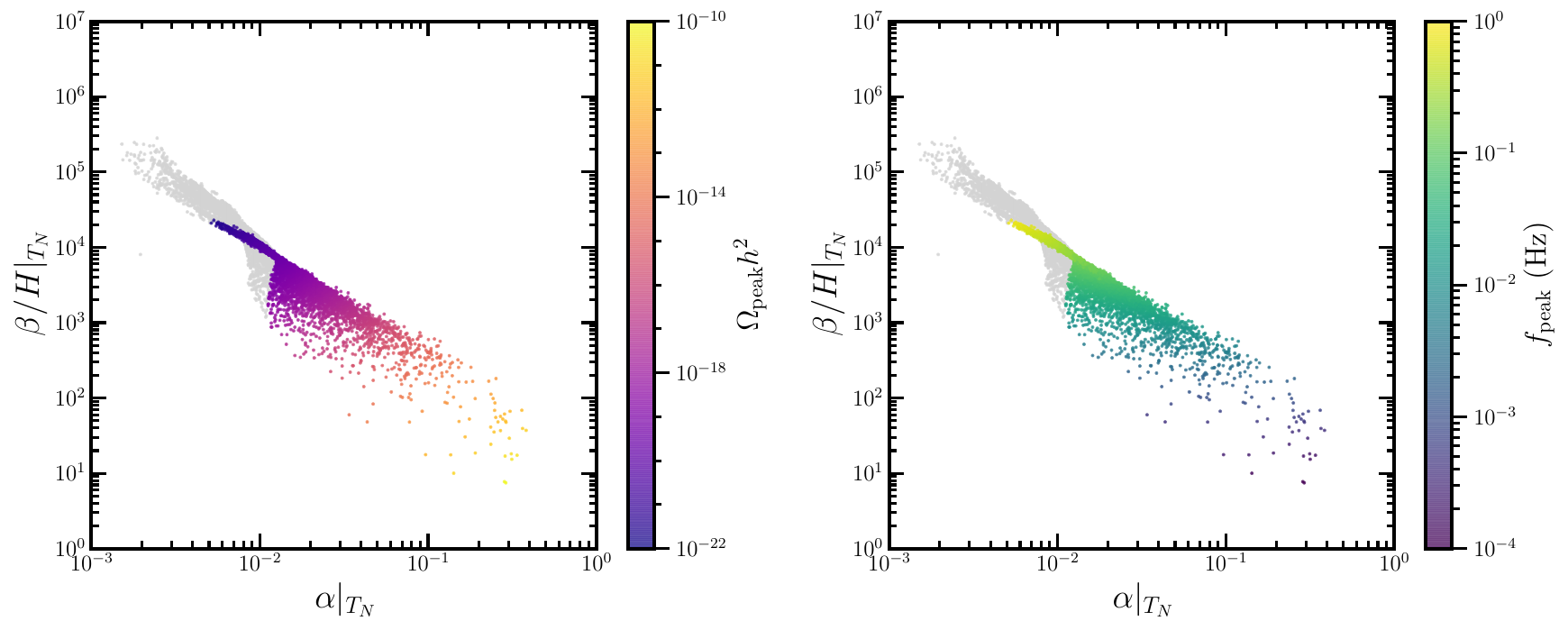}
\caption{Dependence of peak amplitude $\Omega_\text{peak}$ and peak frequency $f_\text{peak}$ on $\alpha|_{T_N}$ and $\beta/H|_{T_N}$. The gray points indicate parameter points with $\xi_N<1$. }
\label{fig:ab-Af}
\end{figure}

Provided that the phase transition is sufficiently strong, a stochastic GW background can be generated. For all parameter points that exhibit a FOPT, we compute the relevant GW parameters, including the strength parameter $\alpha$, the inverse duration parameter $\beta/H$, the peak amplitude $\Omega_{\rm peak} h^2$, and the corresponding peak frequency $f_{\rm peak}$. The distribution of viable parameter points in the $(\alpha,\beta/H)$ plane is shown in figure~\ref{fig:ab-Af}, together with the associated values of $\Omega_{\rm peak} h^2$ and $f_{\rm peak}$. We observe that larger values of $\alpha$ and smaller values of $\beta/H$ favor the production of stronger GW signals. This behavior can be understood from Eqs.~\eqref{eq:Omg-sw} and~\eqref{eq:Omg-tur}, where the GW amplitude increases with the released vacuum energy (encoded in $\alpha$) and with the duration of the phase transition (inversely proportional to $\beta$). For benchmark points satisfying $\xi_N>1$, the predicted peak amplitude spans more than ten orders of magnitude, $10^{-22} \lesssim \Omega_{\rm peak} h^2 \lesssim 10^{-9}$, while the corresponding peak frequencies lie in the range $10^{-5}~\mathrm{Hz} \lesssim f_{\rm peak} \lesssim 1~\mathrm{Hz}$. The points yielding the large GW signals are mainly concentrated in the region $1\lesssim \beta/H \lesssim 10^4$, and $10^{-3} \lesssim \alpha \lesssim 10^{-1}$. 

\begin{figure}[htbp!]
\centering
\includegraphics[scale=0.55]{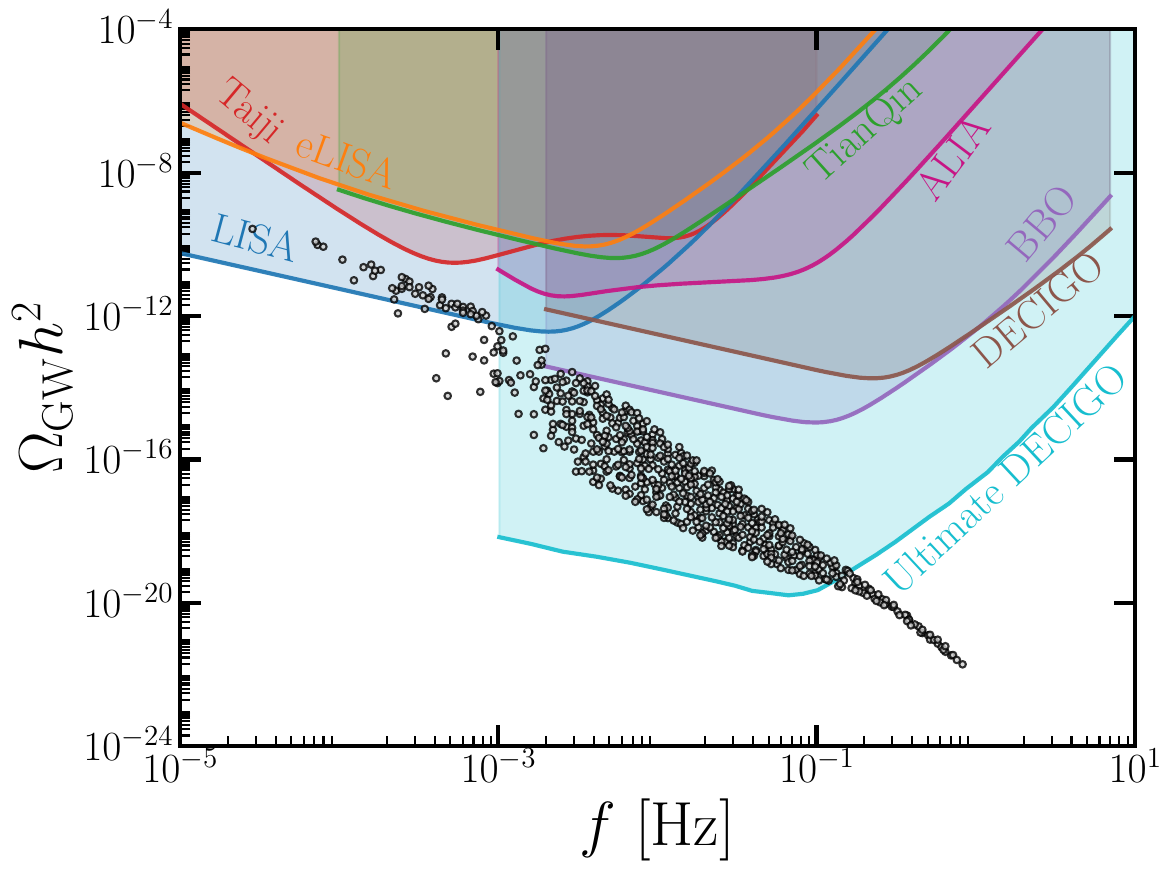}
\caption{Relation between the peak frequency $f_{\rm peak}$ and the peak GW energy density $\Omega_{\rm peak}h^2$ in the singlet--triplet scotogenic model. Each hollow marker corresponds to a different combination of $\mu_1/v_\Omega$ and $\lambda_1^\Omega$. The corresponding $\Omega_{\rm GW}h^2$ sensitivity curves for LISA, eLISA, BBO, DECIGO, TianQin, and ALIA are taken from \cite{Moore:2014lga}, while the Taiji sensitivity curve is adopted from Ref.~\cite{Liu:2023qap}. The Ultimate DECIGO sensitivity curve is adopted from Ref.~\cite{Takahashi:2004yr}. Only the points satisfying $\xi_N>1$ are shown here.}
\label{fig:pkAf-dtt}
\end{figure}

Having established the typical ranges of the GW parameters and identified the regions yielding sizable signals, we proceed to evaluate the experimental prospects for detecting these predictions. To this end, we compare the predicted GW spectra with the sensitivity curves of future GW observatories. In figure~\ref{fig:pkAf-dtt}, the peak amplitude $\Omega_{\rm peak} h^2$ and peak frequency $f_{\rm peak}$ are shown together with the projected sensitivities of LISA~\cite{Sathyaprakash:2009xs}, eLISA~\cite{Amaro-Seoane:2012aqc}, ALIA~\cite{Berry:2012im}, BBO~\cite{Yagi:2011wg}, DECIGO~\cite{Yagi:2011wg}, TianQin~\cite{Hu:2024vvu}, Taiji~\cite{Luo:2021qji} and Ultimate DECIGO \cite{Takahashi:2004yr}. The viable parameter points populate a narrow band in the $(\Omega_{\rm peak}h^2,\,f_{\rm peak})$ plane, spanning approximately $10^{-22} \lesssim \Omega_{\rm peak} h^2 \lesssim 10^{-9}$ and $10^{-5}~\mathrm{Hz} \lesssim f_{\rm peak} \lesssim 1~\mathrm{Hz}$. This distribution indicates that a subset of the predicted GW signals falls within the projected sensitivity reach of space-based interferometers. In particular, the points accessible to LISA are mainly concentrated in the frequency range $10^{-4}$--$10^{-3}~\mathrm{Hz}$, where the predicted amplitudes are sufficiently large to be probed. A smaller subset of points extends toward higher frequencies, where partial overlap with the sensitivity region of BBO may occur. The Ultimate DECIGO sensitivity, corresponding to an idealized sensitivity limit of DECIGO, could potentially probe the predicted GW signals with $10^{-3}\mathrm{Hz}<f_{\rm peak}\lesssim10^{-1}\mathrm{Hz}$. By contrast, the projected sensitivities of Taiji, TianQin, eLISA, ALIA, and DECIGO exhibit limited overlap with the predicted signal region in this setup, and are therefore less effective in probing the present parameter space.

\begin{figure}[t!]
\centering
\includegraphics[scale=0.50]{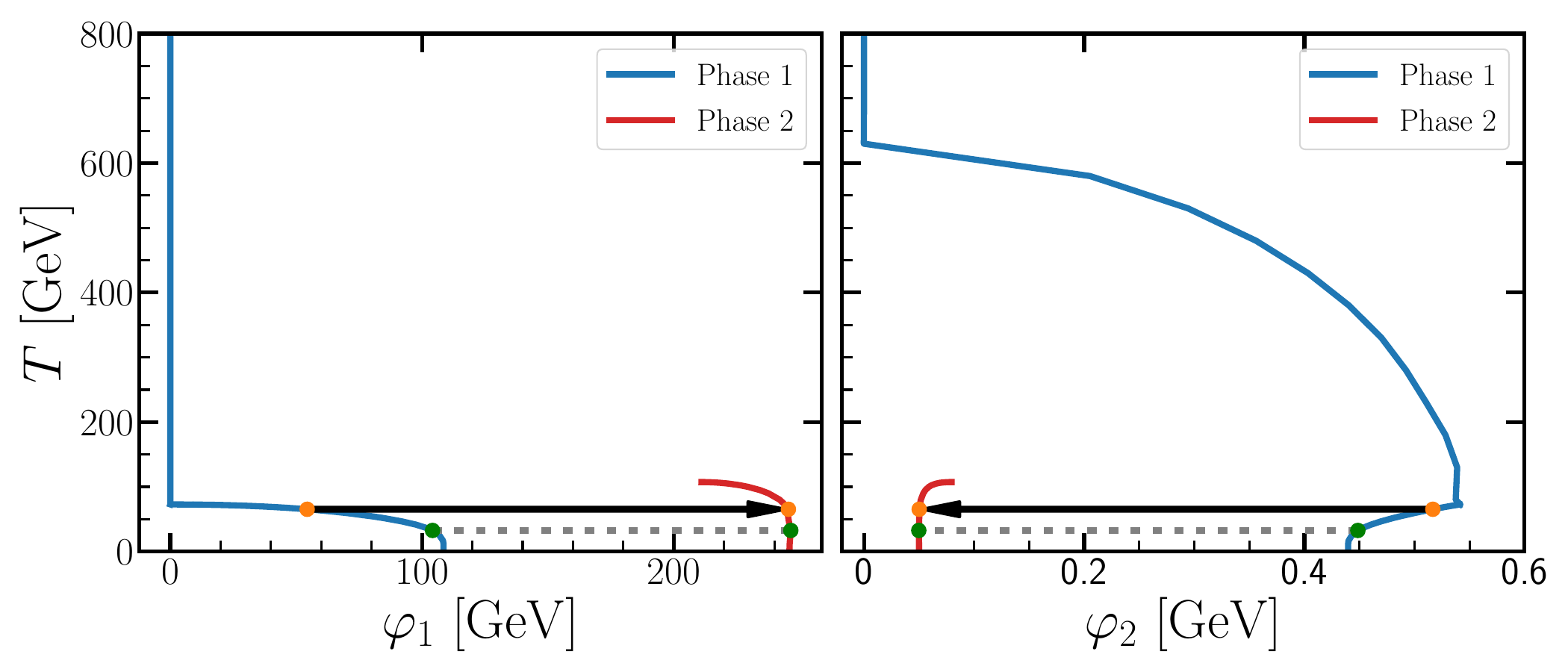}
\includegraphics[scale=0.50]{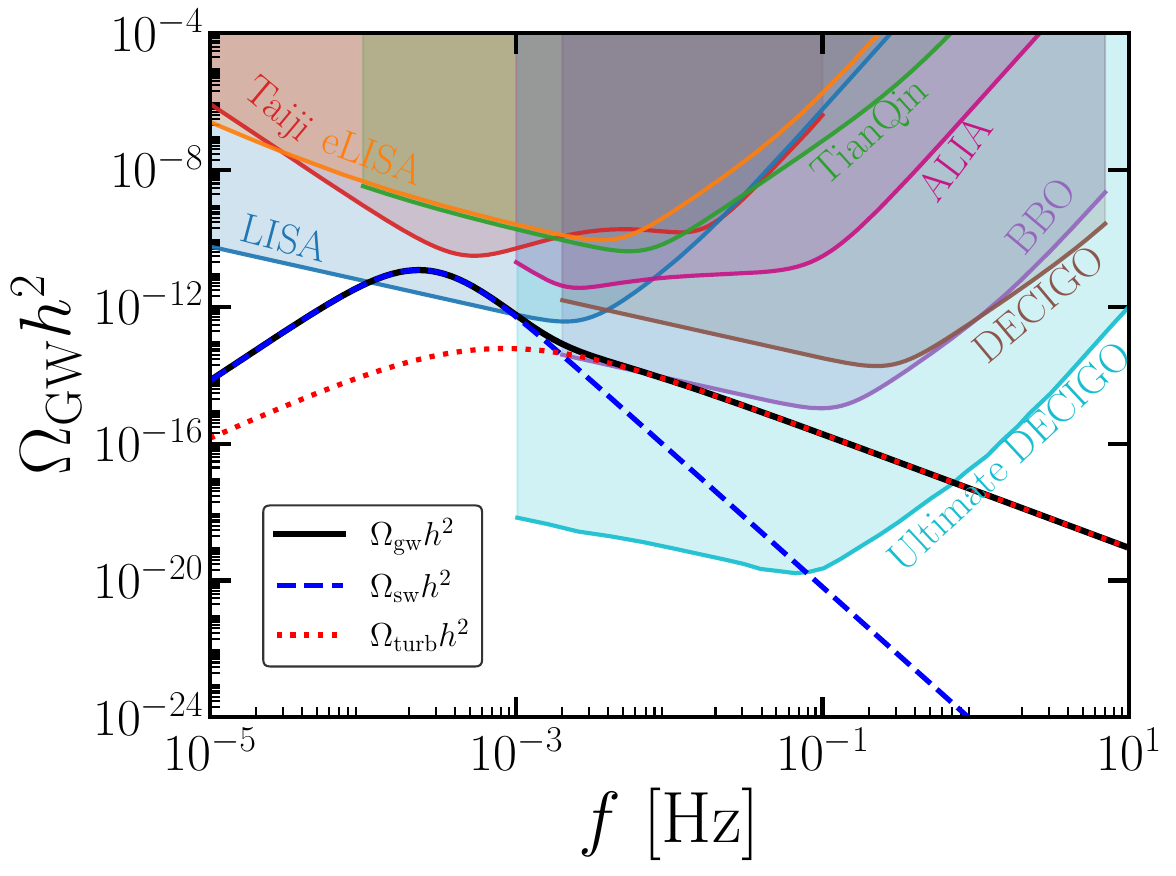}
\caption{Phase-transition profiles at the critical temperature $T_C$ and the nucleation temperature $T_N$ for the benchmark point $\lambda_{1}^{\Omega}=5.01$ and $\mu_1=0.16~\mathrm{GeV}$, together with the corresponding GW spectrum. The black arrow indicates the phase-transition from the false vacuum to the true vacuum at the critical temperature $T_C$, while the gray dotted line denotes the nucleation temperature $T_N$. 
\label{fig:spGW} }
\end{figure}

To further illustrate the phase-transition dynamics of the model, we select a representative benchmark point. The concrete values of the model parameters are listed in  table~\ref{tab:benchmark_point}. 
Note that the parameters $\lambda_4$, $\lambda_3$, $\lambda_3^\Omega$, $\mu_2$, $v_\Omega$, and $v_{\phi}$ are fixed to the benchmark values given in Eq.~\eqref{eq:LFV-numvalue} for the LFV analysis. The neutrino mixing parameters are taken to be the best-fit values from NuFIT v6.1~\cite{Esteban:2024eli}. In addition, the Majorana CP-violating phase $\alpha_{21}$ and the free angle $\varrho$ are set to zero. The mass parameters are chosen as $m_\eta = 637.11~\text{GeV}$, $M_N = 2210.16~\text{GeV}$, and $M_\Sigma = 4642.88~\text{GeV}$ satisfying the constraints from both collider and charged LFV processes. Furthermore, we adopt $\lambda_2 = \lambda_2^\Omega = 0.1$, consistent with the values used in our previous analysis of GW signals from phase transitions.  The portal parameters are chosen as $\lambda_{1}^{\Omega} = 5.01$ and $\mu_{1} = 0.16~\text{GeV}$.
The corresponding critical temperature $T_C$, nucleation temperature $T_N$, phase-transition strength $\xi_N$, and GW parameters $\alpha|_{T_N}$ and $\beta/H|_{T_N}$ are also summarized in table~\ref{tab:benchmark_point}. The thermal evolution of the symmetric and broken phases are illustrated in figure~\ref{fig:spGW}.

\begin{table}[t!]
\centering
\begingroup
\setlength{\extrarowheight}{0.08cm}
\renewcommand{\arraystretch}{0.92}
    \centering
    \begin{tabularx}{\linewidth}{!{\vrule width 1pt}>{\centering\arraybackslash}X|>{\centering\arraybackslash}X|>{\centering\arraybackslash}X|>{\centering\arraybackslash}X|>{\centering\arraybackslash}X|>{\centering\arraybackslash}X|>{\centering\arraybackslash}X!{\vrule width 1pt}}
        \Xhline{1.5pt}
        \multicolumn{7}{!{\vrule width 1pt}c!{\vrule width 1pt}}{Input parameters} \\
        \hline
        $\lambda_{2}$ & $\lambda_{3}$ & $\lambda_{4}$ & $\lambda_{5}$ & $\lambda_{1}^{\Omega}$ & $\lambda_{2}^{\Omega}$ & $\lambda_{3}^{\Omega}$ \\
        \hline
        $0.10$ & $7\times10^{-2}$ & $5.0\times10^{-4}$ & $2\times10^{-5}$ & $5.01$ & $0.1$ & $0.1$ \\
        \hline\hline
        $Y_\Omega$ & $\mu_1\,[\mathrm{GeV}]$ & $\mu_2\,[\mathrm{GeV}]$ & $v_\Omega\,[\mathrm{GeV}]$ & $m_\eta\,[\mathrm{GeV}]$ & $m_N\,[\mathrm{GeV}]$ & $m_\Sigma\,[\mathrm{GeV}]$ \\
        \hline
        $7.0\times10^{-3}$ & $0.16$ & $300$ & $0.05$ & $637.11$ & $2210.16$ & $4642.88$\\
        \hline\hline
    \end{tabularx}

    \vspace{0.2em}
    
    \begin{tabularx}{\linewidth}{!{\vrule width 1pt}>{\hsize=1.0\hsize\centering\arraybackslash}X|>{\hsize=1\hsize\centering\arraybackslash}X|>{\hsize=1\hsize\centering\arraybackslash}X!{\vrule width 1pt}}
        \Xhline{1.5pt}
        \multicolumn{3}{!{\vrule width 1pt}c!{\vrule width 1pt}}{LFV} \\
        \hline
         & NO & IO \\
        \hline
        BR$(\mu\to e\gamma)$ & $4.51\times10^{-21}$ & $1.81\times10^{-21}$ \\
        \hline
        BR$(\mu\to e\overline{e}e)$ & $2.48\times10^{-23}$ & $6.11\times10^{-23}$ \\
        \hline
        CR$(\mu^-,\text{Ti}\to e^-,\text{Ti})$ & $5.76\times10^{-24}$ & $4.84\times10^{-22}$ \\
        \Xhline{1.5pt}
    \end{tabularx}

    \vspace{0.2em}
    
\begin{tabularx}{\linewidth}{!{\vrule width 1pt}>{\centering\arraybackslash}X|>{\centering\arraybackslash}X|>{\centering\arraybackslash}X|>{\centering\arraybackslash}X!{\vrule width 1pt}}\Xhline{1.5pt}
\multicolumn{4}{!{\vrule width 1pt}c!{\vrule width 1pt}}{GW} \\
        \hline
        $T_C\,[\mathrm{GeV}]$ & $T_N\,[\mathrm{GeV}]$ & $\xi_N$ &  $\alpha|_{T_N}$ \\
        \hline
        $63.37$ & $31.37$ & $4.10$ &    $0.33$ \\
        \hline    
$\beta/H|_{T_N}$
 & $\Omega_{\rm peak} h^2$ & $f_{\rm peak}\,[\mathrm{Hz}]$ & \\ \hline
 $24.57$ & $1.18\times10^{-11}$ & $2.28\times10^{-4}$ & \\  \Xhline{1.5pt}
\end{tabularx}

\vspace{0.2em}

\begin{tabularx}{\linewidth}{!{\vrule width 1pt}>{\hsize=0.8\hsize\centering\arraybackslash}X|>{\hsize=1.2\hsize\centering\arraybackslash}X!{\vrule width 1pt}}
\Xhline{1.5pt}
\multicolumn{2}{!{\vrule width 1pt}c!{\vrule width 1pt}}{DM} \\ \hline
$\Omega h^2$ & 0.1229 \\ \hline
$\sigma^{\rm SI}\,[\mathrm{cm}^2]$ & $1.69\times10^{-47}$\\
\Xhline{1.5pt}
\end{tabularx}
\endgroup
\caption{ The values of the input parameters for the benchmark point and the corresponding predictions for the LFV observables, GW signals, and DM observables.}
\label{tab:benchmark_point}
\end{table}

The upper-left panel of figure~\ref{fig:spGW} shows the temperature evolution of the VEV of $\varphi_1$. Two distinct phases are present: the Phase~1, represented by the blue dashed line, and the Phase~2, shown by the red solid line. In Phase~1, the vacuum remains at $\langle\varphi_1\rangle=0$ from high temperature ($T\sim 1~\mathrm{TeV}$) down to $T\simeq82.71~\mathrm{GeV}$, where a crossover transition occurs and $\varphi_1$ develops a nonzero VEV, eventually reaching $\langle\varphi_1\rangle=120.68~\mathrm{GeV}$ at zero temperature. The Phase~2 appears at $T\simeq104.96~\mathrm{GeV}$ and coexists with Phase~1 over a finite temperature interval. As the temperature decreases, the VEV of $\varphi_1$ in Phase~2 evolves toward the electroweak vacuum value $\langle\varphi_1\rangle=246~\mathrm{GeV}$ at zero temperature. At the critical temperature $T_C=63.37~\mathrm{GeV}$, the two phases become degenerate, indicating the occurrence of a FOPT. Bubble nucleation subsequently takes place at $T_N=31.37~\mathrm{GeV}$, marking the transition to the true vacuum and the onset of GW production.

The upper-right panel displays the corresponding temperature evolution of the VEV of $\varphi_2$. In Phase~1, $\varphi_2$ undergoes a crossover transition at $T\simeq630~\mathrm{GeV}$, where it acquires a nonzero VEV. As the temperature decreases, the VEV initially increases, reaches a maximum, and then gradually decreases, approaching $\langle\varphi_2\rangle=0.39~\mathrm{GeV}$ at zero temperature. In contrast, in Phase~2, the VEV of $\varphi_2$ decreases monotonically with decreasing temperature and eventually reaches $\langle\varphi_2\rangle=0.05~\mathrm{GeV}$ at zero temperature. At the critical temperature $T_C$, the VEVs of $\varphi_2$ are $0.44~\mathrm{GeV}$ and $0.05~\mathrm{GeV}$ in Phase~1 and Phase~2, respectively.  This benchmark point exhibits a two-step thermal history with a strong FOPT, which is crucial for generating observable GWs.

The parameters $\alpha$ and $\beta/H$, which characterize the GW signal from the FOPT for this benchmark point, are also given in table~\ref{tab:benchmark_point}. Using Eqs.~\eqref{eq:Omg-sw} and~\eqref{eq:Omg-tur}, we show the resulting GW spectrum in the lower panel of figure~\ref{fig:spGW}. The total GW energy density $\Omega_{\rm gw}h^2$ receives contributions from both sound waves and magnetohydrodynamic turbulence, denoted by $\Omega_{\rm sw}h^2$ and $\Omega_{\rm turb}h^2$, respectively. The contribution from bubble collisions is typically subdominant and consequently it is neglected. At low frequencies ($\lesssim 10^{-3}~\mathrm{Hz}$), the spectrum is dominated by the sound-wave contribution. The GW spectrum peaks at $f_{\rm peak} \simeq 2.28\times10^{-4}~\mathrm{Hz}$, with a corresponding amplitude $\Omega_{\rm peak} h^2 \simeq 1.18\times10^{-11}$, which lies within the projected sensitivity range of LISA~\cite{Sathyaprakash:2009xs}. At higher frequencies, the turbulence contribution dominates over the sound-wave contribution, and the total GW spectrum monotonically decreases due to the high-frequency suppression of the spectrum. It is notable that the turbulence-induced GW signal is potentially accessible to BBO~\cite{Yagi:2011wg} for the frequency around $f\sim2\times10^{-3}\,\mathrm{Hz}$, and the Ultimate DECIGO~\cite{Takahashi:2004yr} allows us to explore the model up to $f\sim1\,\mathrm{Hz}$.

For the benchmark point specified in table~\ref{tab:benchmark_point}, we not only study the GW phenomenology but also evaluate the LFV observables. These include the branching ratios $\text{BR}(\mu\to e\gamma)$ and $\text{BR}(\mu\to e\overline{e}e)$, as well as the $\text{CR}(\mu^-,\text{Ti}\to e^-,\text{Ti})$, computed using the formulae presented in section~\ref{sec:LFV}. The corresponding numerical results are summarized in table~\ref{tab:benchmark_point} for both NO and IO neutrino mass spectra. For the radiative LFV decay $\mu\to e\gamma$, the predicted branch ratios are $4.51\times10^{-21}$ for NO and $1.81\times10^{-21}$ for IO, respectively. Both values are below the projected MEG~II upper bound of $6\times10^{-14}$~\cite{Venturini:2024keu}. The predicted branch ratios for the three-body decay $\mu\to e\overline{e}e$ are $2.48\times10^{-23}$ (NO) and $6.11\times10^{-23}$ (IO), which are significantly below the SINDRUM limit~\cite{SINDRUM:1987nra} and Mu3e limit~\cite{MartinPerez:2023vfv}. For coherent $\mu$--$e$ conversion in titanium, the predicted conversion rates are $5.76\times10^{-24}$ (NO) and $4.84\times10^{-22}$ (IO), both of which satisfy the current experimental constraint~\cite{SINDRUMII:1993gxf} and targeted experiments, such as PRIME, which is expected to reach sensitivities around $\mathcal{O}(10^{-18})$~\cite{prime}. Therefore, the benchmark point is fully consistent with current LFV constraints.

The strong suppression of the LFV observables,
particularly $\mathrm{BR}(\mu\to e\gamma)$,
is closely related to the choice of $\lambda_5$.
As shown in Eq.~\eqref{eq:BR_lambda5},
$\mathrm{BR}(\mu\to e\gamma)\propto\lambda_5^{-2}$
for fixed new-particle masses and neutrino parameters. Achieving an LFV branching ratio near the MEG~II sensitivity requires a lepton-number-violating coupling $\lambda_5$ of order $10^{-8}$. However, such a small coupling substantially reduces the mass splitting between the inert neutral scalars $\eta_I$ and $\eta_R$, potentially enhancing $Z$-mediated inelastic DM--nucleus scattering beyond the limits set by direct-detection experiments. For the benchmark choice $\lambda_5=2\times10^{-5}$, the resulting mass splitting is sufficiently large to suppress this scattering and satisfy the dark matter constraints. The corresponding LFV branching ratios, however, are strongly suppressed, illustrating the interplay between LFV phenomenology and dark matter direct-detection constraints. The corresponding DM phenomenology is discussed in the following section. 

\subsection{Dark matter}

In the preceding sections, we have investigated the LFV phenomenology, the electroweak FOPT, and the associated GW signals in this model.  In the following, we further investigate the DM phenomenology of this model. In the singlet--triplet scotogenic model, the DM candidate is stabilized by the imposed $Z_{2}$ symmetry and simultaneously plays a crucial role as the mediator in the radiative generation of neutrino masses. 

Depending on the mass spectra, the model can accommodate either scalar or fermionic DM. In the scalar scenario, the lightest neutral component of the inert doublet, $\eta_{R}$ or $\eta_{I}$, serves as the DM candidate. Alternatively, in the fermionic scenario, DM can be identified with the lightest neutral mass eigenstate $\chi_{1}$, which arises from the mixture between the neutral component of the fermion triplet $\Sigma^{0}$ and the fermion singlet $N$.

The DM phenomenology of this model has been extensively investigated from several complementary perspectives. Scalar-DM realizations have been studied in Refs.~\cite{Diaz:2016udz,Avila:2019hhv,Batra:2022org}, including the constraints from the observed relic abundance, direct and indirect detection, electroweak precision observables, and collider searches. These analyses have identified viable regions in both the low- and high-mass regimes. In the high-mass region, annihilation into electroweak gauge and Higgs bosons, together with co-annihilation involving the other components of the inert doublet, can play an important role in reproducing the observed relic abundance. Fermionic-DM realizations have been explored in Refs.~\cite{Hirsch:2013ola,Choubey:2017yyn,Restrepo:2019ilz,Karan:2023adm}. These studies have shown that the relic abundance and detection prospects depend sensitively on the singlet--triplet composition of $\chi_{1}$ and on fermion--fermion and fermion--scalar co-annihilation processes. They have also demonstrated the complementarity among constraints from DM searches, charged LFV, neutrino observables, and collider experiments.

Although the DM phenomenology of this model has already been studied comprehensively in the literature, it remains important to verify that the LFV and GW phenomenology discussed above is also consistent with the observed DM relic abundance and the current constraints from DM searches. We do not aim to perform another comprehensive scan of the DM parameter space. Instead, we evaluate the DM observables for the benchmark point presented in table~\ref{tab:benchmark_point}, which was adopted in the preceding analyses of LFV and GW phenomenology.

For this benchmark point, the lightest neutral fermionic state $\chi_{1}$ is heavier than the neutral components of the inert scalar doublet, since $m_{N},m_{\Sigma}>1~\text{TeV}$ and $m_{\eta}<1~\text{TeV}$. This expectation is confirmed by the loop-corrected mass spectra, which gives
\begin{equation}
m_{\eta_I}\simeq639.092~\mathrm{GeV},
\qquad
m_{\chi_1}\simeq2210.195~\mathrm{GeV}.
\end{equation}
The positive value of $\lambda_5$ ensures
$m_{\eta_I}<m_{\eta_R}$, as follows from
Eqs.~\eqref{eq:metaR} and~\eqref{eq:metaI}.
Consequently, $\eta_I$ is the lightest
$\mathbb{Z}_2$-odd particle and serves as the
scalar DM candidate.

The mass of $\eta_I$ places this benchmark point in
the high-mass scalar-DM regime, where annihilation
into electroweak gauge and scalar bosons can play
an important role in thermal freeze-out. The masses of the other inert scalar components
are close to that of $\eta_I$, with mass splittings
given by
\begin{equation}
m_{\eta^\pm}-m_{\eta_I}\simeq 0.323~\mathrm{GeV}\,,\quad 
m_{\eta_R}-m_{\eta_I}\simeq \frac{\lambda_5v_{\phi}^2}{2m_{\eta_I}}\simeq 0.95~\mathrm{MeV}\,,
\end{equation}
where $\lambda_5=2\times10^{-5}$.
For a weakly interacting thermal relic, the
characteristic freeze-out temperature is typically
\begin{equation}
T_f\sim\frac{m_{\eta_I}}{20\text{--}30}
\simeq21\text{--}32~\mathrm{GeV}\,,
\end{equation}
for the present benchmark point.
Since both mass splittings are much smaller than
$T_f$, the equilibrium abundances of $\eta_R$
and $\eta^\pm$ are not significantly Boltzmann
suppressed during freeze-out.
Coannihilation processes involving these nearly
degenerate states are therefore expected to
contribute appreciably to the effective annihilation cross section, together with the annihilation of
$\eta_I$ into electroweak gauge and scalar bosons.

For direct detection, the spin independent elastic scattering of $\eta_I$ off nuclei is mediated by the CP-even scalar bosons and depends on the corresponding
scalar couplings. The off-diagonal $Z\eta_I\eta_R$
interaction also allows the inelastic process
$\eta_I\mathcal{N}\to\eta_R\mathcal{N}$,
where $\mathcal{N}$ denotes a target nucleus.
However, this process requires sufficient kinetic
energy to overcome the neutral-scalar mass splitting
and can therefore be suppressed by a sufficiently
large splitting~\cite{Avila:2019hhv}.
It was found that direct-detection
constraints exclude mass splittings below approximately
$250~\mathrm{keV}$ and that a dedicated IceCube
analysis could probe splittings up to about
$500~\mathrm{keV}$~\cite{deBoer:2021pon}. For the present benchmark point, $m_{\eta_R}-m_{\eta_I}\simeq0.95~\mathrm{MeV}$ exceeds both
scales. Consequently, the $Z$-mediated
inelastic scattering channel is kinematically suppressed,
and scalar-mediated elastic scattering provides the
dominant tree-level contribution to direct detection.

In the numerical analysis, we implement the model in
\texttt{SARAH}~\cite{Staub:2015kfa} and generate the
corresponding \texttt{SPheno}~\cite{Porod:2011nf}
module to calculate the particle mass spectra
and the relevant interactions.
The resulting spectrum is then used as input
for \texttt{micrOMEGAs-5.3.41}~\cite{Belanger:2018ccd}
to evaluate the thermal relic abundance and
the spin-independent (SI) DM--nucleon scattering
cross section.
The numerical results are summarized in
Table~\ref{tab:benchmark_point}.
The predicted thermal relic abundance is
\begin{equation}
\Omega_{\eta_I}h^2=0.1229\,,
\end{equation}
which is consistent at the $3\sigma$ level with the Planck measurement $\Omega_{\rm DM}h^2=0.1200\pm0.0012$ ($68\%$ CL)~\cite{Planck:2018vyg}. This indicates that $\eta_I$ can account for the observed DM abundance through standard thermal freeze-out.

We further examine the compatibility of this benchmark point with current direct-detection constraints. The predicted elastic SI DM--nucleon scattering cross section is
\begin{equation}
\sigma_{\eta_I}^{\rm SI}
=1.69\times10^{-47}~\mathrm{cm}^2\,.
\end{equation}
At $m_{\eta_I}=639.09~\mathrm{GeV}$, this prediction lies below the $90\%$ CL upper limit from the LUX--ZEPLIN (LZ) experiment, $\sigma_{\rm LZ}^{\rm SI}\simeq1.9\times10^{-47}~\mathrm{cm}^2$ at the same DM mass~\cite{LZ:2024zvo}, and also satisfies the corresponding, weaker limits reported by XENONnT~\cite{XENON:2025vwd} and PandaX-4T~\cite{PandaX:2024qfu}. Thus, with respect to elastic SI scattering, the benchmark point is compatible with the current experimental limits. Since the predicted cross section lies close to the present LZ exclusion boundary, further improvements in the sensitivity of multi-tonne direct-detection experiments can decisively test this benchmark point.

\section{Conclusion\label{sec:con}}

The existence of nonzero neutrino masses and DM provides compelling evidence for BSM physics, while searches for LFV and GWs offer complementary probes of the underlying new physics. In this work, we have performed a combined phenomenological study of LFV, the electroweak phase transition, GWs, and scalar DM in the singlet--triplet scotogenic model. The model extends the SM by introducing a fermion singlet $N$, a fermion triplet $\Sigma$, an inert scalar doublet $\eta$, and a real scalar triplet $\Omega$, along with an exact $\mathbb{Z}_2$ symmetry that forbids tree-level neutrino masses and stabilizes the lightest $\mathbb{Z}_2$-odd particle. The new Yukawa interactions generate neutrino masses radiatively and induce LFV processes, while the scalar triplet modifies the Higgs potential and can give rise to a strong electroweak FOPT accompanied by potentially observable GW signals.

For the LFV analysis presented in section~\ref{sec:LFV}, we first derived general analytical expressions for the one-loop Wilson coefficients  induced by the penguin and box diagrams with scalar and fermion mediators. The explicit results are collected in appendices~\ref{app:general-penguins} and~\ref{app:general-box}. 
These expressions are formulated independently of the detailed flavor structure of a particular model and can therefore be applied to a broad class of extensions of the SM. We then specialized the general results to the concerned singlet--triplet scotogenic model and calculated the observables associated with $e_j\to e_i\gamma$, $e_j\to e_i\overline{e}_i e_i$, and coherent $\mu$--$e$ conversion, as detailed in section~\ref{sec:LFV}. For $e_j\to e_i\gamma$, we included both the $\chi_\alpha$--$\eta^\pm$ and $\Sigma^\pm$--$\eta_{R,I}$ loops and obtained the dominant dipole coefficient $K_2^R$, while $K_2^L$ is suppressed by the final-state lepton mass. The decay $e_j\to e_i\overline{e}_i e_i$ receives photon-penguin, $Z$-penguin, and box contributions. In the nearly degenerate limit $m_{\eta_R}\simeq m_{\eta_I}$, the Wilson coefficient $F_L$ is strongly suppressed by the small coupling constant $\lambda_5$ which is responsible for the tiny neutrino mass, allowing $F_R$ to become relevant despite its chirality suppression. For coherent $\mu$--$e$ conversion, we derived the photon- and $Z$-penguin contributions to the quark-level vector operators, and the box contributions are absent because all the new fields are color singlets.

Based on these analytical results, we then performed a numerical investigation of the LFV phenomenology in section~\ref{subsec:ph-lfv}. The Yukawa couplings $Y_N$ and $Y_\Sigma$ were reconstructed by using the parametrization in Eq.~\eqref{eq:CI} for both NO and IO, with the neutrino oscillation parameters varied within their respective experimental $3\sigma$ ranges. We find that $\mathrm{BR}(\mu\to e\gamma)$ is governed primarily by $\lambda_5$ and the new-particle masses $m_\eta$, $M_N$, and $M_\Sigma$. In the quasi-degenerate regime $m_\eta\simeq M_\Sigma\simeq M_N\equiv M$, the numerical results reproduce the approximate scaling $\mathrm{BR}(\mu\to e\gamma)\propto(M\lambda_5)^{-2}$, as shown in figure~\ref{fi:RP-meg-lam5}.

We further examined the dependence of $\mathrm{BR}(\mu\to e\gamma)$, $\mathrm{BR}(\mu\to e\overline{e}e)$, and $\mathrm{CR}(\mu^-,\mathrm{Ti}\to e^-,\mathrm{Ti})$ on the new-particle masses $m_\eta$, $M_N$, and $M_\Sigma$. Each mass parameter was varied independently over the range $100~\mathrm{GeV}$--$100~\mathrm{TeV}$, while the neutrino mixing parameters were limited in their respective $3\sigma$ intervals. The corresponding results are presented in figures~\ref{fi:RP-meg}, \ref{fi:RP-m3e}, and~\ref{fi:RP-mec}, respectively. Among the three LFV observables, we find that the current experimental limit on $\mu\to e\gamma$ provides the strongest overall constraint on the model. For NO, satisfying the current MEG~II limit on the predicted $\mathrm{BR}(\mu\to e\gamma)$ generally requires $M_N$ and $M_\Sigma$ to be above approximately $1~\mathrm{TeV}$ and $m_\eta$ to be below a few TeV. For IO, the same limit excludes $m_\eta$ above approximately $1~\mathrm{TeV}$, $M_\Sigma$ below a few TeV, and the entire scanned range of $M_N$. The current SINDRUM limit on $\mathrm{BR}(\mu\to e\overline{e}e)$ places little constraint on the scanned mass ranges for NO, whereas for IO it excludes $m_\eta$ above approximately $10~\mathrm{TeV}$ and $M_\Sigma$ below $1~\mathrm{TeV}$. Similarly, the current SINDRUM~II limit on $\mathrm{CR}(\mu^-,\mathrm{Ti}\to e^-,\mathrm{Ti})$ leaves nearly all of the scanned mass ranges allowed for NO and excludes only $M_\Sigma$ below $1~\mathrm{TeV}$ for IO. The projected sensitivity of MEG~II could further strengthen the constraints on these masses, while Mu3e and PRIME are expected to probe substantial portions of the mass ranges that remain allowed by current limits.

Among the neutrino oscillation parameters, the uncertainty in $\delta_{\rm CP}$ is the dominant source of the spread in the predicted LFV rates. Moreover, varying the Casas-Ibarra parametrization angle $\varrho$ can change the LFV observables by several orders of magnitude, as illustrated in figure~\ref{fi:lfv-gamma}. The Majorana phase $\alpha_{21}$, through its interplay with $\varrho$, can also have a sizable impact on $\mathrm{BR}(\mu\to e\gamma)$, as shown in figure~\ref{fi:Majvr-meg}. The improved sensitivities of MEG~II to $\mu\to e\gamma$, Mu3e to $\mu\to e\overline{e}e$, and PRIME to $\mu$--$e$ conversion, combined with a more precise determination of $\delta_{\rm CP}$ in future neutrino oscillation experiments, will sharpen the LFV predictions and place stronger constraints on $\varrho$ and $\alpha_{21}$.

For the phase-transition analysis, in section~\ref{sec:FTV} we constructed the complete one-loop finite-temperature effective potential in the two-dimensional background-field space spanned by the neutral components of $\phi$ and $\Omega$, assuming that the $\mathbb{Z}_2$ symmetry remains unbroken throughout the thermal evolution. As given in Eq.~\eqref{eq:Vtot}, the effective potential comprises the tree-level potential, the Coleman--Weinberg correction, counter terms, finite-temperature corrections, and daisy resummation. We subsequently established the framework for studying bubble nucleation and the associated stochastic GW signals in section~\ref{eq:GW_FOPT}. The nucleation temperature $T_N$ was determined from the finite-temperature bounce action, and the phase transition parameters $\alpha$ and $\beta/H$ were evaluated at $T_N$ as the key inputs for calculating the GW spectrum. We retained the contributions from sound waves and magnetohydrodynamic turbulence, which are expected to dominate for a FOPT in a thermal plasma, while neglecting the subdominant contribution from bubble collisions.

In section~\ref{sec:PH_GW}, we numerically investigated the phase-transition dynamics and the resulting GW signals. We find that the thermal evolution is governed primarily by the Higgs--triplet interactions controlled by $\lambda_1^\Omega$ and $\mu_1$. Accordingly, we scanned over $\lambda_1^\Omega$, $\mu_1/v_\Omega$, and $m_\eta$, while fixing the remaining input parameters to the same values as those adopted in Eq.~\eqref{eq:LFV-numvalue}, thereby maintaining consistency with the LFV analysis. Strong FOPTs are favored by relatively large values of both $\lambda_1^\Omega$ and $\mu_1/v_\Omega$, with most viable points satisfying $\lambda_1^\Omega\gtrsim\mu_1/v_\Omega$, as shown in figure~\ref{fig:lamfr-xcxn}. Figure~\ref{fig:ab-Af} displays the distribution of viable points in the $(\alpha,\beta/H)$ plane, where we find that larger $\alpha$ and smaller $\beta/H$ generally lead to stronger GW signals. The points satisfying $\xi_N>1$ populate the ranges $10^{-22}\lesssim\Omega_{\rm peak}h^2\lesssim10^{-9}$ and $10^{-5}~\mathrm{Hz}\lesssim f_{\rm peak}\lesssim 1~\mathrm{Hz}$. Comparing these predictions with the projected sensitivity curves in figure~\ref{fig:pkAf-dtt}, we find that a subset of the signals can be probed by LISA  in the frequency range $10^{-5}$--$10^{-3}~\mathrm{Hz}$. BBO has partial sensitivity to signals extending toward higher frequencies, while Ultimate DECIGO could probe a broader part of the predicted region over approximately $10^{-3}$--$10^{-1}~\mathrm{Hz}$.

Finally, the representative benchmark point summarized in table~\ref{tab:benchmark_point} demonstrates the simultaneous viability of the LFV, GW, and DM phenomenology. As shown in figure~\ref{fig:spGW}, it exhibits a two-step thermal history with a strong FOPT and produces a GW spectrum peaking at $f_{\rm peak}\simeq2.28\times10^{-4}~\mathrm{Hz}$ with $\Omega_{\rm peak}h^2\simeq1.18\times10^{-11}$, within the projected sensitivity of LISA. All three LFV observables satisfy the current experimental bounds. The predictions for $\mu\to e\gamma$, $\mu\to e\overline{e}e$ and coherent $\mu$--$e$ conversion in Ti lie far below the projected sensitivities of MEG~II, Mu3e, and PRIME, respectively. At the same parameter point, the scalar field $\eta_I$ constitutes a viable high-mass scalar DM candidate whose thermal relic abundance is compatible with the cosmological measurement and whose SI scattering cross section remains below the current direct-detection limits.

An important feature of the singlet--triplet scotogenic model is the central role of the lepton-number-violating coupling $\lambda_5$ in connecting neutrino
masses, LFV, and DM phenomenology. This coupling controls the radiative generation of neutrino masses and determines the mass
splitting between the inert neutral scalars $\eta_R$ and $\eta_I$. For fixed neutrino mass-squared differences and other relevant parameters, a larger $\lambda_5$ generally leads to smaller Yukawa couplings $Y_{N,i}$ and $Y_{\Sigma,i}$, thereby suppressing LFV rates. At the same time, it increases the mass splitting between the inert neutral scalars $\eta_R$ and $\eta_I$, suppressing $Z$-mediated inelastic DM scattering and helping to satisfy direct-detection constraints. Consequently, DM direct-detection constraints can indirectly restrict LFV predictions through
their common dependence on $\lambda_5$. 

In summary, the singlet--triplet scotogenic model provides a unified and testable framework connecting radiative neutrino masses, LFV, a strong electroweak FOPT, stochastic GWs, and scalar DM. Our combined analysis shows that these phenomena can be accommodated simultaneously in the singlet--triplet scotogenic model. The resulting predictions provide concrete targets for coordinated tests through MEG~II, Mu3e, PRIME, space-based GW observatories, and next-generation DM direct-detection experiments.

\section*{Acknowledgements}

This work is supported by the National Natural Science Foundation of China under Grant Nos. 12505133, 12375104, 12547106, 12335005, and Guizhou Provincial Major Scientific and Technological Program XKBF (2025)010. 

\newpage
\appendix
\section{Effective amplitudes and generic expressions for LFV observables\label{app:lfv-observables}}

This appendix is devoted to the conventions and analytical expressions for the Wilson coefficients associated with the LFV observables considered in the main text. We begin by reviewing the general effective amplitudes and decay rates for the radiative process $e_j\to e_i\gamma$ and the three-body decay $e_j\to e_i\overline e_i e_i$. We subsequently summarize the quark-level effective operators relevant to coherent $\mu$--$e$ conversion,  and employ them to calculate the corresponding CR. The general one-loop expressions for the Wilson coefficients of the penguin and box contributions mediated by scalar and fermion fields are derived in Appendices~\ref{app:general-penguins} and~\ref{app:general-box}, respectively; the corresponding loop functions are collected in Appendix~\ref{ap:loop-function}. 

\subsection{$e_j\to e_i\gamma$}

The amplitude of $e_j\to e_i\gamma$ compatible with both Lorentz symmetry and gauge symmetry can be generally parametrized as
\begin{eqnarray}
\mathcal{M}_{e_j \to e_i \gamma} 
&=& e\,\overline{u}_i(p_i) 
\big[\left( \gamma^\mu q^2 - q^\mu \slashed{q} \right)
\left( K_1^L P_L+K_1^R P_R\right) + i m_{e_j}\, \sigma^{\mu\nu} q_\nu
\left( K_2^L P_L + K_2^R P_R \right)\big]u_j(p_j) 
\epsilon_\mu^* \,,\label{eq:M-meg}\nonumber\\ 
\end{eqnarray}
where $e$ is the electric charge of the positron, $q=p_j-p_i$ is the photon momentum, $P_{L,R} = \frac{1\mp\gamma_5}{2}$ are the usual chirality projectors and $\epsilon_\mu$ is the polarization vector of the photon. $K_1^{L,R}$ and $K_2^{L,R}$ are the so called monopole and dipole Wilson coefficients, respectively. For the radiative decay $e_j\to e_i\gamma$, the photon is on shell so that only the dipole contributions are relevant, and the corresponding  decay width is given by \cite{Hisano:1995cp}
\begin{eqnarray}
\Gamma\left(e_j\to e_i\gamma\right)=\frac{\alpha m_{e_j}^5}{4}\left(\left|K_2^L\right|^2+\left|K_2^R\right|^2\right)\,.
\end{eqnarray}
Here $\alpha=e^2/4\pi$ is the fine structure constant.

\subsection{$e_j\to e_i\overline{e}_ie_i$}

The decay $e_j\to e_i\overline{e}_ie_i$ receives contributions from $\gamma$-penguin, $Z$-penguin, box, and Higgs-penguin diagrams, the corresponding amplitudes are denoted by $\mathcal{M}_{\gamma\text{-penguin}}$, $\mathcal{M}_{Z\text{-penguin}}$, $\mathcal{M}_{\text{box}}$, and $\mathcal{M}_{H\text{-penguin}}$, respectively.  As shown in Ref.~\cite{Hisano:1995cp},  the Higgs-penguin contribution can be combined with the box amplitude through definition of the Wilson coefficients. Therefore the total amplitude for the three-body decay $e_j\to e_i\overline{e}_ie_i$ can be written as
\begin{eqnarray}
\mathcal{M}_{\text{total}}=\mathcal{M}_{\gamma\text{-penguin}}+\mathcal{M}_{Z\text{-penguin}}+\mathcal{M}_{\text{box}}\,,
\end{eqnarray}
where the contribution of Higgs-penguin has been  absorbed in the last term $\mathcal{M}_{\text{box}}$. To be more specific, the $\gamma$-penguin contribution is given by
\begin{eqnarray}
        \mathcal{M}_{\gamma\text{-penguin}}&=&\frac{e^2}{q^2}\overline{u}_i(p_1)\left[\left( \gamma^\mu q^2 - q^\mu \slashed{q} \right)
        \left( K_1^L P_L + K_1^R P_R \right)
        + i m_{e_j}\, \sigma^{\mu\nu} q_\nu
        \left( K_2^L P_L + K_2^R P_R \right)\right]u_j(p)\nonumber\\
        &&\times\overline{u}_i(p_2)\gamma^\mu v_i(p_3)-\left(p_1\leftrightarrow p_2\right)\,.\label{eq:M-Apen}
\end{eqnarray}
The  amplitude from the $Z$-diagram can be generally parametrized as follows,
\begin{eqnarray}
        \mathcal{M}_{Z\text{-penguin}}&=&\frac{1}{m_Z^2}\left[\overline{u}_i(p_1)\gamma_\mu \left(F_LP_L+F_RP_R\right)u_j(p)\right]
        \left[\overline{u}_i(p_2)\gamma^\mu\left(Z_L^{(e)}P_L+Z_R^{(e)}P_R\right)v_i(p_3)\right]\nonumber\\
        &&-(p_1\leftrightarrow p_2)\,.\label{eq:M-Zpen}
\end{eqnarray}
where $F_{L,R}$ denote the $Z$-penguin Wilson coefficients. Here, $Z_L^{(e)}=\frac{g}{\cos\theta_W}\left(1/2-\sin^2\theta_W \right)$ and $Z_R^{(e)}=-\frac{g\sin^2\theta_W}{\cos\theta_W}$ denote the couplings of the $Z$ boson to the left- and right-handed charged leptons in the SM, respectively.

The amplitude of the box diagrams can be decomposed into the products of the vector, scalar, and tensor currents as,
\begin{eqnarray}
        \mathcal{M}_{\text{box}} &=&e^2B_1^L\left[\overline{u}_{e_i}(k_1)\gamma_\mu P_Lu_{e_j}(p)\right]\left[\overline{u}_{e_i}(k_3)\gamma^\mu P_Lv_{e_i}(k_2)\right]\nonumber\\
        &&+e^2B_1^R\left[\overline{u}_{e_i}(k_1)\gamma_\mu P_Ru_{e_j}(p)\right]\left[\overline{u}_{e_i}(k_3)\gamma^\mu P_Rv_{e_i}(k_2)\right]\nonumber\\
        &&+e^2B_2^L\left\{\left[\overline{u}_{e_i}(k_1)\gamma_\mu P_Lu_{e_j}(p)\right]\left[\overline{u}_{e_i}(k_3)\gamma^\mu P_Rv_{e_i}(k_2)\right]-(k_1\leftrightarrow k_3)\right\}\nonumber\\
        &&+e^2B_2^R\left\{\left[\overline{u}_{e_i}(k_1)\gamma_\mu P_Ru_{e_j}(p)\right]\left[\overline{u}_{e_i}(k_3)\gamma^\mu P_Lv_{e_i}(k_2)\right]-(k_1\leftrightarrow k_3)\right\}\nonumber\\
        &&+e^2B_3^L\left\{\left[\overline{u}_{e_i}(k_1)P_Lu_{e_j}(p)\right]\left[\overline{u}_{e_i}(k_3)P_Lv_{e_i}(k_2)\right]-(k_1\leftrightarrow k_3)\right\}\nonumber\\
        &&+e^2B_3^R\left\{\left[\overline{u}_{e_i}(k_1)P_Ru_{e_j}(p)\right]\left[\overline{u}_{e_i}(k_3)P_Rv_{e_i}(k_2)\right]-(k_1\leftrightarrow k_3)\right\}\nonumber\\
        &&+e^2B_4^L\left\{\left[\overline{u}_{e_i}(k_1)\sigma_{\mu\nu}P_Lu_{e_j}(p)\right]\left[\overline{u}_{e_i}(k_3)\sigma_{\mu\nu}P_Lv_{e_i}(k_2)\right]-(k_1\leftrightarrow k_3)\right\}\nonumber\\
        &&+e^2B_4^R\left\{\left[\overline{u}_{e_i}(k_1)\sigma_{\mu\nu}P_Ru_{e_j}(p)\right]\left[\overline{u}_{e_i}(k_3)\sigma_{\mu\nu}P_Rv_{e_i}(k_2)\right]-(k_1\leftrightarrow k_3)\right\}\,.\label{eq:M-box}
\end{eqnarray}
where $B_{1,2,3,4}^{L,R}$ denote the corresponding box Wilson coefficients. After integrating over the three-body phase space, the above amplitudes yield the following decay width~\cite{Arganda:2005ji}:  
\begin{eqnarray}
        \Gamma(e_j^-\rightarrow e_i^-e_i^-e_i^+)&=&\frac{e^4}{512\pi^3}m_{e_j}^5\Big[|K_1^L|^2+|K_1^R|^2-2(K_1^LK_2^{R*}+K_2^LK_1^{R*}+h.c.)\nonumber\\
        &&+(|K_2^L|^2+|K_2^R|^2)\left(\frac{16}{3}\ln\frac{m_{e_j}}{m_{e_i}}-\frac{22}{3}\right)\nonumber\\
        &&+\frac{1}{6}(|B_1^L|^2+|B_1^R|^2)+\frac{1}{3}(|B_2^L|^2+|B_2^R|^2)+\frac{1}{24}(|B_3^L|^2+|B_3^R|^2)\nonumber\\
        &&+6(|B_4^L|^2+|B_4^R|^2)-\frac{1}{2}(B_3^LB_4^{L*}+B_3^RB_4^{R*}+\text{h.c.})\nonumber\\
        &&+\frac{1}{3}(K_1^LB_1^{L*}+K_1^RB_1^{R*}+K_1^LB_2^{L*}+K_1^RB_2^{R*}+\text{h.c.})\nonumber\\
        &&-\frac{2}{3}(K_2^RB_1^{L*}+K_2^LB_1^{R*}+K_2^LB_2^{R*}+K_2^RB_2^{L*}+\text{h.c.})\nonumber\\
        &&+\frac{1}{3}\Big\{2(|F_{LL}|^2+|F_{RR}|^2)+|F_{LR}|^2+|F_{RL}|^2\nonumber\\
        &&+(B_1^LF_{LL}^*+B_1^RF_{RR}^*+B_2^LF_{LR}^*+B_2^RF_{RL}^*+\text{h.c.})\nonumber\\
        &&+2(K_1^LF_{LL}^*+K_1^RF_{RR}^*+h.c.)+(K_1^LF_{LR}^*+K_1^RF_{RL}^*+\text{h.c.})\nonumber\\
        &&-4(K_2^RF_{LL}^*+K_2^LF_{RR}^*+\text{h.c.})-2(K_2^LF_{RL}^*+K_2^RF_{LR}^*+\text{h.c.})\Big\}\Big]\label{eq:Gamma-mu3e1}\,,
\end{eqnarray}
where
\begin{eqnarray}
        F_{LL}=\frac{F_LZ_L^{(e)}}{g^2\sin^2\theta_Wm_Z^2}\,,~
    F_{LR}=\frac{F_LZ_R^{(e)}}{g^2\sin^2\theta_Wm_Z^2}\,,\quad 
F_{RR}=F_{LL}|_{L\leftrightarrow R}\,,~ F_{RL}=F_{LR}|_{L\leftrightarrow R}~~~\,.
\end{eqnarray}
The masses of the leptons in the final state, $m_{e_i}$, are neglected in Eq.~\eqref{eq:Gamma-mu3e1} \cite{Porod:2014xia}, with the exception of the dipole terms $K_2^{L,R}$, where retaining the lepton masses is necessary to regulate the IR divergence that would otherwise arise from the massless photon propagator.

\subsection{Coherent $\mu$--$e$ conversion in nuclei}

The coherent $\mu$--$e$ conversion process can be described by the quark-level effective operators in Eq.~\eqref{eq:STM-muecon}, which can be converted into an effective Lagrangian at the nucleon level by using the nucleon form factors. The momentum transfer in the $\mu$--$e$ conversion process is small so that the momentum dependence of the nucleon form factors is negligible. Then the resulting nucleon isoscalar and isovector coupling constants $g_{XK}^{(0)}$ and $g_{XK}^{(1)}$ are expressed in terms of the quark-level couplings $g_{XK(q)}$ as follows, 
\begin{eqnarray}
g_{XK}^{(0)}&=&\frac{1}{2}\sum_{q=u,d,s}\left(g_{XK(q)}G_K^{(q,p)}+g_{XK(q)}G_K^{(q,n)}\right)\,,\nonumber\\ 
g_{XK}^{(1)}&=&\frac{1}{2}\sum_{q=u,d,s}\left(g_{XK(q)}G_K^{(q,p)}-g_{XK(q)}G_K^{(q,n)}\right)\,,
\end{eqnarray}
where $X=L,R$ labels the lepton chirality and $K=S,V$ denotes the  Lorentz property of the quark bilinear. The numerical values of the relevant $G_K$ factors are given in Ref.~\cite{Bernabeu:1993ta},
\begin{align}
G_V^{(u,p)}&=G_V^{(d,n)}=2\,,&G_V^{(d,p)}&=G_V^{(u,n)}=1\,,\nonumber\\
G_S^{(u,p)}&=G_S^{(d,n)}=5.1\,,&G_S^{(d,p)}&=G_S^{(u,n)}=4.3\,,\nonumber\\
G_S^{(s,p)}&=G_S^{(s,n)}=2.5\,.&&
\end{align}
In the conventions of Refs.~\cite{Kuno:1999jp,Arganda:2007jw}, the CR normalized to the ordinary muon-capture rate is given by 
\begin{eqnarray}
        \text{CR}(\mu-e,\text{Nucleus})&=&\frac{p_eE_em_\mu^3G_F^2\alpha^3Z_{\text{eff}}^4F_p^2}{8\pi^2Z}\nonumber\\
        &&\times\Big\{\left|\left(Z+N\right)\left(g_{LV}^{(0)}+g_{LS}^{(0)}\right)+\left(Z-N\right)\left(g_{LV}^{(1)}+g_{LS}^{(1)}\right)\right|^2\nonumber\\
        &&\hspace{0.5cm}+\left|\left(Z+N\right)\left(g_{RV}^{(0)}+g_{RS}^{(0)}\right)+\left(Z-N\right)\left(g_{RV}^{(1)}+g_{RS}^{(1)}\right)\right|^2\Big\}\frac{1}{\Gamma_\text{capt}}\,,~~~~~
\end{eqnarray}  
where $p_e$ and $E_e$ denote the magnitude of the three-momentum and the energy of the emitted electron, respectively. In the numerical evaluation, we set $p_e\simeq E_e\simeq m_\mu$. Here, $Z$ and $N$ denote the proton and neutron numbers of the target nucleus, respectively, while $Z_{\mathrm{eff}}$ is the effective atomic charge obtained by averaging the bound-muon wave function over the nuclear density \cite{Chiang:1993xz}. The quantity $F_p$ denotes the  nuclear matrix element, and $\Gamma_{\text{capt}}$ is the total muon-capture rate. The values of $Z_{\mathrm{eff}}$, $F_p$ and $\Gamma_{\text{capt}}$  for the nuclei used in current experiments can be found in Ref.~\cite{Kitano:2002mt}. In the present work,  we are concerned with ${}^{48}_{22}\mathrm{Ti}$ as the target nucleus, with $Z_{\mathrm{eff}}=17.6$, $F_p=0.54$, and $\Gamma_{\text{capt}}=1.70422\times10^{-18}$~GeV, taken from Ref.~\cite{Kitano:2002mt}.

\section{General formulae for  the penguin diagrams of $e_j\to e_i\overline{e}_i e_i$\label{app:general-penguins}}

\begin{figure}[htbp]
\centering
\includegraphics[scale=0.7]{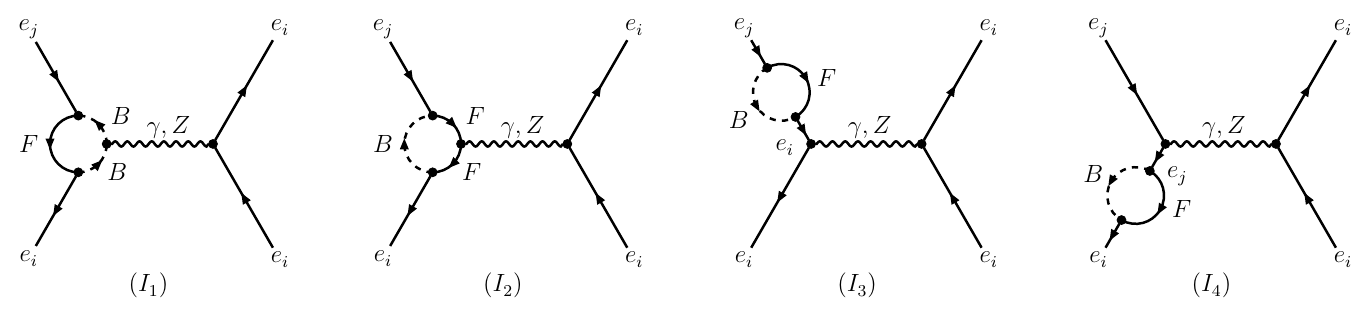}
\caption{ Generic scalar--fermion one-loop diagrams contributing to the LFV process $e_j\to e_i\overline{e}_i e_i$.  \label{fi:Apd-meg}}
\end{figure}

In this appendix, we shall derive the general expressions for the Wilson coefficients governing the $e_j\to e_i\overline{e}_i e_i$ process, which is induced by one-loop diagrams involving a scalar messenger $B$ and a fermion messenger $F$. The relevant diagrams are shown in figure~\ref{fi:Apd-meg}, where the photon or $Z$ boson are off shell.  Both  $(I_1)$ and $(I_2)$ are penguin diagrams that induce LFV through the one-loop vertex corrections, while $(I_3)$ and $(I_4)$ contribute to LFV through the external-leg self-energy corrections associated with the transition $e_j\rightarrow e_i$. 
Here, we neglect the contribution from the penguin diagram analogous to $(I_1)$, with $B$ and $F$ replaced by the $W$ boson and neutrino, respectively, since it is highly suppressed by the Glashow--Iliopoulos--Maiani (GIM) mechanism and the tiny neutrino masses~\cite{Cheng:1980tp}.  

Following the strategy of Ref.~\cite{Lavoura:2003xp}, we first specify the interactions between the new fields and SM fields, and subsequently introduce a set of basic one-loop integrals, and then  present the analytic expressions of the Wilson coefficients relevant to $e_j\to e_i\overline{e}_i e_i$.  In addition to verifying the one-loop expressions of $K_2^{L,R}$ given in Ref.~\cite{Lavoura:2003xp}, we provide the corresponding one-loop analytic expressions of $K_1^{L,R}$ and $F_{L,R}$.  We further verify that the Ward identity is satisfied and that all ultraviolet divergences cancel exactly, ensuring that the Wilson coefficients derived in this appendix are finite at one-loop order.
        
\subsection{Interaction Lagrangian}

Our model extends the SM particle content by two fermions, $N$ and $\Sigma$, and two scalars, $\eta$ and $\Omega$, while the SM gauge symmetry is left unmodified. Consequently the model contains no new gauge bosons. Here, following Ref.~\cite{Lavoura:2003xp}, we consider a BSM scalar $B$ with mass $m_B$ and a BSM fermion $F$ with mass $m_F$, which couple to the charged lepton $e_i$ of the $i$-th generation through the general chiral interaction
\begin{eqnarray}
        \mathcal{L}_{\text{Yukawa}} = \sum_{i=1}^3 \left[
        B \, \overline{F}\left(L_i P_L + R_i P_R\right)e_i
        + B^* \, \overline{e}_i\left(L_i^* P_R + R_i^* P_L\right)F
        \right]\,, \label{eq:ykw-def}
\end{eqnarray}
where $L_i$ and $R_i$ are complex dimensionless couplings, corresponding to the left- and right-handed components of the charged lepton $e_i$ as distinguished in Eq.~\eqref{eq:ykw-def}. At the low energy effective field theory stage, no assumption concerning the electroweak representations of $B$ and $F$ is required. For the transition from $e_j$ to $e_i$, it is convenient to define the coupling combinations

\begin{eqnarray}
\lambda &=& L_i^* L_j\,, ~~~  \rho = R_i^* R_j\,, ~~~
\zeta = L_i^* R_j\,,~~~ \upsilon= R_i^* L_j\,. \label{eq:coeff}
\end{eqnarray}
 In the  case where only a left-handed or right-handed interaction is present, the chirality-flipping combinations $\zeta$ and $\upsilon$ would vanish. Therefore, non-zero $\zeta$ and $\upsilon$ require both left- and right-chiral coupling structures to be present simultaneously. The interactions of $B$, $F$, and $e_i$ with the gauge bosons $Z_\mu$ and $A_\mu$ relevant for the penguin diagrams are given by

\begin{eqnarray}
\mathcal{L}_\text{gauge} &=& 
\sum_{i=1}^3 \overline{e}_i \gamma^\mu \left(Z_L^{(e)} P_L + Z_R^{(e)} P_R\right) e_i Z_\mu+ \overline{F} \gamma^\mu \left(Z_L^{(F)} P_L + Z_R^{(F)} P_R\right) F Z_\mu\nonumber\\
&&+ iZ_B \left(B^* \partial^\mu B - \partial^\mu B^* B\right) Z_\mu- e\sum_{i=1}^3 \overline{e}_i \gamma^\mu Q_e e_i A_\mu
- e\overline{F} \gamma^\mu Q_F F A_\mu \nonumber\\
&& - ieQ_B \left(B^* \partial^\mu B - \partial^\mu B^* B\right) A_\mu\,.\label{eq:BF-ZA}
\end{eqnarray}
The electric charges of the scalar $B$, the fermion $F$, and the charged lepton $e_i$ are denoted by $Q_B$, $Q_F$, and $Q_e$, respectively. Obviously $Q_e=-1$ for charged leptons, and electric-charge conservation of the Yukawa interactions in Eq.~\eqref{eq:ykw-def} then requires
\begin{eqnarray}
Q_B - Q_F + Q_e = 0\,. \label{eq:q-con}
\end{eqnarray}
The couplings of $B$, $F$, and $e_i$ to the $Z$ boson are denoted by $Z_B$, $Z_{L,R}^{(F)}$, and $Z_{L,R}^{(e)}$, respectively. In the convention $Q=T_3+Y$, they are given by
\begin{eqnarray}
\nonumber Z_B&=&-\frac{g}{\cos\theta_W}\left(\cos^2\theta_W Q_B - Y_B\right)\,,\\
\nonumber Z_{L,R}^{(F)}&=&-\frac{g}{\cos\theta_W}\left(\cos^2\theta_W Q_F - Y_{F_{L,R}}\right)\,,\\
Z_{L,R}^{(e)}&=&-\frac{g}{\cos\theta_W}\left(\cos^2\theta_W Q_e - Y_{e_{L,R}}\right)\,,\label{eq:ZLRe}
\end{eqnarray}
where $g$ is the $SU(2)_L$ gauge coupling, $\theta_W$ is the Weinberg angle, and $Y$ denotes hypercharge, with $Y_{F_{L,R}}$ and $Y_{e_{L,R}}$ corresponding to the hypercharges of the left- and right-handed fermions, respectively.

\subsection{Basic one-loop integrals\label{sec:BC}}

We employ dimensional regularization in $\mathbf{d}=4-2\epsilon$ dimensions and define
$1/\widetilde{\epsilon}\equiv1/\epsilon-\gamma_E+\log4\pi$, where $\gamma_E\approx 0.577$ is the Euler-Mascheroni constant. Throughout this appendix, we adopt the convention for loop integrals used in
Refs.~\cite{tHooft:1978jhc,Passarino:1978jh,Denner:1991kt}.
The $B$ functions denote the one-loop two-point integrals relevant for the self-energy calculations and are defined as
\begin{eqnarray}
\frac{i}{16\pi^2}B_0(p,m_1,m_2) &=& \mu^{4-\mathbf{d}} \int \frac{d^\mathbf{d}k}{(2\pi)^\mathbf{d}}\frac{1}{\left[(k-p)^2-m_1^2\right]\left(k^2-m_2^2\right)},\\
\frac{i}{16\pi^2}B_\mu(p,m_1,m_2)&=&\mu^{4-\mathbf{d}} \int \frac{d^\mathbf{d}k}{(2\pi)^\mathbf{d}}\frac{k_\mu}{\left[(k-p)^2-m_1^2\right]\left(k^2-m_2^2\right)}\,.
\end{eqnarray}
For the limit $p^2\ll \{m_1^2,m_2^2\}$, following Ref.~\cite{Goto:2026FormulaeSUSY}, we can expand $B_0(p,m_1,m_2)$ as
\begin{eqnarray}
B_0(p,m_1,m_2)=B_0(0,m_1,m_2)+\sum_{n=1}^\infty\frac{1}{n}\left(\frac{p^2}{m_2^2}\right)^nJ(n,n,n;z)\,,
\end{eqnarray}
where $z\equiv m_1^2/m_2^2$.
In the following, we  keep terms up to $\mathcal{O}((p^2/m_2^2)^2)$, corresponding to truncating the expansion at $n=2$. The analytic expressions for $B_0(0,m_1,m_2)$, $J(1,1,1;z)$, and $J(2,2,2;z)$ are
\begin{eqnarray}
B_0(0,m_1,m_2)&=&\frac{1}{\widetilde{\epsilon}}-\log\frac{m_2^2}{\mu^2}+1-\frac{z\ln z}{z-1}\,,\nonumber\\
J(1,1,1;z)&=&-\frac{z\ln z}{(z-1)^3}+\frac{z+1}{2(z-1)^2}\,,\nonumber\\ 
J(2,2,2;z)&=&-\frac{2z(z+1)\ln z}{(z-1)^5}+\frac{z^2+10z+1}{3(z-1)^4}\,.
\end{eqnarray}
The vector two-point integral $B_\mu(p,m_1,m_2)$ can be decomposed as
\begin{eqnarray}
B_\mu(p,m_1,m_2)=p_\mu B_1(p,m_1,m_2)\,.
\end{eqnarray}
The corresponding scalar coefficient $B_1(p,m_1,m_2)$ is expanded as
\begin{eqnarray}
  B_1(p,m_1,m_2)=B_1(0,m_1,m_2)+\sum_{n=1}^\infty\frac{1}{n}\left(\frac{p^2}{m_2^2}\right)^nJ(n,n,n+1;z)\,.
\end{eqnarray}
Similarly, with the expansion truncated at $n=2$, the relevant analytic expressions for $B_1(0,m_1,m_2)$, $J(1,1,2;z)$, and $J(2,2,3;z)$ are given by
\begin{eqnarray}
B_1(0,m_1,m_2)&=&\frac{1}{2}\left[\frac{1}{\widetilde{\epsilon}}-\ln\frac{m_2^2}{\mu^2}\right]-\frac{z(z-2)\ln z}{2(z-1)^2}+\frac{z-3}{4(z-1)}\,, \nonumber\\
J(1,1,2;z)&=&\frac{z\ln z}{(z-1)^4}+\frac{z^2-5z-2}{6(z-1)^3}\,, \nonumber\\ 
J(2,2,3;z)&=&\frac{z(3z+2)\ln z}{(z-1)^6}+\frac{z^3-11z^2-47z-3}{12(z-1)^5}\,.
\end{eqnarray}
The $C$ functions correspond to the one-loop three-point integrals and are defined as~\cite{tHooft:1978jhc,Passarino:1978jh,Denner:1991kt}

\begin{eqnarray}
\frac{i}{16\pi^2}C_0(p,p',m_1,m_2,m_3)&=& \mu^{4-\mathbf{d}} \int \frac{d^\mathbf{d}k}{(2\pi)^\mathbf{d}}\frac{1}{\left(k^2-m_1^2\right)\left[(k-p)^2-m_2^2\right]\left[(k-p')^2-m_3^2\right]}\,,\nonumber\\ 
\frac{i}{16\pi^2}C_\mu(p,p',m_1,m_2,m_3)&=& \mu^{4-\mathbf{d}} \int \frac{d^\mathbf{d}k}{(2\pi)^\mathbf{d}}\frac{k_\mu}{\left(k^2-m_1^2\right)\left[(k-p)^2-m_2^2\right]\left[(k-p')^2-m_3^2\right]}\,,\nonumber\\ 
\frac{i}{16\pi^2}C_{\mu\nu}(p,p',m_1,m_2,m_3)&=& \mu^{4-\mathbf{d}} \int \frac{d^\mathbf{d}k}{(2\pi)^\mathbf{d}}\frac{k_\mu k_\nu}{\left(k^2-m_1^2\right)\left[(k-p)^2-m_2^2\right]\left[(k-p')^2-m_3^2\right]}\,.\nonumber\\ 
\end{eqnarray}
In the concerned limit $\{p^2, p'^2, p\cdot p'\}\ll \{m^2_1, m^2_2=m^2_3\}$,  the scalar three-point function $C_0(p,p',m_1,m_2,m_2)$ can be expanded as
\begin{eqnarray}
C_0(p,p',m_1,m_2,m_2)&=&-\frac{1}{m_2^2}\left[C_{01}(z)+\frac{p^2+p'^2}{6m_2^2}C_{02}(z)+\frac{p\cdot p'}{6m_2^2}C_{03}(z)\right]+\mathcal{O}\left(\frac{p^4}{m_2^4}\right)\,,~~~
\end{eqnarray}
where
\begin{eqnarray}
C_{01}(z)&=&\frac{1}{1-z}+\frac{z}{(1-z)^2}\ln z\,,\nonumber\\ 
C_{02}(z)&=&\frac{2+5z-z^2}{(1-z)^3}+\frac{6z}{(1-z)^4}\ln z\,,\nonumber \\ 
C_{03}(z)&=&\frac{-1+5z+2z^2}{(1-z)^3}+\frac{6z^2}{(1-z)^4}\ln z\,.
\end{eqnarray}
The three-point integral $C_\mu$ can be decomposed as
\begin{eqnarray}
C_{\mu}(p,p',m_1,m_2,m_3)=\left[p_\mu C_{11}+p'_\mu C_{12}\right](p,p',m_1,m_2,m_3)\,.
\end{eqnarray}
In the limit $\{p^2, p'^2, p\cdot p'\}\ll \{m^2_1, m^2_2=m^2_3\}$, the expressions of $C_{11}$ and $C_{12}$ are given by
\begin{eqnarray}
C_{11}(p,p',m_1,m_2,m_2)&=& -\frac{1}{m_2^2}\Bigg\{\frac{1-3z}{4(1-z)^2}-\frac{z^2\ln z}{2(1-z)^3}+\frac{p^2}{m_2^2}\left[\frac{1-7z-7z^2+z^3}{12(1-z)^4}-\frac{z^2\ln z}{(1-z)^5}\right]\nonumber\\ 
&&+\frac{p'^2}{m_2^2}\left[\frac{2-13z-4z^2+3z^3}{36(1-z)^4}-\frac{(3z^2-z^3)\ln z}{6(1-z)^5}\right]\nonumber\\ 
&&+\frac{p\cdot p'}{m_2^2}\left[\frac{-1+5z-13z^2-3z^3}{18(1-z)^4}-\frac{2z^3\ln z}{3(1-z)^5}\right]+\mathcal{O}\left(\frac{p^4}{m_2^4}\right)\Bigg\}\,,\nonumber\\ 
C_{12}(p,p',m_1,m_2,m_2)&=&C_{11}(p',p,m_1,m_2,m_2)\,.
\end{eqnarray}
Furthermore, the three-point integral $C_{\mu\nu}$ can be decomposed as
\begin{eqnarray}
C_{\mu\nu}(p,p',m_1,m_2,m_3)=\left[p_\mu p_\nu C_{21}+p'_\mu p'_\nu C_{22}+\left(p_\mu p'_\nu+p'_\mu p_\nu\right)C_{23}+g_{\mu\nu} C_{24}\right](p,p',m_1,m_2,m_3)\,,\nonumber\\ 
\end{eqnarray}
In the limit $\{p^2, p'^2, p\cdot p'\}\ll \{m^2_1, m^2_2=m^2_3\}$, the coefficients $C_{21}$, $C_{22}$, $C_{23}$, and $C_{24}$ are given by
\begin{eqnarray}
C_{21}&=&C_{22}=2C_{23}=\frac{1}{18m_2^2}\left[\frac{-2+7z-11z^2}{(1-z)^3}-\frac{6z^3}{(1-z)^4}\ln z+\mathcal{O}\left(\frac{p^2}{m_2^2}\right)\right]\,,\nonumber\\ 
C_{24}&=&\frac{1}{4}\left(\frac{1}{\widetilde{\epsilon}}-\log\frac{m_2^2}{\mu^2}\right)+\frac{1-3z}{8(1-z)}-\frac{z^2}{4(1-z)^2}\ln z\nonumber\\ 
&&+\frac{p^2+p'^2}{72m_2^2}\left[\frac{5-22z+5z^2}{(1-z)^3}-\frac{18z^2-6z^3}{(1-z)^4}\ln z\right]\nonumber \\ 
&&-\frac{p\cdot p'}{36m_2^2}\left[\frac{2-7z+11z^2}{(1-z)^3}+\frac{6z^3}{(1-z)^4}\ln z\right]+\mathcal{O}\left(\frac{p^4}{m_2^4}\right)\,.
\end{eqnarray}

\subsection{Expressions of Wilson coefficients\label{st:caldia}}

The Wilson coefficients $K_1^{L,R}$, $K_2^{L,R}$, and $F_{L,R}$ are determined by calculating the one-loop Feynman diagrams shown in figure~\ref{fi:Apd-meg}. Using the relevant masses, the loop functions defined in Sec.~\ref{sec:BC}, and the coefficient combinations in Eq.~\eqref{eq:coeff}, we derive their analytic expressions below.

\subsubsection{$\gamma$-penguin diagrams}

The effective amplitude in Eq.~\eqref{eq:M-Apen} arising from the $\gamma$-penguin diagrams is expressed in terms of the monopole coefficients $K_1^{L,R}$ and the dipole coefficients $K_2^{L,R}$. The monopole coefficients $K_1^{L,R}$ receive contributions from the diagrams $(I_1)$--$(I_4)$ shown in figure~\ref{fi:Apd-meg}. Denoting the corresponding contributions from the diagram $(I_1)$, the diagram $(I_2)$, and the diagrams $(I_3)$ and $(I_4)$ by $\overline C_{L,R}$, $C_{L,R}$, and $C'_{L,R}$, respectively, we obtain
\begin{eqnarray}
  q^2K_1^L&\equiv&Q_B\overline{C}_L+Q_FC_L+\frac{Q_e}{m_{e_j}^2-m_{e_i}^2}C'_L\,,\label{eq:K1L-def}\nonumber\\
  q^2K_1^R&\equiv&Q_B\overline{C}_R+Q_FC_R+\frac{Q_e}{m_{e_j}^2-m_{e_i}^2}C'_R\,,\label{eq:K1R-def}
\end{eqnarray}
where
\begin{eqnarray}
                \overline{C}_L&=&\big\{\lambda\left[m_{e_j}^2\left(C_{21}+C_{23}-C_{11}\right)+m_{e_i}^2\left(C_{22}+C_{23}-C_{12}\right)+2C_{24}\right]\nonumber\\ 
    &&+\rho m_{e_j}m_{e_i}\left(C_{21}+C_{22}+2C_{23}-C_{11}-C_{12}\right)\nonumber\\ 
    &&+\zeta m_{e_j}m_F\left(C_{11}+C_{12}-C_0\right)+\upsilon m_{e_i}m_F\left(C_{11}+C_{12}-C_0\right)\big\}(p_j,p_i,m_F,m_B,m_B)\,,\label{eq:bC-gammaL}\nonumber\\
    C_L&=&\big\{\lambda\left[m_F^2C_0+q^2C_{23}-(\mathbf{d}-2)C_{24}\right]\nonumber\\ 
    &&+\rho m_{e_j}m_{e_i}\left(C_0+C_{21}+C_{22}+2C_{23}-2C_{11}-2C_{12}\right)\nonumber\\ 
    &&+\zeta m_{e_j}m_F\left(C_0-C_{11}-C_{12}\right)+\upsilon m_{e_i}m_F\left(C_0-C_{11}-C_{12}\right)\big\}(p_j,p_i,m_B,m_F,m_F)\,,\label{eq:C-gammaL}\nonumber\\
        C'_L&=&\left[\lambda m_j^2 B_1+\rho m_jm_i B_1+\zeta m_jm_F B_0+\upsilon m_i m_F B_0\right](p_j,m_B,m_F)\nonumber\\ 
    &&-\left[\lambda m_i^2 B_1+\rho m_jm_i B_1+\zeta m_jm_F B_0+\upsilon m_i m_F B_0\right](p_i,m_B,m_F)\,,\label{eq:C'-gammaL}\nonumber\\
                \overline{C}_R&=&\overline{C}_L|_{\lambda\leftrightarrow\rho,\zeta\leftrightarrow\upsilon}\,,\qquad C_R=C_L|_{\lambda\leftrightarrow\rho,\zeta\leftrightarrow\upsilon}\,,\qquad C'_R=C'_L|_{\lambda\leftrightarrow\rho,\zeta\leftrightarrow\upsilon}\label{eq:C'-gammaR}\,.
\end{eqnarray}
For the process $e_j\to e_i\overline{e}_ie_i$, the incoming and outgoing charged leptons are on shell, with $p_j^2=m_{e_j}^2$ and $p_i^2=m_{e_i}^2$.  The gauge-boson momentum is given by $q=p_j-p_i$, with $q^2$ of the same order as $m_{e_j}^2$. A hierarchy of scales is assumed, $m_{e_j}^2,m_{e_i}^2,q^2\ll m_B^2,m_F^2$. Substituting the analytic expressions for the loop functions given in section~\ref{sec:BC} into Eq.~\eqref{eq:C'-gammaR} and retaining the leading order, the analytic expressions for $K_1^{L,R}$ are obtained.

It is worth noting that the monopole coefficients $K_1^{L,R}$ enter the amplitude through $q^2\gamma^\mu-\slashed{q}q^\mu$, as required by the Ward identity. Upon expanding the right-hand side of Eqs.~\eqref{eq:K1L-def}, the terms involving the charged-lepton masses cancel at leading order, leaving only terms proportional to $q^2$. The resulting leading-order expressions for $K_1^{L,R}$ are given by
\begin{eqnarray}
        K_1^L&\simeq&\frac{\lambda}{96\pi^2m_B^2}\left[Q_BF_1(t)-Q_FG_1(t)\right]\,,\label{eq:K1L-res}\nonumber\\
        K_1^R&\simeq&\frac{\rho}{96\pi^2m_B^2}\left[Q_BF_1(t)-Q_FG_1(t)\right]\,,\label{eq:K1R-res}
\end{eqnarray}
where the exact cancellation of the ultraviolet divergences is ensured by the charge conservation condition in Eq.~\eqref{eq:q-con}. We note that the leading non-vanishing contributions to $K_1^{L,R}$ are proportional to $\lambda$ and $\rho$, respectively. The loop functions $F_1(t)$ and $G_1(t)$ are defined in Eqs.~\eqref{eq:F1} and \eqref{eq:G1}, respectively. The dipole coefficients $K_2^{L,R}$ receive contributions only from the diagrams $(I_1)$ and $(I_2)$ shown in figure~\ref{fi:Apd-meg}, and are therefore given by
\begin{eqnarray}
        m_{e_j}K_2^L&\equiv&Q_B\overline{S}_L+Q_FS_L\,,\nonumber\\
        m_{e_j}K_2^R&\equiv&Q_B\overline{S}_R+Q_FS_R\,,
\end{eqnarray}
where
\begin{eqnarray}
                \overline{S}_L&=&\Big[m_F\upsilon(C_{11}+C_{12}-C_0)+m_{e_j}\rho(C_{21}+C_{23}-C_{11})\nonumber\\ &&+m_{e_i}\lambda(C_{22}+C_{23}-C_{12})\Big](p_j,p_i,m_F,m_B,m_B)\,,\nonumber\\ 
        S_L&=&\Big[-m_F\upsilon\left(C_{11}+C_{12}\right)+m_{e_j}\rho\left(C_{21}+C_{23}-C_{11}\right)\nonumber\\ &&+m_{e_i}\lambda\left(C_{22}+C_{23}-C_{12}\right)\Big](p_j,p_i,m_B,m_F,m_F)\,,\nonumber\\ 
        \overline{S}_R&=&\overline{S}_L|_{\lambda\leftrightarrow\rho,\zeta\leftrightarrow\upsilon}\,,\qquad S_R=S_L|_{\lambda\leftrightarrow\rho,\zeta\leftrightarrow\upsilon}\,.\label{eq:CsgmR}
\end{eqnarray}
For the dipole coefficients $K_2^{L,R}$, we retain the expansion up to the subleading order, keeping terms proportional to the charged-lepton masses $m_{e_j}$ or $m_{e_i}$, which are given by
\begin{eqnarray}
        m_{e_j}K_2^L&\simeq&\frac{1}{32\pi^2m_B^2}\Big[m_F\upsilon\left(Q_FF_3(t)+Q_BG_3(t)\right)\nonumber\\
        &&\qquad+\left(m_{e_j}\rho+m_{e_i}\lambda\right)\left(Q_FG_2(t)+Q_BF_2(t)\right)\Big]\,,\label{eq:K2L-res}\nonumber\\
        m_{e_j}K_2^R&\simeq&\frac{1}{32\pi^2m_B^2}\Big[m_F\zeta\left(Q_FF_3(t)+Q_BG_3(t)\right)\nonumber\\
        &&\qquad+\left(m_{e_j}\lambda+m_{e_i}\rho\right)\left(Q_FG_2(t)+Q_BF_2(t)\right)\Big]\,.\label{eq:K2R-res}
\end{eqnarray}
As a consistency check, Eq.~\eqref{eq:K2R-res} reproduces the general expression for the dipole form factor given in Ref.~\cite{Lavoura:2003xp}. The leading-order contributions to $m_{e_j}K_2^{L,R}$ are proportional to $\upsilon$ and $\zeta$, respectively, while the subleading terms proportional to the charged-lepton masses involve $\lambda$ and $\rho$. The loop functions appearing in these expressions, $F_2(t)$, $F_3(t)$,  $G_2(t)$ and $G_3(t)$, are defined in Eqs.~\eqref{eq:F2}, \eqref{eq:F3}, \eqref{eq:G2}, and \eqref{eq:G3}.

\subsubsection{$Z$-penguin diagrams}

For the $Z$-penguin diagrams, the left- and right-handed components of the $F$ may couple differently to the $Z$ boson, as shown in Eq.~\eqref{eq:BF-ZA}. It is therefore useful to keep the contributions proportional to $Z_B$, $Z_{L,R}^{(F)}$, and $Z_{L,R}^{(e)}$ separately. The form factors $F_{L,R}$ receive contributions from diagrams $(I_1)$--$(I_4)$ shown in figure~\ref{fi:Apd-meg}. Specifically, the term proportional to $Z_B$ originates from $(I_1)$, those proportional to $Z_{L,R}^{(F)}$ arise from $(I_2)$, and the terms proportional to $Z_{L,R}^{(e)}$ correspond to the contributions $(I_3)$ and $(I_4)$. They can then be expressed as
\begin{eqnarray}
-F_L&=&Z_B\overline{N}_L+Z_L^{(F)}N_L+Z_R^{(F)}\widetilde{N}_L+\frac{Z_L^{(e)}}{m_{e_j}^2-m_{e_i}^2}N'_L,\label{eq:FL-def}\nonumber\\
-F_R&=&Z_B\overline{N}_R+Z_L^{(F)}N_R+Z_R^{(F)}\widetilde{N}_R+\frac{Z_R^{(e)}}{m_{e_j}^2-m_{e_i}^2}N'_R\,,\label{eq:FR-def}
\end{eqnarray}
where
\begin{eqnarray}
\overline{N}_L&=&-\overline{C}_L\,,\qquad N'_L=-C'_L\,,\nonumber\\
N_L&=&\big[-\lambda m_F^2C_0+\rho m_{e_j}m_{e_i}\left(-C_0+2C_{11}+2C_{12}-C_{21}-C_{22}-2C_{23}\right)\nonumber\\ 
&&+\zeta m_{e_j}m_F\left(-C_0+C_{11}+C_{12}\right)+\upsilon m_{e_i}m_F\left(-C_0+C_{11}+C_{12}\right)\big](p_j,p_i,m_B,m_F,m_F)\,,\nonumber\\
\widetilde{N}_L&=&\lambda\left[(\mathbf{d}-2)C_{24}-q^2C_{23}\right](p_j,p_i,m_B,m_F,m_F)\,,\nonumber\\
\overline{N}_R&=&\overline{N}_L\big|_{\lambda\leftrightarrow\rho,\zeta\leftrightarrow\upsilon}\,,\quad N_R=\widetilde{N}_L\big|_{\lambda\leftrightarrow\rho,\zeta\leftrightarrow\upsilon}\,,\quad\widetilde{N}_R=N_L\big|_{\lambda\leftrightarrow\rho,\zeta\leftrightarrow\upsilon}\,,\quad N'_R=N'_L\big|_{\lambda\leftrightarrow\rho,\zeta\leftrightarrow\upsilon}\,.\label{eq:FLR-N}
\end{eqnarray}
Substituting the leading-order expressions of the two-point and three-point one-loop functions given in section~\ref{sec:BC} into Eq.~\eqref{eq:FR-def}, we obtain the following expressions for $F_{L,R}$:
\begin{eqnarray}
F_L&\simeq&\frac{Z_B}{64\pi^2}\Bigg[\lambda F_5\left(t\right)+\frac{2(\zeta m_{e_j}+\upsilon m_{e_i})}{m_B}\sqrt{t}G_3\left(t\right)+\frac{4\rho m_{e_j}m_{e_i}}{m_B^2}F_2\left(t\right)\Bigg]\nonumber\\
&&+\frac{Z_L^{(F)}}{64\pi^2}\Bigg[4\lambda G_5\left(t\right)-\frac{2(\zeta m_{e_j}+\upsilon m_{e_i})}{m_B}\sqrt{t}G_3\left(t\right)-\frac{4\rho m_{e_j}m_{e_i}}{m_B^2}F_2\left(t\right)\Bigg]+\frac{Z_R^{(F)}}{64\pi^2}\lambda G_6\left(t\right)\nonumber\\
&&+\frac{Z_L^{(e)}}{64\pi^2}\Bigg[\lambda F_5\left(t\right)+\frac{2(\zeta m_{e_j}+\upsilon m_{e_i})}{m_B}\sqrt{t}G_3\left(t\right)+\frac{4\rho m_{e_j}m_{e_i}}{m_B^2}F_2\left(t\right)\Bigg]\,,\nonumber\\
F_R&=&F_L\big|_{\lambda\leftrightarrow\rho,\zeta\leftrightarrow\upsilon,L\leftrightarrow R}\,.\label{eq:FL-res}
\end{eqnarray}
The cancellation of the ultraviolet divergences in $F_L$ and $F_R$ relies on the charge-conservation conditions $Z_{e_L}+Z_B-Z_{F_R}=0$ and $Z_{e_R}+Z_B-Z_{F_L}=0$, respectively, which are ensured by the invariance of the interactions under the $U(1)_Q$ and $U(1)_Y$ gauge symmetries.  

If the neutral complex scalar is decomposed as $B=\frac{1}{\sqrt{2}}(B_R+iB_I)$, where $B_R$ and $B_I$ are the CP-even and CP-odd mass eigenstates with masses $m_R$ and $m_I$, respectively, both states contribute to the loop amplitudes and should therefore be included. For non-degenerate masses, $m_R\neq m_I$, the corresponding expression is modified to
\begin{eqnarray}
F_L&\simeq&\frac{Z_B}{64\pi^2}\Bigg\{\lambda\Big[3+F_{13}\left(t_R,t_I\right)\Big]+\frac{8(\zeta m_{e_j}+\upsilon m_{e_i})}{m_F}G_{14}\left(t_R,t_I\right)-\frac{2\rho m_{e_j}m_{e_i}}{m_F^2}F_{14}\left(t_R,t_I\right)\Bigg\}\nonumber\\
&&+\Bigg\{\frac{Z_L^{(F)}}{64\pi^2}\Bigg[2\lambda G_5\left(t_R\right)-\frac{\zeta m_{e_j}+\upsilon m_{e_i}}{m_F}t_RG_3\left(t_R\right)-\frac{2\rho m_{e_j}m_{e_i}}{m_F^2}t_RF_2\left(t_R\right)\Bigg]+\frac{Z_R^{(F)}}{128\pi^2}\lambda G_6\left(t_R\right)\nonumber\\
&&+\frac{Z_L^{(e)}}{128\pi^2}\Bigg[\lambda F_5\left(t_R\right)+\frac{2(\zeta m_{e_j}+\upsilon m_{e_i})}{m_F}t_RG_3\left(t_R\right)+\frac{4\rho m_{e_j}m_{e_i}}{m_F^2}t_RF_2(t_R)\Bigg]+\left(t_R\leftrightarrow t_I\right)\Bigg\}\,,\nonumber\\
F_R&=&F_L\big|_{\lambda\leftrightarrow\rho,\zeta\leftrightarrow\upsilon,L\leftrightarrow R}\,,\label{eq:FL-nd-res}
\end{eqnarray}
where $t_R\equiv m_F^2/m_R^2$ and $t_I\equiv m_F^2/m_I^2$. These expressions reduce continuously to the degenerate result when $m_R=m_I$ and are directly applicable to models with a split neutral-scalar sector. The relevant loop functions in Eqs.~(\ref{eq:FL-res}, \ref{eq:FL-nd-res}) are defined in section~\ref{ap:loop-function}.

The general formulae derived above can be directly applied to the concerned scotogenic model. The loop contributions to the $e_j\to e_i\overline{e}_ie_i$ process arise from the scalar-fermion pairs $(B, F)=(\eta^{+}, \chi_{\alpha})$ or $(B, F)=(\eta_{R, I}, \Sigma^{-})$, as can be seen from figures~\ref{fi:Apenguin} and \ref{fi:Zpenguin}. Because the Yukawa interactions only couple to left-handed charged leptons, the right-handed coefficients in Eq.~\eqref{eq:ykw-def} vanish and only $L_i$ remains non-zero. For the one-loop diagrams with  $(B, F)=(\eta^{+}, \chi_{\alpha})$,  we have $L_i=Y_{\chi,i\alpha}^*$ and $t=\zeta_\alpha$ defined in Eq.~\eqref{eq:zeta-rho-def}. For  the cases of $(B, F)=(\eta_{R}, \Sigma^{-})$ and $(B, F)=(\eta_{I}, \Sigma^{-})$, we have $L_i=Y_{\Sigma,i}/\sqrt{2}$, $t=\rho_{R}$ and $L_i=i{Y_\Sigma,i}/\sqrt{2}$, $t=\rho_{I}$, respectively. The Wilson coefficients $K_2^R$, $K_1^L$, and $F_{L,R}$ then follow from Eqs.~\eqref{eq:K1L-res}, \eqref{eq:K2R-res}, and \eqref{eq:FL-nd-res}, and are explicitly presented in Eqs.~\eqref{eq:K2R-exp}, \eqref{eq:K1L-exp}, \eqref{eq:FL}, and \eqref{eq:FR}. 

\section{General formulae for the box diagrams of $e_j\to e_i\overline{e}_i e_i$\label{app:general-box}}

\begin{figure}[htbp]
\centering
\includegraphics[width=1\textwidth]{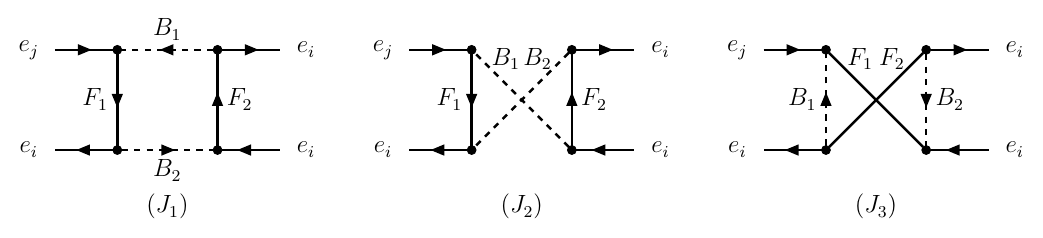}
\caption{One-loop box Feynman diagrams contributing to the LFV process $e_j\to e_i\overline{e}_i e_i$.}
\label{fi:Apd_box}  
\end{figure}

In addition to the penguin diagrams, box diagrams generally contribute to the decay $e_j\to e_i\overline{e}_i e_i$. The three relevant one-loop box diagrams are shown in figure~\ref{fi:Apd_box}, where $B_{1,2}$ denote new scalar fields with masses $m_{B_{1,2}}$, and $F_{1,2}$ denote new fermions with masses $m_{F_{1,2}}$. Diagram $(J_1)$ represents the generic box contribution. If $B_1$ and $B_2$ are chargeless and real scalar fields, diagram $(J_2)$ provides an additional contribution. Similarly, if $F_1$ and $F_2$ are Majorana fermions, diagram $(J_3)$ must be included.  Following the discussion in section~\ref{app:general-penguins}, we neglect the contributions from box diagrams involving $W$ bosons and neutrinos, as these contributions are suppressed by the small neutrino mass-squared differences relative to $m_W^2$~\cite{Petcov:1976ff,Hernandez-Tome:2018fbq}.

The relevant Yukawa interactions can be written in the general form
\begin{eqnarray}
\mathcal{L}_{\text{Yukawa}} = \sum_{i=1}^3\sum_{a,b=1}^2\left[B_b\, \overline{F}_a\left(L_i^{ab} P_L + R_i^{ab} P_R\right)e_i+ B_b^* \, \overline{e}_i\left(L_i^{ab*} P_R + R_i^{ab*} P_L\right)F_a
\right]\,, \label{eq:ykwbox-def}
\end{eqnarray}
where $a$ and $b$ label the fermion and scalar fields, respectively, and $L_i^{ab}$ and $R_i^{ab}$ are arbitrary dimensionless coefficients.

For convenience, we define the following combinations of coupling coefficients:
\begin{align}
\lambda_1&=L_j^{11}L_i^{12*}\,,&\lambda_2&=L_i^{21}L_i^{22*}\,,&\rho_1&=R_j^{11}R_i^{12*}\,,&\rho_2&=R_i^{21}R_i^{22*}\,,\nonumber\\ 
\zeta_1&=R_j^{11}L_i^{12*}\,,&\zeta_2&=R_i^{21}L_i^{22*}\,,&\upsilon_1&=L_j^{11}R_i^{12*}\,,&\upsilon_2&=L_i^{21}R_i^{22*}\,,\nonumber\\
\lambda'_1&=L_j^{11}L_i^{12}\,,&\lambda'_2&=L_i^{21*}L_i^{22*}\,,&\rho'_1&=R_j^{11}R_i^{12}\,,&\rho'_2&=R_i^{21*}R_i^{22*}\,,\nonumber\\ 
\zeta'_1&=R_j^{11}L_i^{12}\,,&\zeta'_2&=R_i^{21*}L_i^{22*}\,,&\upsilon'_1&=L_j^{11}R_i^{12}\,,&\upsilon'_2&=L_i^{21*}R_i^{22*}\,.
\end{align}
Following the convention for four-point loop integrals in Ref.~\cite{Denner:1991kt}, we define the four-point functions $D_0$, $D_\mu$, and $D_{\mu\nu}$ as
\begin{eqnarray}
\frac{i}{16\pi^2}D_0(p_1,p_2,p_3,m_0,m_1,m_2,m_3)&=& \mu^{4-\mathbf{d}} \int \frac{d^\mathbf{d}k}{(2\pi)^\mathbf{d}}\frac{1}{\Delta_0\Delta_1\Delta_2\Delta_3}\,,\nonumber\\ 
\frac{i}{16\pi^2}D_\mu(p_1,p_2,p_3,m_0,m_1,m_2,m_3)&=& \mu^{4-\mathbf{d}} \int \frac{d^\mathbf{d}k}{(2\pi)^\mathbf{d}}\frac{k_\mu}{\Delta_0\Delta_1\Delta_2\Delta_3}\,,\nonumber\\ 
\frac{i}{16\pi^2}D_{\mu\nu}(p_1,p_2,p_3,m_0,m_1,m_2,m_3)&=& \mu^{4-\mathbf{d}} \int \frac{d^\mathbf{d}k}{(2\pi)^\mathbf{d}}\frac{k_\mu k_\nu}{\Delta_0\Delta_1\Delta_2\Delta_3}\,,
\end{eqnarray}
where
\begin{eqnarray}
\Delta_0=k^2-m_0^2\,,\qquad \Delta_i=(k+p_i)^2-m_i^2\,,\qquad i=1,\cdots,3\,.
\end{eqnarray}
The explicit Lorentz decompositions of the integrals $D_0$, $D_\mu$ and $D_{\mu\nu}$ are given by
\begin{eqnarray}
D_\mu&=&\sum_{i=1}^3p_{i\mu}D_i\,,\nonumber\\ 
D_{\mu\nu}&=&g_{\mu\nu}D_{00}+\sum_{i,j=1}^3p_{i\mu}p_{j\nu}D_{ij}\,,
\end{eqnarray}
with $D_{ji}=D_{ij}$. The masses and momenta of the external particles are neglected in the evaluation of $\mathcal{M}_{\mathrm{box}}$, which considerably simplifies the calculation~\cite{Petcov:1976ff}. As a result, only the four-point loop functions $D_0$ and $D_{00}$ contribute. Applying the Fierz identities, the resulting amplitudes can be cast into the general form given in Eq.~\eqref{eq:M-box}. The corresponding Wilson coefficients from diagram $(J_1)$ are
\begin{eqnarray}
e^2B_1^{(J_1)L}&=&\frac{\lambda_1\lambda_2^*}{8\pi^2}D_{00}(0,0,0,m_{F_1},m_{B_1},m_{B_2},m_{F_2})\,,\nonumber\\ 
e^2B_1^{(J_1)R}&=&\frac{\rho_1\rho_2^*}{8\pi^2}D_{00}(0,0,0,m_{F_1},m_{B_1},m_{B_2},m_{F_2})\,,\nonumber\\ 
e^2B_2^{(J_1)L}&=&\frac{1}{16\pi^2}\left[\lambda_1\rho_2^*D_{00}
-\frac{1}{2}\upsilon_1\upsilon_2^*m_{F_1}m_{F_2}D_0\right](0,0,0,m_{F_1},m_{B_1},m_{B_2},m_{F_2})\,,\nonumber\\ 
e^2B_2^{(J_1)R}&=&\frac{1}{16\pi^2}\left[\rho_1\lambda_2^*D_{00}
-\frac{1}{2}\zeta_1\zeta_2^*m_{F_1}m_{F_2}D_0\right](0,0,0,m_{F_1},m_{B_1},m_{B_2},m_{F_2})\,,\nonumber\\ 
e^2B_3^{(J_1)L}&=&\frac{\upsilon_1\zeta_2^*}{16\pi^2}m_{F_1}m_{F_2}D_0(0,0,0,m_{F_1},m_{B_1},m_{B_2},m_{F_2})\,,\nonumber\\ 
e^2B_3^{(J_1)R}&=&\frac{\zeta_1\upsilon_2^*}{16\pi^2}m_{F_1}m_{F_2}D_0(0,0,0,m_{F_1},m_{B_1},m_{B_2},m_{F_2})\,,\nonumber\\ 
e^2B_4^{(J_1)L}&=&0\,,\nonumber\\ 
e^2B_4^{(J_1)R}&=&0\,.\label{eq:box-J1}
\end{eqnarray}

For diagram $(J_2)$, the corresponding contributions are
\begin{eqnarray}
e^2B_1^{(J_2)L}&=&-\frac{\lambda_1\lambda_2}{8\pi^2}D_{00}(0,0,0,m_{F_1},m_{B_1},m_{B_2},m_{F_2})\,,\nonumber\\ 
e^2B_1^{(J_2)R}&=&-\frac{\rho_1\rho_2}{8\pi^2}D_{00}(0,0,0,m_{F_1},m_{B_1},m_{B_2},m_{F_2})\,,\nonumber\\ 
e^2B_2^{(J_2)L}&=&-\frac{1}{16\pi^2}\left[\lambda_1\rho_2D_{00}
+\frac{1}{2}\zeta_2\upsilon_1m_{F_1}m_{F_2}D_0\right](0,0,0,m_{F_1},m_{B_1},m_{B_2},m_{F_2})\,,\nonumber\\ 
e^2B_2^{(J_2)R}&=&-\frac{1}{16\pi^2}\left[\lambda_2\rho_1D_{00}
+\frac{1}{2}\zeta_1\upsilon_2m_{F_1}m_{F_2}D_0\right](0,0,0,m_{F_1},m_{B_1},m_{B_2},m_{F_2})\,,\nonumber\\ 
e^2B_3^{(J_2)L}&=&\frac{\upsilon_1\upsilon_2}{16\pi^2}m_{F_1}m_{F_2}D_0(0,0,0,m_{F_1},m_{B_1},m_{B_2},m_{F_2})\,,\nonumber\\ 
e^2B_3^{(J_2)R}&=&\frac{\zeta_1\zeta_2}{16\pi^2}m_{F_1}m_{F_2}D_0(0,0,0,m_{F_1},m_{B_1},m_{B_2},m_{F_2})\,,\nonumber\\ 
e^2B_4^{(J_2)L}&=&0\,,\nonumber\\ 
e^2B_4^{(J_2)R}&=&0\,.\label{eq:box-J2}
\end{eqnarray}

The $(J_3)$ contributions read
\begin{eqnarray}
e^2B_1^{(J_3)L}&=&\frac{\lambda'_1\lambda'_2}{16\pi^2}m_{F_1}m_{F_2}D_0(0,0,0,m_{F_1},m_{B_1},m_{B_2},m_{F_2})\,,\nonumber\\ 
e^2B_1^{(J_3)R}&=&\frac{\rho'_1\rho'_2}{16\pi^2}m_{F_1}m_{F_2}D_0(0,0,0,m_{F_1},m_{B_1},m_{B_2},m_{F_2})\,,\nonumber\\ 
e^2B_2^{(J_3)L}&=&\frac{1}{16\pi^2}\left[\zeta'_2\upsilon'_1
-\upsilon'_1\upsilon'_2\right]D_{00}(0,0,0,m_{F_1},m_{B_1},m_{B_2},m_{F_2})\,,\nonumber\\ 
e^2B_2^{(J_3)R}&=&\frac{1}{16\pi^2}\left[\zeta'_1\upsilon'_2
-\zeta'_1\zeta'_2\right]D_{00}(0,0,0,m_{F_1},m_{B_1},m_{B_2},m_{F_2})\,,\nonumber\\ 
e^2B_3^{(J_3)L}&=&-\frac{\lambda'_1\rho'_2}{32\pi^2}m_{F_1}m_{F_2}D_0(0,0,0,m_{F_1},m_{B_1},m_{B_2},m_{F_2})\,,\nonumber\\ 
e^2B_3^{(J_3)R}&=&-\frac{\lambda'_2\rho'_1}{32\pi^2}m_{F_1}m_{F_2}D_0(0,0,0,m_{F_1},m_{B_1},m_{B_2},m_{F_2})\,,\nonumber\\ 
e^2B_4^{(J_3)L}&=&\frac{\lambda'_1\rho'_2}{128\pi^2}m_{F_1}m_{F_2}D_0(0,0,0,m_{F_1},m_{B_1},m_{B_2},m_{F_2})\,,\nonumber\\ 
e^2B_4^{(J_3)R}&=&\frac{\lambda'_2\rho'_1}{128\pi^2}m_{F_1}m_{F_2}D_0(0,0,0,m_{F_1},m_{B_1},m_{B_2},m_{F_2})\,.\label{eq:box-J3}
\end{eqnarray}

The corresponding expressions for $D_0$ and $D_{00}$ are given by
\begin{align}
&D_0(0,0,0,m_{F_1},m_{B_1},m_{B_2},m_{F_2})
=-\frac{m_{B_1}^2\ln\!\left(\dfrac{m_{B_1}^2}{m_{B_2}^2}\right)}
{(m_{B_1}^2-m_{B_2}^2)(m_{B_1}^2-m_{F_1}^2)(m_{B_1}^2-m_{F_2}^2)}
\nonumber\\
&\quad+\frac{m_{F_1}^2\ln\!\left(\dfrac{m_{B_2}^2}{m_{F_1}^2}\right)}
{(m_{F_1}^2-m_{B_1}^2)(m_{F_1}^2-m_{B_2}^2)(m_{F_1}^2-m_{F_2}^2)}
+\frac{m_{F_2}^2\ln\!\left(\dfrac{m_{B_2}^2}{m_{F_2}^2}\right)}
{(m_{F_2}^2-m_{B_1}^2)(m_{F_2}^2-m_{B_2}^2)(m_{F_2}^2-m_{F_1}^2)}\,,
\label{eq:D0-zero-momentum}\\[2mm]
&D_{00}(0,0,0,m_{F_1},m_{B_1},m_{B_2},m_{F_2})
=\frac{m_{F_2}^4\ln\!\left(\dfrac{m_{B_2}^2}{m_{F_2}^2}\right)}
{4(m_{F_2}^2-m_{B_1}^2)(m_{F_2}^2-m_{B_2}^2)(m_{F_2}^2-m_{F_1}^2)}
\nonumber\\
&\quad+\frac{m_{F_1}^4\ln\!\left(\dfrac{m_{B_2}^2}{m_{F_1}^2}\right)}
{4(m_{F_1}^2-m_{B_1}^2)(m_{F_1}^2-m_{B_2}^2)(m_{F_1}^2-m_{F_2}^2)}
-\frac{m_{B_1}^4\ln\!\left(\dfrac{m_{B_1}^2}{m_{B_2}^2}\right)}
{4(m_{B_1}^2-m_{B_2}^2)(m_{B_1}^2-m_{F_1}^2)(m_{B_1}^2-m_{F_2}^2)}\,.
\label{eq:D00-zero-momentum}
\end{align}

The general formulae derived above are readily applicable to the scotogenic model under consideration. Analogous to the $\gamma$- and $Z$-penguin diagrams discussed in section~\ref{app:general-penguins}, the box contributions to the process $e_j\to e_i\overline{e}_i e_i$ arise from two types of scalar and fermion mediator combinations: $(B_{1,2},F_{1,2})=(\eta^+,\chi_{\alpha,\beta})$  and $(B_{1,2},F_{1,2})=(\eta_{\alpha,\beta},\Sigma^-)$ , as shown in figure~\ref{fi:box}. 
Since the Yukawa interactions couple exclusively to left-handed charged leptons, the right-handed coefficients $R_i^{ab}$ defined in Eq.~\eqref{eq:ykwbox-def} vanish, whereas the left-handed coefficients $L_i^{ab}$ may be nonzero. It follows that the only potentially nonvanishing coefficients are $B_1^{(J_1)L}$, $B_1^{(J_2)L}$, and $B_1^{(J_3)L}$, as can be seen from Eqs.~(\ref{eq:box-J1}, \ref{eq:box-J2}, \ref{eq:box-J3}). 

For the case of $(B_{1,2},F_{1,2})=(\eta^+,\chi_{\alpha,\beta})$, we have $L_i^{a1}=L_i^{a2}=Y_\chi^{ia*}$ with $a=\alpha, \beta$. Thus, only diagrams $(J_1)$ and $(J_3)$ contribute to $B_1^L$, and all other Wilson coefficients vanish. For $(B_{1,2},F_{1,2})=(\eta_{\alpha,\beta},\Sigma^-)$, the corresponding couplings are $L_i^{1R}=L_i^{2R}=Y_\Sigma^i/\sqrt{2}$ and $L_i^{1I}=L_i^{2I}=iY_\Sigma^i/\sqrt{2}$, respectively. In this case, only diagrams $(J_1)$ and $(J_2)$ contribute to $B_1^L$.  Applying Eqs.~\eqref{eq:box-J1}--\eqref{eq:box-J3} to these mediator combinations gives the Wilson coefficient $B_1^L$, whose explicit form is presented in Eq.~\eqref{eq:B1L-exp}.

\section{Loop functions\label{ap:loop-function}}

In the following, we present the expressions for the one-loop functions entering the LFV amplitudes. The variables $x$ and $y$ denote ratios of the squared masses of the messenger fields.  These functions are used in the calculations presented in section~\ref{sec:LFV} and Appendix~\ref{app:general-penguins}. 

\subsection{Functions of one mass ratio}

The relevant one-loop functions involving a single squared mass ratio $x$ are given by
\begin{eqnarray}
F_1(x)&=&\frac{2-9x+18x^2-11x^3+6x^3\ln x}{6(1-x)^4}\,,\label{eq:F1}\\
F_2(x)&=&\frac{1-6x+3x^2+2x^3-6x^2\ln x}{6(1-x)^4}\,,\label{eq:F2}\\
F_3(x)&=&\frac{-3+4x-x^2-2\ln x}{(1-x)^3}\,,\label{eq:F3}\\
F_4(x)&=&\frac{-1+x-x^2\ln x}{(1-x)^2}\,,\label{eq:F4}\\
F_5(x)&=&\frac{1-4x+3x^2-2x^2\ln x}{(1-x)^2}\,,\label{eq:F5}\\
F_6(x)&=&\frac{5x-4x^2-x^3+x(4x+2)\ln x}{6(1-x)^4}\,,\label{eq:F6}\\
F_7(x)&=&\frac{-1+4x-3x^2+(-2+4x)\ln x}{(1-x)^2}\,,\label{eq:F7}\\
G_1(x)&=&\frac{-16+45x-36x^2+7x^3+6(-2+3x)\ln x}{6(1-x)^4}\,,\label{eq:G1}\\
G_2(x)&=&\frac{2+3x-6x^2+x^3+6x\ln x}{6(1-x)^4}\,,\label{eq:G2}\\
G_3(x)&=&\frac{1-x^2+2x\ln x}{(1-x)^3}\,,\label{eq:G3}\\
G_4(x)&=&\frac{x-x^2+\ln x}{(1-x)^2}\,,\label{eq:G4}\\
G_5(x)&=&\frac{x-x^2+x\ln x}{(1-x)^2}\,,\label{eq:G5}\\
G_6(x)&=&\frac{-1+x^2+2x(x-2)\ln x}{(1-x)^2}\,,\label{eq:G6}\\
G_7(x)&=&\frac{1-x^2+2\ln x}{(1-x)^2}\,.\label{eq:G7}
\end{eqnarray}

\subsection{Functions of two mass ratios}

When two independent internal mass ratios are involved, we introduce the following symmetric two-variable functions,
\begin{eqnarray}
F_{13}(x,y)&=&\frac{2x\ln y}{(1-y)(x-y)}-\ln y+(x\leftrightarrow y)\,, \label{eq:F13}\\
F_{14}(x,y)&=&\frac{xy}{3}\left[\frac{1-xy}{(1-x)^2(1-y)^2}+\frac{(3x-1)\ln x}{(1-x)^3(x-y)}+(x\leftrightarrow y)\right]\,, \label{eq:F14}\\
F_{15}(x,y)&=&xy\left(-\frac{x}{\left(x-y\right)^2\left(1-y\right)}-\frac{xy\left(3x-y-2\right)\ln x}{\left(x-1\right)^2\left(x-y\right)^3}+\left(x\leftrightarrow y\right)\right)\,, \\
G_{12}(x,y)&=&\frac{\sqrt{xy}}{2}\Big[\frac{1}{(1-x)(1-y)}+\frac{x\ln x}{(1-x)^2(x-y)}+\frac{y\ln y}{(1-y)^2(y-x)}\Big]\,, \label{eq:G12}\\
G_{13}(x,y)&=&\frac{1}{4}\Big[\frac{1}{(1-x)(1-y)}+\frac{x^2\ln x}{(1-x)^2(x-y)}+\frac{y^2\ln y}{(1-y)^2(y-x)}\Big]\,,\label{eq:G13}\\
G_{14}(x,y)&=&\frac{xy}{4}\Big[\frac{1}{(1-x)(1-y)}+\frac{\ln x}{(1-x)^2(x-y)}+\frac{\ln y}{(1-y)^2(y-x)}\Big]\,. \label{eq:G14}
\end{eqnarray}
  

\providecommand{\href}[2]{#2}\begingroup\raggedright\endgroup

\end{document}